\documentclass[nofootinbib,superscriptaddress,aps,preprint,prc]{revtex4}
\usepackage[utf8]{inputenc}

\usepackage{amsmath}
\usepackage{mathrsfs}
\usepackage{amssymb}
\usepackage{bbm}
\usepackage{graphicx}
\usepackage{graphics}
\usepackage[small,compact]{titlesec}
\usepackage{xspace}
\usepackage{colordvi}
\usepackage{color}
\usepackage{units}
\usepackage{stackrel}
\usepackage{hyperref}
\usepackage{slashed}

\usepackage{ulem}

\renewcommand{\emph}[1]{\textit{#1}}

\newcommand{\vect}[1]{\vec{\mathbf{#1}}}
\newcommand{\vectS}[1]{\vec{\boldsymbol{#1}}}

\newcommand{\EFT}{$\mathrm{EFT}(\slashed{\pi})$\xspace}

\newcommand{\comment}[1]{}
\newcommand{\srutTypeOne}[1]{\vrule width0pt height0pt depth #1\relax}
\newcommand{\jjvHe}{{}^3\mathrm{He}}
\newcommand{\jjvH}{{}^3\mathrm{H}}
\newcommand{\nnlo}{NNLO\xspace}

\newcommand{\id}{\boldsymbol{1}}

\newcommand{\G}{\mathcal{G}}
\newcommand{\Gb}{\boldsymbol{\mathcal{G}}}
\newcommand{\Dwd}{\widetilde{D}}
\newcommand{\NN}{$N\!N$\xspace}

\makeatletter

\newcommand{\Rmnum}[1]{\expandafter\@slowromancap\romannumeral #1@}
\newcommand{\vast}{\bBigg@{4}}
\newcommand{\Vast}{\bBigg@{5}}
\makeatother

\begin{document}

\title{The Triton Charge Radius to Next-to-next-to-leading order in Pionless Effective Field Theory}

\author{Jared Vanasse}
\email{jjv9@phy.duke.edu}
\email{vanasse@ohio.edu}
\affiliation{Department of Physics, Duke University, Durham, NC 27708, USA
}
\affiliation{Department of Physics and Astronomy Ohio University, Athens OH 45701, USA
}

\date{\today}

\begin{abstract}
The triton point charge radius is calculated to next-to-next-to-leading order (\nnlo) in pionless effective field theory (\EFT), yielding a prediction  of $1.14\pm0.19$~fm (leading order), $1.59\pm0.08$~fm (next-to leading order), and $1.62\pm0.03$~fm (\nnlo) in agreement with the current experimental extraction of $1.5978\pm0.040$~fm~\cite{Angeli201369}.  The error at \nnlo is due to cutoff variation ($\sim$ 1\%) within a reasonable range of calculated cutoffs and from a \EFT error estimate ($\sim$ 1.5\%).  In addition new techniques are introduced to add perturbative corrections to bound and scattering state calculations for short range effective field theories, but with a focus on their use in \EFT. 
\end{abstract}

\keywords{latex-community, revtex4, aps, papers}

\maketitle

\section{Introduction}

If a system is probed at length scales, $\ell$, larger than the range of the underlying interaction, $r$, then its interactions can be expanded in a series of contact interactions known as short-range effective field theory (EFT)~\cite{vanKolck:1998bw}, and its applicability to any system for which $\ell>r$ is known as universality~\cite{Braaten:2004rn}.  Short range EFT has been used in cold atom systems, halo nuclei using halo EFT, and for low-energy few-body nuclear systems using pionless EFT (\EFT).  For all of these systems the scattering length, $a$, is unnaturally large ($a\gg r$).\footnote{Note, for nuclear systems the scattering length is fixed, but for cold atom systems the scattering length can be made large by tuning a magnetic field near a Feshbach resonance.}  Thus at leading order (LO) the scattering length contribution is treated nonperturbatively, and higher order range corrections ($[r/a]^{n}$) are added perturbatively~\cite{Kaplan:1998tg,Kaplan:1998we}.

Nucleon-nucleon (\NN) interactions are dominated by one pion exchange at large length scales.  Thus for length scales $\ell>1/m_{\pi}$ (or energies $E<m_{\pi}^{2}/M_{N}$) \NN interactions can be expanded in a short-range EFT known as \EFT.  The series of contact interactions in \EFT can be written down as a Lagrangian of nucleon terms and possible external currents.  These terms are ordered by the power counting of \EFT~\cite{Kaplan:1998tg,Kaplan:1998we,vanKolck:1998bw} which has the expansion  $(1/(M_{N}Q))(Q/\Lambda_{\slashed{\pi}})^{n}$, where $(Q/\Lambda_{\slashed{\pi}})\!\sim\!1/3$, $\Lambda_{\slashed{\pi}}\sim m_{\pi}$, $Q\sim\gamma_{t}$, $n\geq 0$, and $\gamma_{t}\approx 45$~MeV is the deuteron binding momentum.\footnote{In the two-body sector the factor of $1/(M_{N}Q)$ only occurs for two-body resonant $S$-wave interactions, which are a leading contribution in the three-body sector.  However, for higher two-body partial waves the factor of $1/(M_{N}Q)$ will not occur and $n\geq 1$ in the power counting since these partial waves are not resonant for physical systems in \EFT.}  In addition to making \EFT tractable (one only needs a finite number of terms to a given order) the power counting also allows for an estimation of the error in calculations.

LO \EFT has two low energy constants (LECs) in the two-body sector fit to the ${}^{3}\!S_{1}$ and ${}^{1}\!S_{0}$ bound and virtual bound state poles respectively, and one three-body LEC fit to a three-body datum.  At next-to-leading order (NLO) there are two more LECs in the two-body sector fit to the effective ranges in the ${}^{3}\!S_{1}$ and ${}^{1}\!S_{0}$ channels. Next-to-next-to-leading order (NNLO) has a two-body LEC parametrizing the mixing between the \NN ${}^{3}\!S_{1}$ and ${}^{3}\!D_{1}$ channels and an energy dependent three-body LEC~\cite{Bedaque:2002yg}.  Thus to NNLO in \EFT two- and three-body systems are characterized by seven LECs and predict observables to roughly $6\%$ accuracy.  However, certain observables, such as the neutron-deuteron ($nd$) polarization observable $A_{y}$, are sensitive to higher order interactions and are three orders of magnitude smaller than experiment at NNLO, which is the first order at which $A_{y}$ is non-zero.  The $A_{y}$ observable is sensitive to two-body $P$-wave contact interactions that occur at N$^{3}$LO~\cite{Margaryan:2015rzg}.

\EFT (see e.g. Ref.~\cite{Beane:2000fx} for a review) has been used with great success in the two-body sector calculating deuteron electromagnetic form factors~\cite{Chen:1999tn,Ando:2004mm}, \NN scattering~\cite{Chen:1999tn,Kong:1999sf,Ando:2007fh}, neutron-proton ($np$) capture~\cite{Chen:1999tn,Chen:1999bg,Ando:2004mm} to ($<\sim$1\%) \cite{Rupak:1999rk}, proton-proton fusion~\cite{Kong:2000px,Ando:2008va,Chen:2012hm}, and neutrino deuteron scattering~\cite{Butler:2000zp}.  Progress has also been made in the three-body sector with calculations of $nd$ scattering~\cite{Bedaque:1998mb,Bedaque:1999ve,Gabbiani:1999yv,Bedaque:2002yg,Griesshammer:2004pe,Vanasse:2013sda,Margaryan:2015rzg}, $pd$ scattering~\cite{Rupak:2001ci,Konig:2011yq,Vanasse:2014kxa,Konig:2014ufa}, $nd$ capture~\cite{Sadeghi:2006fc,Arani:2014qsa}, and the energy difference between $\jjvH$ and $\jjvHe$~\cite{Ando:2010wq,Konig:2011yq,Konig:2015aka}.  Previous three-body calculations of $nd$ scattering in \EFT made use of the partial resummation technique~\cite{Bedaque:2002yg}.  This method has the advantage of being able to calculate diagrams that contain full off-shell scattering amplitudes without needing to calculate the full off-shell scattering amplitude.  However, this method suffers the drawback that it contains an infinite subset of higher order diagrams and although correct up to the order one is working is not strictly perturbative.  This work was improved upon in Ref.~\cite{Vanasse:2013sda} where a new technique no more numerically complicated than the partial resummation technique but strictly perturbative was introduced.  This technique makes higher order strictly perturbative numerical calculations in $nd$ scattering much simpler~\cite{Margaryan:2015rzg}.  However, this method initially suffered the drawback that it could not be used to calculate perturbative corrections to three-body bound-state systems such as the triton.  This work corrects that drawback.  Using the new perturbative method developed here for bound states I will show that the triton charge radius has excellent agreement with experiment at \nnlo in \EFT.

Hagen \emph{et al.}~\cite{Hagen:2013xga} calculated the point charge radius of halo nuclei to LO in halo EFT and introduced the concept of a trimer field to calculate vertex functions for bound-state calculations.  Building on that work a technique similar to Hagen \emph{et al.}~is introduced, but one that can also calculate perturbative corrections to three-body bound states.  This technique introduces a triton auxiliary field and thus treats three-body forces in the doublet $S$-wave channel differently, but analytically equivalent to previous approaches to all orders~\cite{Vanasse:2013sda}.  In addition it is shown how this technique improves the calculation of the LO three-body force by removing the need for iterative numerical schemes.  One can also now calculate the \nnlo energy dependent three-body force without the need for a numerical limiting procedure~\cite{Ji:2012nj}. The new technique also leads to slight numerical simplifications in the calculation of $nd$ scattering.

Using this new technique for perturbative corrections to bound states the calculation of the triton charge form factor to \nnlo and the resulting point charge radius for the triton are considered.  The charge form factor of the triton is reproduced well by potential model calculations (PMC)~\cite{Schiavilla:1990zz} including chiral EFT ($\chi$EFT)~\cite{Piarulli:2012bn} potentials which give diffraction minima at the correct values of $Q^{2}$.  From experimental data the triton point charge radius has been extracted, most recently with a value of $1.5978\pm0.040$~fm~\cite{Angeli201369}.  A \nnlo \EFT calculation of the triton point charge radius is accurate to roughly 1.5\%.  However, as I will show cutoff variation gives an additional source of error leading to an overall error estimate of 2\%.  This cutoff variation is either a signal of slow divergence or convergence.  Either a careful asymptotic analysis or a numerical calculation to higher cutoffs will be needed to answer this unambiguously.  However, reliable calculations to very large cutoffs ($\Lambda>10^{6}$~MeV) are currently unfeasible, due to numerical instabilities.


This paper is organized as follows.  In Sec.~\ref{sec:twobody} properties of the two-body system in \EFT necessary for three-body calculations are reviewed. Sec.~\ref{sec:threebody} introduces new techniques for $nd$ scattering, the connection between the auxiliary triton and non-auxiliary triton field approach for three-body forces, and the calculation of perturbative corrections to the triton vertex function.  In Sec.~\ref{sec:doubleSwave} it is shown how the triton auxiliary field is used to calculate three-body forces in the doublet $S$-wave channel.  Discussion of the calculation of the triton charge form factor to \nnlo is given in Sec.~\ref{sec:chargeform}, results are shown in Sec.~\ref{sec:results}, and conclusions given in Sec.~\ref{sec:conclusions}.


\section{\label{sec:twobody} Two-Body System}

The two-body Lagrangian in \EFT is 
\begin{align}
\mathcal{L}_{2}=\ &\hat{N}^{\dagger}\left(i\partial_{0}+\frac{\vect{\nabla}^2}{2M_{N}}\right)\hat{N}+\,\hat{t}_{i}^{\dagger}\left[\Delta_{t}-c_{0t}\left(i\partial_{0}+\frac{\vect{\nabla}^{2}}{4M_{N}}+\frac{\gamma_{t}^{2}}{M_{N}}\right)\right]\hat{t}_{i}\\\nonumber
&+\hat{s}_{a}^{\dagger}\left[\Delta_{s}-c_{0s}\left(i\partial_{0}+\frac{\vect{\nabla}^{2}}{4M_{N}}+\frac{\gamma_{s}^{2}}{M_{N}}\right)\right]\hat{s}_{a}\\\nonumber
&+y_{t}\left[\hat{t}_{i}^{\dagger}\hat{N} ^{T}P_{i}\hat{N} +\mathrm{H.c.}\right]+y_{s}\left[\hat{s}_{a}^{\dagger}\hat{N}^{T}\bar{P}_{a}\hat{N}+\mathrm{H.c.}\right],
\end{align}
where $\hat{t}_{i}$ ($\hat{s}_{a}$) is the spin-triplet iso-singlet (spin-singlet iso-triplet) dibaryon auxiliary field.  The projector $P_{i}=\frac{1}{\sqrt{8}}\sigma_{2}\sigma_{i}\tau_{2}$ ($\bar{P}_{a}=\frac{1}{\sqrt{8}}\tau_{2}\tau_{a}\sigma_{2}$) projects out the spin-triplet iso-singlet (spin-singlet iso-triplet) combination of nucleons.

\begin{figure}[hbt]
\includegraphics[width=110mm]{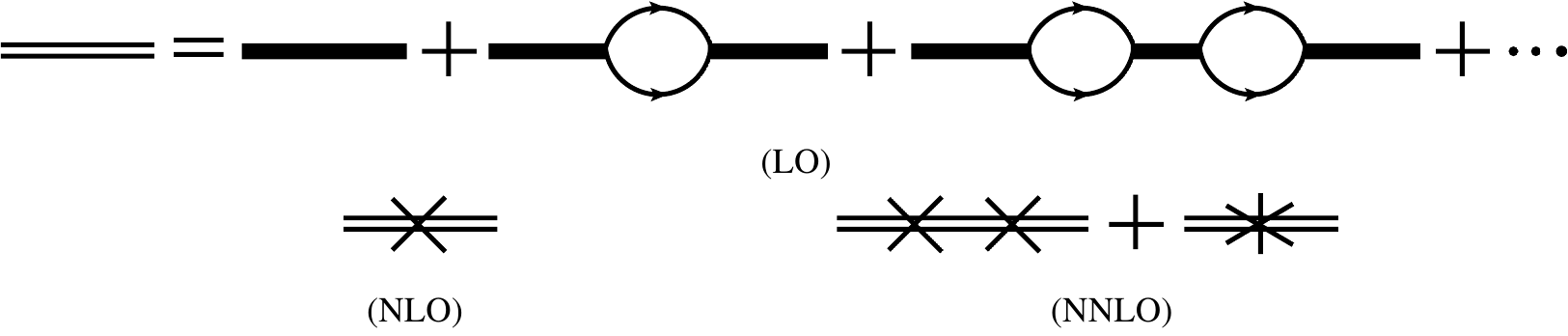}
\caption{The top equation shows the LO dressed spin-triplet dibaryon propagator, which can be solved analytically via a geometric series.  The solid bar is the bare dibaryon propagator $i/\Delta_{t}$, the single lines with arrows are nucleon propagators, the cross represents a NLO effective range insertion from $c_{0t}^{(0)}$, and the star a \nnlo correction from $c_{0t}^{(1)}$.\label{fig:DeutProp}}
\end{figure}

At LO the bare deuteron propagator, $i/\Delta_{t}$, is dressed by the infinite sum of bubble diagrams in Fig.~\ref{fig:DeutProp}.
The parameters are then fit to reproduce the deuteron pole at the physical position.  At NLO the parameters are chosen to fix the deuteron pole at the same position and give the correct residue about the deuteron pole.  This parametrization is known as the $Z$-parametrization~\cite{Phillips:1999hh,Griesshammer:2004pe} and is advantageous because it reproduces the correct residue about the deuteron pole at NLO instead of being approached perturbatively order-by-order as in the effective range expansion (ERE) parametrization.  The same procedure is carried out in the ${}^{1}\!S_{0}$ channel except the virtual bound-state pole and its residue is fit to.  Carrying out this procedure the coefficients are given by~\cite{Griesshammer:2004pe}
\begin{align}
&y_{t}^{2}=\frac{4\pi}{M_{N}},\quad \Delta_{t}=\gamma_{t}-\mu,\quad c_{0t}^{(n)}=(-1)^{n}(Z_{t}-1)^{n+1}\frac{M_{N}}{2\gamma_{t}},\\\nonumber
&y_{s}^{2}=\frac{4\pi}{M_{N}},\quad \Delta_{s}=\gamma_{s}-\mu,\quad c_{0s}^{(n)}=(-1)^{n}(Z_{s}-1)^{n+1}\frac{M_{N}}{2\gamma_{s}},
\end{align}
where $\gamma_{t}=45.7025$~MeV is the deuteron binding momentum, $Z_{t}=1.6908$ is the residue about the deuteron pole, $\gamma_{s}=-7.890$~MeV is the ${}^{1}\!S_{0}$ virtual bound-state momentum, and $Z_{s}=0.9015$ is the residue about the ${}^{1}\!S_{0}$ pole \cite{deSwart:1995ui}.  The non-physical scale $\mu$ is introduced by using dimensional regularization with the power divergence subtraction scheme~\cite{Kaplan:1998tg,Kaplan:1998we}.  All physical observables are $\mu$-independent.

After fitting the coefficients, the spin-triplet and spin-singlet dibaryon propagators up to and including \nnlo are given by
\begin{align}
\label{eq:dib}
&iD_{\{t,s\}}^{\mathrm{NNLO}}(p_{0},\vect{p})=\frac{i}{\gamma_{\{t,s\}}-\sqrt{\frac{\vect{p}^{2}}{4}-M_{N}p_{0}-i\epsilon}}\\\nonumber
&\times\left[\underbrace{\srutTypeOne{.5cm} 1}_{\mathrm{LO}}+\underbrace{\frac{Z_{\{t,s\}}-1}{2\gamma_{\{t,s\}}}\left(\gamma_{\{t,s\}}+\sqrt{\frac{\vect{p}^{2}}{4}-M_{N}p_{0}-i\epsilon}\,\right)}_{\mathrm{NLO}}\right.\\\nonumber
&\left.\hspace{.5cm}+\underbrace{\left(\frac{Z_{\{t,s\}}-1}{2\gamma_{\{t,s\}}}\right)^{2}\left(\frac{\vect{p}^{2}}{4}-M_{N}p_{0}-\gamma_{\{t,s\}}^{2}\right)}_{\mathrm{NNLO}}+\cdots\right].\\\nonumber
\end{align}
The deuteron wavefunction renormalization is given by the residue about the deuteron pole of the spin-triplet dibaryon, which to \nnlo yields
\begin{equation}
\label{eq:ZD}
Z_{D}=\frac{2\gamma_{t}}{M_{N}}\left[\underbrace{\srutTypeOne{.1cm}1}_{\mathrm{LO}}+\underbrace{(Z_{t}-1)}_{\mathrm{NLO}}+\underbrace{\srutTypeOne{.1cm}0}_{\mathrm{NNLO}}+\cdots\right].
\end{equation}
In the formalism used here higher-order corrections to the deuteron wavefunction renormalization will be built into the integral equation and do not need to be added separately.  The LO deuteron wavefunction renormalization is defined by
\begin{equation}
Z_{\mathrm{LO}}=\frac{2\gamma_{t}}{M_{N}}.
\end{equation}
%


\section{\label{sec:threebody}Three-Body System}

\subsection{Doublet Channel Scattering}
\begin{figure}[hbt]
\includegraphics[width=100mm]{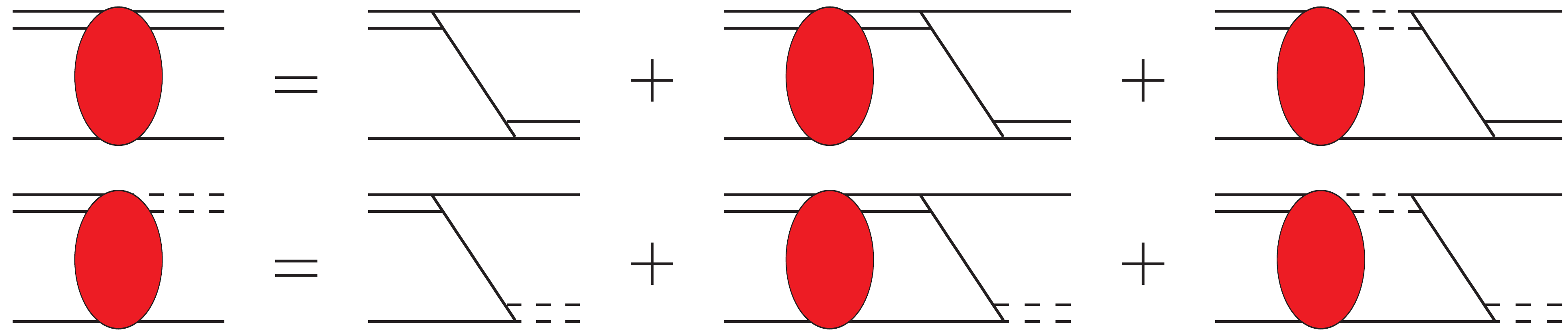}
\caption{The coupled-channel integral equations for the LO doublet channel $nd$ scattering amplitude.  Single lines represent nucleons and double lines (dashed double lines) spin-triplet (spin-singlet) dibaryons.\label{fig:DoubletLO}}
\end{figure}

The LO $nd$ scattering amplitude in the doublet channel is given by an infinite sum of diagrams represented by the coupled-channel integral equations in Fig.~\ref{fig:DoubletLO}.  Single lines are nucleons and the double line (dashed double line) is the spin-triplet (spin-singlet) dibaryon.  For the doublet $S$-wave channel there is also a contribution from a LO three-body force.  However, in the approach used here three-body forces will be treated in separate diagrams discussed later.
By projecting out the diagrams of Fig.~\ref{fig:DoubletLO} in the doublet channel and in a partial wave basis the integral equations can be written as an infinite set of matrix equations in cluster configuration (c.c.) space~\cite{Griesshammer:2004pe}, which gives 
\begin{align} 
\label{eq:LOndScatt}
&\mathbf{t}^{\ell}_{0,d}(k,p)=\mathbf{B}^{\ell}_{0}(k,p)+\mathbf{K}^{\ell}_{0}(q,p,E)\otimes\mathbf{t}^{\ell}_{0,d}(k,q),
\end{align}
where the subscript ``$d$" refers to the doublet channel, and the superscript ``$\ell$" to the partial wave.  The ``$\otimes$" notation is shorthand for the integration 
\begin{equation}
\label{eq:otimes}
A(q)\!\otimes\! B(q)=\frac{1}{2\pi^{2}}\int_{0}^{\Lambda}dqq^{2}A(q)B(q),
\end{equation}
where $\Lambda$ is a cutoff imposed to regulate divergences.  Physical results should be $\Lambda$-independent for sufficiently large $\Lambda$.  In the integral equation $k$ is the magnitude of the incoming  on-shell momentum in the $nd$ center of mass (c.m.) frame and $p$ is the magnitude of the off-shell outgoing momentum.  Since $k$ is on-shell it is related to the total energy of the three-body system by $E=\frac{3}{4}\frac{k^{2}}{M_{N}}-\frac{\gamma_{t}^{2}}{M_{N}}$.  $\mathbf{t}^{\ell}_{m,d}(k,p)$  and the inhomogeneous term $\mathbf{B}_{0}^{\ell}(k,p)$ are vectors in c.c.~space, defined as
\begin{equation}
\hspace{-.85cm}\label{eq:tDef}
\mathbf{t}^{\ell}_{m,d}(k,p)=\left(
\begin{array}{c}
t^{\ell}_{m,Nt\to Nt}(k,p)\\
t^{\ell}_{m,Nt\to Ns}(k,p)
\end{array}\right),\quad \mathbf{B}^{\ell}_{0}(k,p)=\left(\!\!\!\begin{array}{c}
\frac{2\pi}{pk}Q_{\ell}\left(\frac{p^{2}+k^{2}-M_{N}E-i\epsilon}{pk}\right)\\
-\frac{6\pi}{pk}Q_{\ell}\left(\frac{p^{2}+k^{2}-M_{N}E-i\epsilon}{pk}\right)
\end{array}\!\!\!\right).
\end{equation}
Here the subscript ``$m$'' refers to the order of the calculation ($m=0$ is LO, $m=1$ is NLO, and etc.), $t^{\ell}_{m,Nt\to Nt}(k,p)$ is the $nd$ scattering amplitude, and $t^{\ell}_{m,Nt\to Ns}(k,p)$  is the unphysical amplitude of a neutron and deuteron going to a nucleon and spin-singlet dibaryon.  In this formalism $\mathbf{B}_{1}^{\ell}(k,p)=\mathbf{B}_{2}^{\ell}(k,p)=0$, even for $\ell=0$, unlike in Ref.~\cite{Vanasse:2013sda}.  The function $Q_{\ell}(a)$ is a Legendre function of the second kind and is related to standard Legendre polynomials by\footnote{This definition of the Legendre functions of the second kind differs from the normal convention by a phase of $(-1)^{\ell}$.}
\begin{equation}
Q_{\ell}(a)=\frac{1}{2}\int_{-1}^{1}\frac{P_{\ell}(x)}{a+x}dx.
\end{equation}
The homogeneous term $\mathbf{K}^{\ell}_{0}(q,p,E)$ is a matrix in c.c.~space defined by
\begin{equation}
\hspace{-.5cm}\mathbf{K}^{\ell}_{0}(q,p,E)=\mathbf{R}_{0}(q,p,E)\,\mathbf{D}^{(0)}\!\!\left(E-\frac{q^{2}}{2M_{N}},\vect{q}\right),
\end{equation}
where
\begin{equation}
\label{eq:DibMatrix}
\mathbf{D}^{(n)}(E,\vect{q})=
\left(
\begin{array}{cc}
D_{t}^{(n)}(E,\vect{q}) & 0 \\
0&D_{s}^{(n)}(E,\vect{q})  
\end{array}\right)
\end{equation}
is a matrix of dibaryon propagators with $n=0$ giving the LO dibaryon propagators, $n=1$ the NLO correction to the dibaryon propagators, and $n=2$ the NNLO correction to the dibyaron propagators as in Eq.~(\ref{eq:dib}), and
\begin{align}
\mathbf{R}_{0}(q,p,E)=-\frac{2\pi}{qp}Q_{\ell}\left(\frac{q^{2}+p^{2}-M_{N}E-i\epsilon}{qp}\right)\left(\!\!\!
\begin{array}{rr}
1 & -3 \\[-1.5 mm]
-3 & 1
\end{array}\!\right).
\end{align}

\begin{figure}[hbt]
\includegraphics[width=100mm]{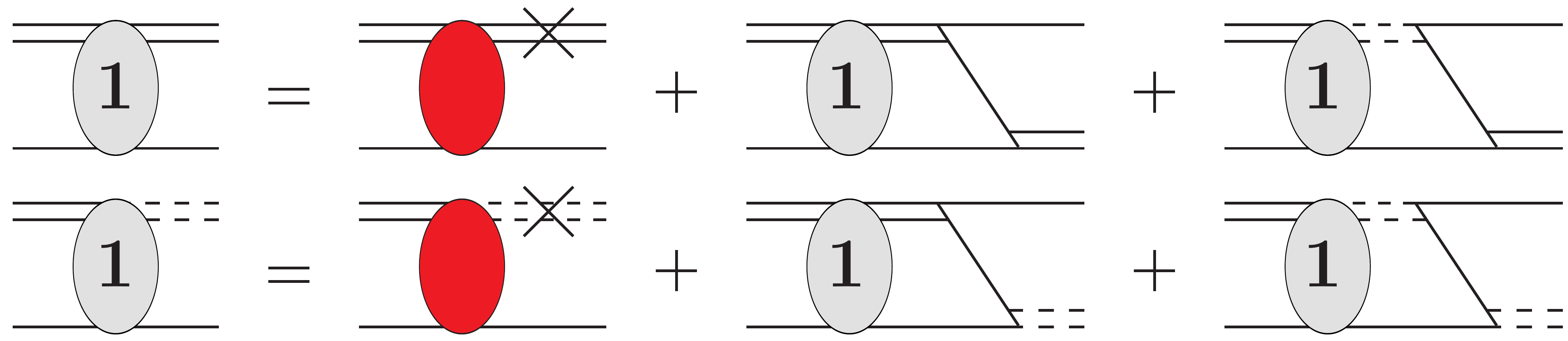}
\caption{The coupled-channel integral equations for the NLO correction to the doublet channel $nd$ scattering amplitude.  The cross refers to a single effective range insertion from $c_{0t}^{(0)}$ or $c_{0s}^{(0)}$ and the number ``1" to the NLO correction to the $nd$ scattering amplitude.\label{fig:DoubletNLO}}
\end{figure}

\begin{figure}[hbt]
\includegraphics[width=120mm]{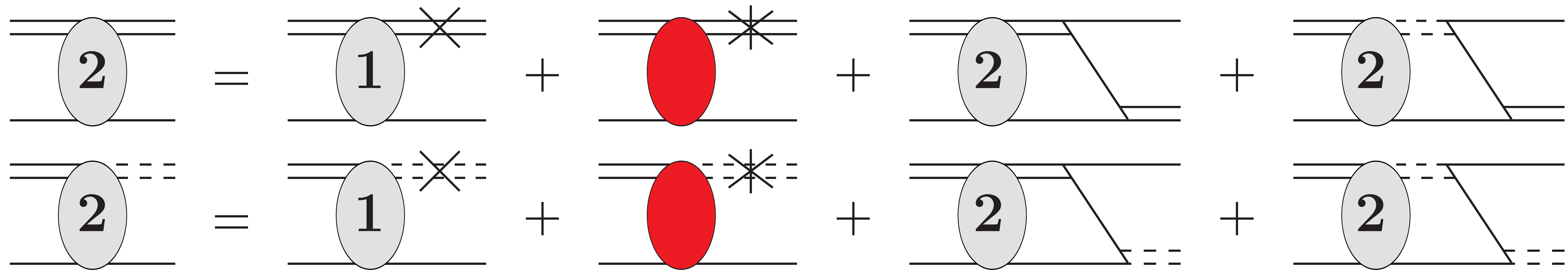}
\caption{The coupled-channel integral equations for the \nnlo correction to the doublet channel $nd$ scattering amplitude.  The star refers to an insertion of $c_{0t}^{(1)}$ or $c_{0s}^{(1)}$ and the number ``2" refers to the \nnlo correction to the doublet channel $nd$ scattering amplitude.\label{fig:DoubletNNLO}}
\end{figure}

The half off-shell NLO correction to the doublet channel $nd$ scattering amplitude is given by the coupled-channel integral equations in Fig.~\ref{fig:DoubletNLO}, where the cross represents an effective range insertion.
Iterating the inhomogeneous piece a single time in the kernel gives the integral equation for the NLO correction to $nd$ scattering as in Ref.~\cite{Vanasse:2013sda} along with an additional diagram where an effective range insertion appears on an external dibaryon leg.  In the on-shell limit the effective range insertion on the external dibaryon leg becomes the NLO wavefunction renormalization, which multiplies the LO $nd$ scattering amplitude. In other words, in the on-shell limit this integral equation gives the NLO correction to the $nd$ scattering amplitude plus the LO $nd$ scattering amplitude times the NLO deuteron wavefunction renormalization, or simply put all NLO contributions.  The integral equation can be written in c.c.~space as
\begin{equation}
\label{eq:NLOamp}
\mathbf{t}^{\ell}_{1,d}(k,p)=\mathbf{t}^{\ell}_{0,d}(k,p)\circ\mathbf{R}_{1}\!\!\left(E-\frac{\vect{p}^{2}}{2M_{N}},\vect{p}\right)+\mathbf{K}^{\ell}_{0}(q,p,E)\otimes \mathbf{t}^{\ell}_{1,d}(k,q),	
\end{equation}
where ``$\circ$" is the Schur product (element wise matrix multiplication) and $\mathbf{R}_{1}(p_{0},\vect{p})$ is a vector in c.c.~space defined by
\begin{equation}
\mathbf{R}_{1}(p_{0},\vect{p})=
\left(\begin{array}{c}
\frac{Z_{t}-1}{2\gamma_{t}}\left(\gamma_{t}+\sqrt{\frac{1}{4}\vect{p}^{2}-M_{N}p_{0}-i\epsilon}\,\right)\\
\frac{Z_{s}-1}{2\gamma_{s}}\left(\gamma_{s}+\sqrt{\frac{1}{4}\vect{p}^{2}-M_{N}p_{0}-i\epsilon}\,\right)
\end{array}\right).
\end{equation}
Choosing the kinematics of the ${}^{3}\!S_{1}$ (${}^{1}\!S_{0}$) bound-state (virtual bound-state) pole for the upper (lower) component of $\mathbf{R}_{1}(p_{0},\vect{p})$, $\mathbf{R}_{1}(p_{0},\vect{p})$ reduces to
%
\begin{equation}
\mathbf{c}_{1}=\left(\begin{array}{c}
Z_{t}-1\\
Z_{s}-1
\end{array}\right),
\end{equation}
which is the NLO correction to the wavefunction renormalization~\cite{Griesshammer:2004pe}.\footnote{Since $t^{\ell}_{m,Nt\to Ns}(k,p)$ is unphysical its normalization can be chosen arbitrarily without affecting physical results.}  Similarly, the half off-shell \nnlo correction to the $nd$ scattering amplitude is given by the coupled-channel integral equations in Fig.~\ref{fig:DoubletNNLO}, where the star represents an insertion of $c_{0t}^{(1)}$ or $c_{0s}^{(1)}$.
In c.c.~space the integral equation is given by
\begin{align}
\label{eq:NNLOamp}
&\mathbf{t}^{\ell}_{2,d}(k,p)=\left[\mathbf{t}^{\ell}_{1,d}(k,p)-\mathbf{c}_{1}\circ\mathbf{t}^{\ell}_{0,d}(k,p)\right]\circ\mathbf{R}_{1}\!\!\left(E-\frac{\vect{p}^{2}}{2M_{N}},\vect{p}\right)\\\nonumber
&\hspace{10cm}+\mathbf{K}^{\ell}_{0}(q,p,E)\otimes \mathbf{t}^{\ell}_{2,d}(k,q).
\end{align}
In the ERE parametrization $\mathbf{c}_{1}=\mathbf{0}$ and the integral equations at NLO and \nnlo look the same.  The presence of $\mathbf{c}_{1}\circ\mathbf{t}^{\ell}_{0,d}(k,p)$ removes the $(Z_{t}-1)^{2}\mathbf{t}^{\ell}_{0,d}(k,k)$ contribution that comes from $\mathbf{t}^{\ell}_{1,d}(k,p)\circ\mathbf{R}_{1}\!\!\left(E-\frac{\vect{p}^{2}}{2M_{N}},\vect{p}\right)$ in the on-shell limit.  Since the wavefunction renormalization in the $Z$-parametrization is exact at NLO by construction, there is no $(Z_{t}-1)^{2}$ correction.  

\subsection{Three-Body Forces}

The above description for doublet channel $nd$ scattering is incomplete since in the $S$-wave channel a three-body force is required at LO~\cite{Bedaque:1999ve}.  The Lagrangian for the three-body force up to \nnlo is 
\begin{align}
&{\mathcal{L}}_{3}=\frac{M_{N}H_{0}(\Lambda)}{3\Lambda^{2}}\left[y_{t}\hat{N}^{\dagger}(\vec{t}\cdot\vectS{\sigma})^{\dagger}-y_{s}\hat{N}^{\dagger}(\vec{s}\cdot\vectS{\tau})^{\dagger}\right]
\left[y_{t}(\vec{t}\cdot\vectS{\sigma})\hat{N}-y_{s}(\vec{s}\cdot\vectS{\tau})\hat{N}\right]\\\nonumber
&+\frac{M_{N}H_{2}(\Lambda)}{3\Lambda^{4}}\frac{4}{3}\left[y_{t}\hat{N}^{\dagger}(\vec{t}\cdot\vectS{\sigma})^{\dagger}-y_{s}\hat{N}^{\dagger}(\vec{s}\cdot\vectS{\tau})^{\dagger}\right]\left(i\vec{\partial}_{0}+\frac{\gamma_{t}^{2}}{M_{N}}\right)\left[y_{t}(\vec{t}\cdot\vectS{\sigma})\hat{N}-y_{s}(\vec{s}\cdot\vectS{\tau})\hat{N}\right].
\end{align}
$H_{0}(\Lambda)$ first occurs at LO and receives higher order corrections that can be written as
\begin{equation}
H_{0}(\Lambda)=\underbrace{H_{0,0}(\Lambda)}_{\mathrm{LO}}+\underbrace{H_{0,1}(\Lambda)}_{\mathrm{NLO}}+\underbrace{H_{0,2}(\Lambda)}_{\mathrm{NNLO}}+\cdots,
\end{equation}
where the first subscript denotes that it is a contribution to $H_{0}(\Lambda)$ and the second subscript gives the order of the contribution.  At \nnlo a new energy-dependent three-body force $H_{2}(\Lambda)$ appears~\cite{Bedaque:2002yg}.  The LO three-body force $H_{0,0}(\Lambda)$ does not renormalize an ultra-violet divergence.  Rather, the solution of the LO doublet $S$-wave $nd$ scattering amplitude is not unique in the limit where $\Lambda\to\infty$ and this causes oscillations in the solution as $\Lambda$ is changed~\cite{Bedaque:2002yg}.  The physical explanation for $H_{0,0}(\Lambda)$ comes from the fact that in the doublet $S$-wave channel there is no Pauli blocking preventing the nucleons from falling to the center.  Thus the doublet $S$-wave channel is sensitive to short range physics, which $H_{0,0}(\Lambda)$ encodes.

The three-body force Lagrangian can be rewritten using a triton auxiliary field $\hat{\psi}$, yielding
\begin{align}
{\mathcal{L}}_{3}=&\hat{\psi}^{\dagger}\left[\Omega-h_{2}(\Lambda)\left(i\partial_{0}+\frac{\vect{\nabla}^{2}}{6M_{N}}+\frac{\gamma_{t}^{2}}{M_{N}}\right)\right]\hat{\psi}+\sum_{n=0}^{\infty}\left[\omega^{(n)}_{t0}\hat{\psi}^{\dagger}\sigma_{i}\hat{N}\hat{t}_{i}-\omega^{(n)}_{s0}\hat{\psi}^{\dagger}\tau_{a}\hat{N}\hat{s}_{a}\right]\\\nonumber
&+\mathrm{H.c.}. 
\end{align}
A matching calculation shows that the parameters from each Lagrangian are related by
\begin{equation}
\label{eq:H00def}
\frac{H_{0,0}(\Lambda)}{\Lambda^{2}}=-\frac{3(\omega^{(0)}_{t0})^{2}}{4\pi\Omega}=-\frac{3(\omega^{(0)}_{s0})^{2}}{4\pi\Omega}=-\frac{3\omega^{(0)}_{t0}\omega^{(0)}_{s0}}{4\pi\Omega},
\end{equation}
\begin{equation}
\frac{H_{0,1}(\Lambda)}{\Lambda^{2}}=-\frac{6\omega^{(0)}_{t0}\omega^{(1)}_{t0}}{4\pi\Omega}=-\frac{6\omega^{(0)}_{s0}\omega^{(1)}_{s0}}{4\pi\Omega}=-\frac{6\omega^{(0)}_{t0}\omega^{(1)}_{s0}}{4\pi\Omega}=-\frac{6\omega^{(1)}_{t0}\omega^{(0)}_{s0}}{4\pi\Omega},
\end{equation}
\begin{align}
\frac{H_{0,2}(\Lambda)}{\Lambda^{2}}&=-\frac{3((\omega^{(1)}_{t0})^{2}+2\omega^{(0)}_{t0}\omega^{(2)}_{t0})}{4\pi\Omega}=-\frac{3((\omega^{(1)}_{s0})^{2}+2\omega^{(0)}_{s0}\omega^{(2)}_{s0})}{4\pi\Omega}\\\nonumber
&=-\frac{3(\omega^{(1)}_{s0}\omega^{(1)}_{t0}+2\omega^{(0)}_{t0}\omega^{(2)}_{s0})}{4\pi\Omega}=-\frac{3(\omega^{(1)}_{s0}\omega^{(1)}_{t0}+2\omega^{(2)}_{t0}\omega^{(0)}_{s0})}{4\pi\Omega},
\end{align}
and
\begin{equation}
\frac{4H_{2,0}(\Lambda)}{\Lambda^{4}}=-\frac{3(\omega^{(0)}_{t0})^{2}}{\pi\Omega^{2}M_{N}}h_{2}(\Lambda)=-\frac{3(\omega^{(0)}_{s0})^{2}}{\pi\Omega^{2}M_{N}}h_{2}(\Lambda)=-\frac{3\omega^{(0)}_{t0}\omega^{(0)}_{s0}}{\pi\Omega^{2}M_{N}}h_{2}(\Lambda).
\end{equation}
It is convenient to make the definitions
\begin{equation}
\label{eq:HLOdef}
H_{\mathrm{LO}}=\frac{4H_{0,0}(\Lambda)}{\Lambda^{2}},\quad H_{\mathrm{NLO}}=\frac{4H_{0,1}(\Lambda)}{\Lambda^{2}},\quad H_{\mathrm{NNLO}}=\frac{4H_{0,2}(\Lambda)}{\Lambda^{2}},
\end{equation}
and
\begin{equation}
\label{eq:H2def}
\widehat{H}_{2}=\frac{4H_{2,0}(\Lambda)}{\Lambda^{4}}.
\end{equation}
From these definitions follow the useful identities
\begin{equation}
\label{eq:HNLOHLOratio}
\frac{H_{\mathrm{NLO}}}{H_{\mathrm{LO}}}=2\frac{\omega^{(1)}_{t0}}{\omega^{(0)}_{t0}},
\end{equation}
and
\begin{equation}
2\frac{\omega^{(2)}_{t0}}{\omega^{(0)}_{t0}}=\frac{H_{\mathrm{NNLO}}H_{\mathrm{LO}}-\frac{1}{4}(H_{\mathrm{NLO}})^{2}}{(H_{\mathrm{LO}})^{2}}.
\end{equation}

\subsection{Triton Vertex Function}
\begin{figure}[hbt]
\includegraphics[width=100mm]{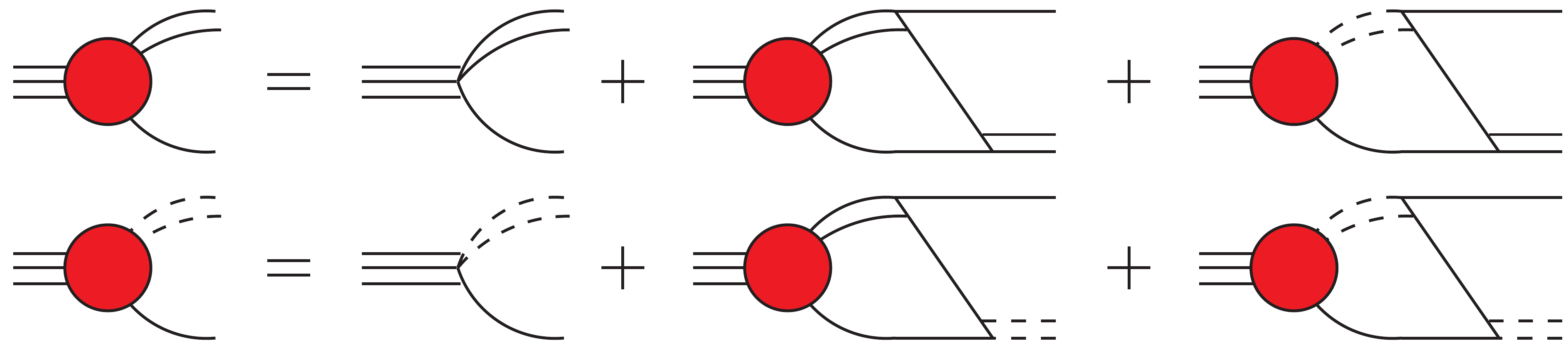}
\caption{The coupled-channel integral equations for the LO triton vertex function, where the triple line is the triton, and the filled circle is the LO triton vertex function.\label{fig:GirrLO}}
\end{figure}

The LO triton vertex function is given by the coupled-channel integral equations in Fig.~\ref{fig:GirrLO}, where the triple line represents the triton propagator.
These integral equations can be written in c.c.~space as
\begin{equation}
\Gb_{0}(E,p)=\mathbf{\widetilde{B}}_{0}+\mathbf{K}^{\ell=0}_{0}(q,p,E)\otimes\Gb_{0}(E,q),
\end{equation}
where the ``0'' subscript indicates LO and $\widetilde{\mathbf{B}}_{0}$ is a c.c space vector defined by
\begin{equation}
\label{eq:BtildeDef}
\mathbf{\widetilde{B}}_{0}=
\left(\begin{array}{c}
1\\[-.2cm]
1
\end{array}\right).
\end{equation}
Note the kernel of these coupled-channel integral equations is the same as in LO $nd$ scattering.  The only difference between the integral equations for the LO triton vertex function $\Gb_{0}(E,p)$ and the LO $nd$ scattering amplitude Eq.~(\ref{eq:LOndScatt}) is the inhomogeneous term. At the energy of the bound state the  matrix $[\id-\mathbf{K}^{\ell=0}_{0}(q,p,E)]$ is invertible for all cutoffs for which $H_{0,0}(\Lambda)\neq0$.  For cutoffs for which $H_{0,0}(\Lambda)=0$ the LO triton vertex is still well defined because the zero of $H_{0,0}(\Lambda)$ and the infinity of $[\id-\mathbf{K}^{\ell=0}_{0}(q,p,E)]^{-1}$ have a well defined limit.  However, this is numerically tricky and therefore such cutoffs are avoided. $\Gb_{0}(E,p)$ is defined in c.c.~space by

\begin{equation}
\label{eq:GDef}
\boldsymbol{\mathcal{G}}_{0}(E,p)=\left(
\begin{array}{c}
\G_{0,\psi\to Nt}(E,p)\\
\G_{0,\psi\to Ns}(E,p)
\end{array}\right),
\end{equation}
where $\G_{0,\psi\to Nt}(E,p)$ ($\G_{0,\psi\to Ns}(E,p)$) is the triton vertex function for an outgoing neutron and deuteron (nucleon and spin-singlet dibaryon) state.  Note $\widetilde{\mathbf{B}}_{0}$ is not the ``physical" inhomogeneous term.  The ``physical" inhomogeneous term $\mathbf{B}_{0}$ is given by
\begin{equation}
\mathbf{B}_{0}=
\left(\!\!\begin{array}{r}
\sqrt{3}\omega^{(0)}_{t0}\\
-\sqrt{3}\omega^{(0)}_{s0}
\end{array}\!\!\right).
\end{equation}
Since an arbitrary normalization can be absorbed into both components of $\Gb_{0}(E,p)$ it is convenient to use $\widetilde{\mathbf{B}}_{0}$ instead of $\mathbf{B}_{0}$.  The ``physical" triton vertex function $\boldsymbol{\Gamma}_{0}(p)$ is related to $\Gb_{0}(E,p)$ by
\begin{equation}
\label{eq:GGPrelation}
\boldsymbol{\Gamma}_{0}(p)=\Gb_{0}(E,p)\circ \mathbf{B}_{0}\sqrt{Z_{\psi}},
\end{equation}
where the value of $E$ is assumed fixed, and here $Z_{\psi}$ is the LO triton wavefunction renormalization to be defined below.  Using $\Gb_{0}(E,p)$ instead of $\boldsymbol{\Gamma}_{0}(p)$ allows three-body forces to be factored out of expressions that would otherwise be absorbed into $\boldsymbol{\Gamma}_{0}(p)$.	

\begin{figure}[hbt]
\includegraphics[width=100mm]{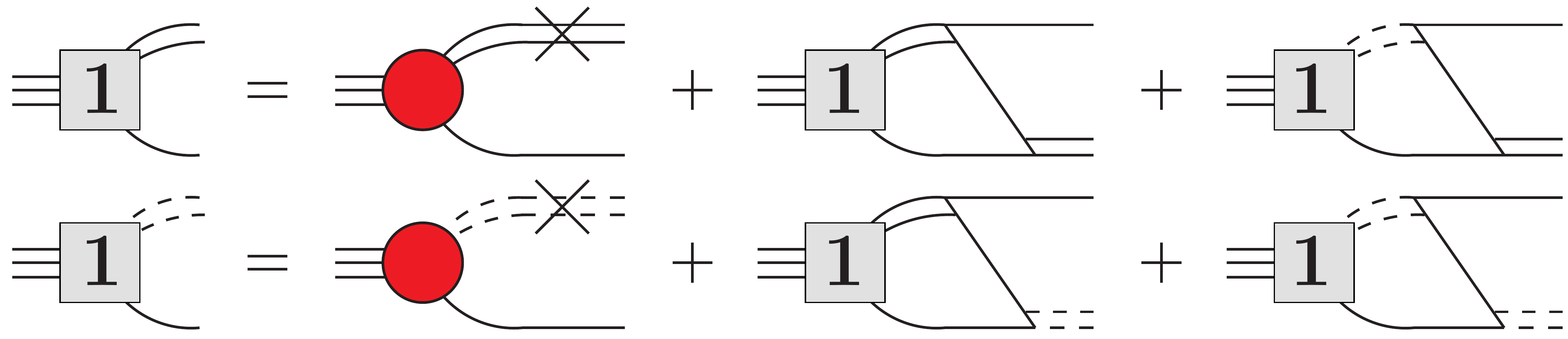}
\caption{The coupled-channel integral equations for the NLO correction to the triton vertex function.\label{fig:GirrNLO}}
\end{figure}

\begin{figure}[hbt]
\includegraphics[width=120mm]{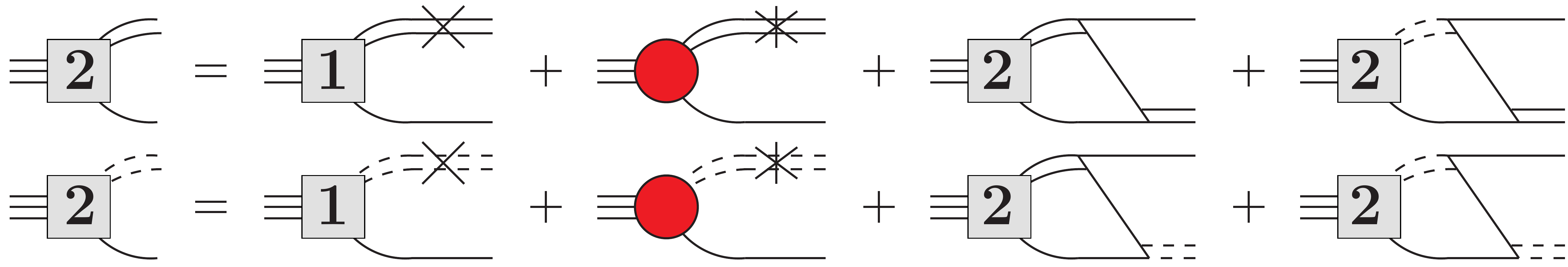}
\caption{The coupled-channel integral equations for the \nnlo correction to the triton vertex function.\label{fig:GirrNNLO}}
\end{figure}

Adding a NLO effective range insertion to the triton vertex function can be achieved via the coupled-channel integral equations in Fig.~\ref{fig:GirrNLO},
which in c.c.~space can be written as
\begin{equation}
\Gb_{1}(E,p)=\Gb_{0}(E,p)\circ\mathbf{R}_{1}\!\!\left(E-\frac{\vect{p}^{2}}{2M_{N}},\vect{p}\right)+\mathbf{K}^{\ell=0}_{0}(q,p,E)\otimes\Gb_{1}(E,q).
\end{equation}
This equation is analogous to the NLO correction to the $nd$ scattering amplitude Eq.~(\ref{eq:NLOamp}).  Two effective range insertions and $c_{0t}^{(1)}$ and $c_{0s}^{(1)}$ corrections to the triton vertex function at \nnlo can be added using the coupled-channel integral equations in Fig.~\ref{fig:GirrNNLO},
which in c.c.~space are
\begin{equation}
\Gb_{2}(E,p)=\Big[\Gb_{1}(E,p)-\mathbf{c}_{1}\circ\Gb_{0}(E,p)\Big]\circ\mathbf{R}_{1}\left(E-\frac{\vect{p}^{2}}{2M_{N}},\vect{p}\right)+\mathbf{K}^{\ell=0}_{0}(q,p,E)\otimes\Gb_{2}(E,q).
\end{equation}
This equation is again entirely analogous to the integral equations for the \nnlo correction to $nd$ scattering Eq.~(\ref{eq:NNLOamp}).  In fact the only difference between the integral equations for the triton vertex function and the $nd$ scattering amplitude up to \nnlo is the LO inhomogeneous term.

\begin{figure}[hbt]
\includegraphics[width=100mm]{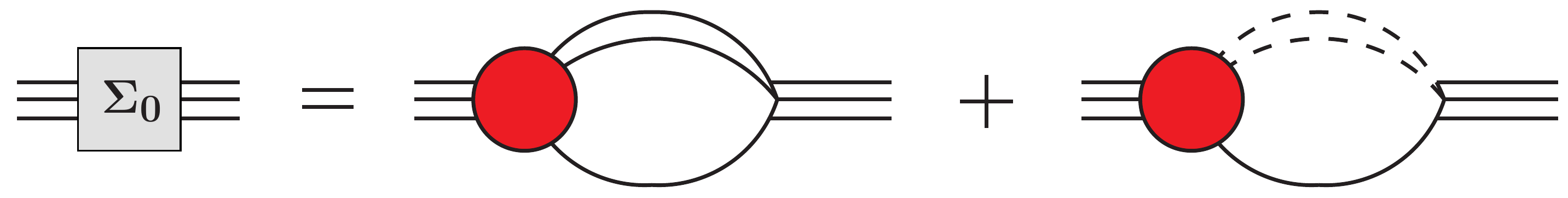}
\caption{Diagrammatic representation of the function $\Sigma_{0}^{P}(E)$.\label{fig:Sigma0}}
\end{figure}

\begin{figure}[hbt]
\includegraphics[width=100mm]{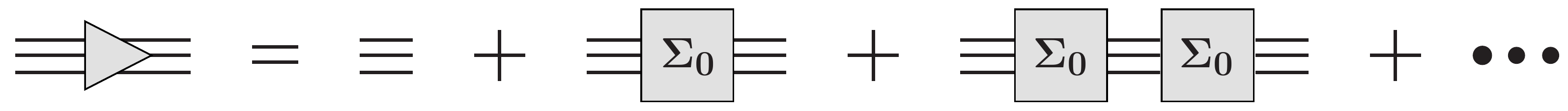}
\caption{LO dressed triton propagator.  The triangle is the dressed triton propagator, and the triple line is the bare triton propagator $i/\Omega$.\label{fig:LOTritonProp}}
\end{figure}

\begin{figure}[hbt]
\includegraphics[width=100mm]{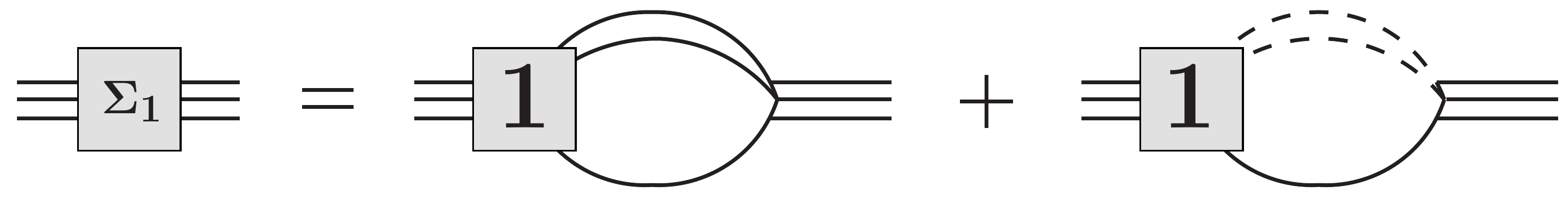}
\caption{Diagrammatic representation of the function $\Sigma_{1}^{P}(E)$.\label{fig:Sigma1}}
\end{figure}

\begin{figure}[hbt]
\includegraphics[width=100mm]{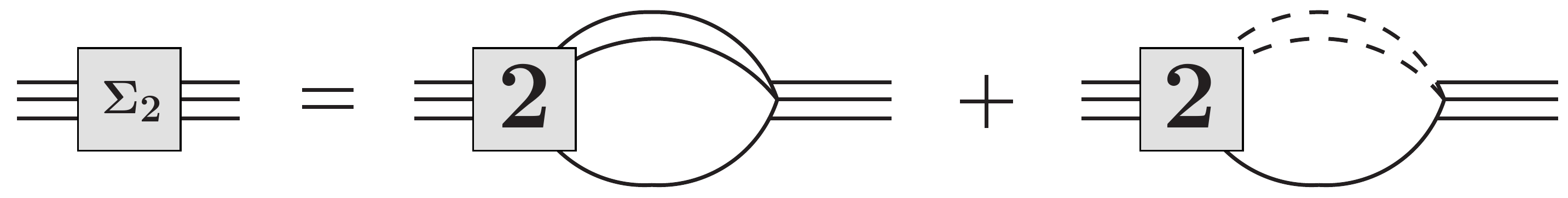}
\caption{Diagrammatic representation of the function $\Sigma_{2}^{P}(E)$.\label{fig:Sigma2}}
\end{figure}

The function $\Sigma_{0}^{P}(E)$ is defined as
\begin{equation}
\Sigma_{0}^{P}(E)=\int\frac{d^{4}q}{(2\pi)^{4}}\frac{i}{E-q_{0}-\frac{q^{2}}{2M_{N}}+i\epsilon}\left[i\mathbf{D}^{(0)}\!\left(E-q_{0},q\right)i\mathbf{B}_{0}\right]\cdot\left[\Gb_{0}(E,q)\circ i \mathbf{B}_{0}\right]
\end{equation}
and describes the sum of all triton-irreducible diagrams in Fig.~\ref{fig:Sigma0}.  Note ``$\cdot$" represents the ordinary dot product of two c.c space vectors.  Subscript ``0" denotes this is LO.
Integrating over the energy pole and angles, the expression for $\Sigma_{0}^{P}(E)$ becomes
\begin{align}
i\Sigma_{0}^{P}(E)=&-i\frac{3(\omega_{t0}^{(0)})^{2}}{\pi}\frac{1}{2\pi}\int_{0}^{\Lambda}dq q^{2} D_{t}^{(0)}\!\!\left(E-\frac{q^{2}}{2M_{N}},q\right)\G_{0,\psi\to Nt}(E,q)\\\nonumber
&-i\frac{3(\omega_{s0}^{(0)})^{2}}{\pi}\frac{1}{2\pi}\int_{0}^{\Lambda}dq q^{2} D_{s}^{(0)}\!\!\left(E-\frac{q^{2}}{2M_{N}},q\right)\G_{0,\psi\to Ns}(E,q).
\end{align}
Defining the functions
\begin{equation}
\Sigma_{n}(E)=-\pi\mathrm{Tr}\left[\mathbf{D}^{(0)}\!\!\left(E-\frac{q^{2}}{2M_{N}},q\right)\otimes\Gb_{n}(E,q)\right],
\end{equation}
and using Eqs.~(\ref{eq:H00def}) and (\ref{eq:HLOdef}) to rewrite $\omega_{s0}^{(0)}$ and $\omega_{t0}^{(0)}$, $\Sigma_{0}^{P}(E)$ becomes
\begin{align}
i\Sigma_{0}^{P}(E)=-i \Omega H_{\mathrm{LO}}\Sigma_{0}(E) .
\end{align}
Using $\Sigma_{0}^{P}(E)$, the LO dressed triton propagator is given by the infinite sum of diagrams in Fig.~\ref{fig:LOTritonProp},
which can be summed as a geometric series giving
\begin{equation}
i\Delta_{3}^{(\mathrm{LO})}(E)=\frac{i}{\Omega}+\frac{i}{\Omega}H_{\mathrm{LO}}\Sigma_{0}(E)+\cdots=\frac{i}{\Omega}\frac{1}{1-H_{\mathrm{LO}}\Sigma_{0}(E)}.
\end{equation}
This is the LO dressed triton propagator in the c.m. frame of the $nd$ system.  Thus the triton propagator always has zero momentum.  The formalism here can be straightforwardly generalized to include a triton propagator with non-zero momentum.  At the bound-state energy $B$ of the triton, the LO dressed triton propagator has a pole, giving the condition
\begin{equation}
\label{eq:Pole}
H_{\mathrm{LO}}=\frac{1}{\Sigma_{0}(B)}.
\end{equation}
Setting $B=E_{(\jjvH)}$ the three-body force can be fit to the triton binding energy $E_{(\jjvH)}=-8.48$~MeV~\cite{Wapstra:1985zz}.  Additionally, the LO triton binding energy can be calculated if a different renormalization condition is used for $H_{\mathrm{LO}}$.  Considering higher orders beyond the work of Hagen \emph{et al.}~\cite{Hagen:2013xga} the triton-irreducible functions  $\Sigma_{1}^{P}(E)$ and $\Sigma_{2}^{P}(E)$ follow the $\Sigma_{0}^{P}(E)$ definition and are given by the sum of diagrams in Fig.~\ref{fig:Sigma1} and \ref{fig:Sigma2} respectively.

\begin{figure}[hbt]
\includegraphics[width=120mm]{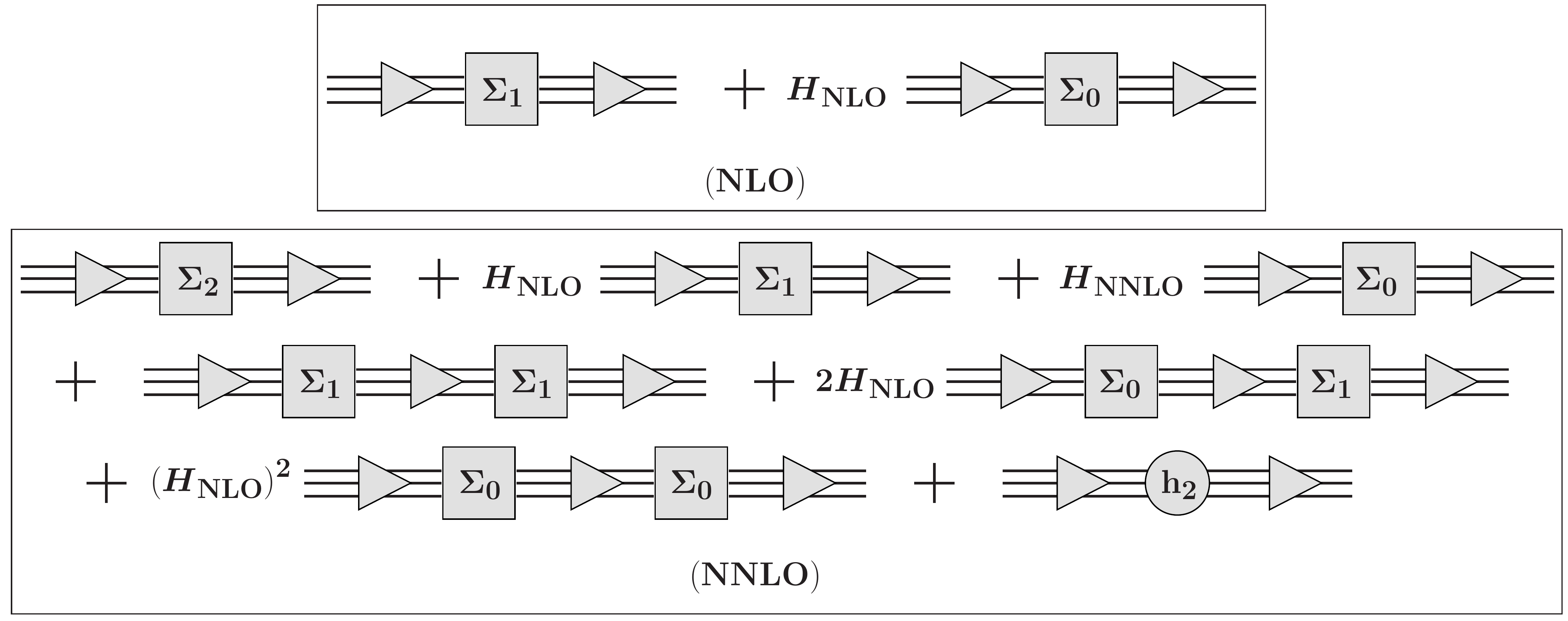}
\caption{NLO and \nnlo corrections to the triton propagator.  The diagram with $h_{2}$ comes from the kinetic term of the triton auxiliary field.\label{fig:NNLOTritonProp}}
\end{figure}

One finds that $\Sigma_{1}^{P}(E)$ and $\Sigma_{2}^{P}(E)$ are defined as
\begin{align}
i\Sigma_{1}^{P}(E)=-i \Omega H_{\mathrm{LO}}\Sigma_{1}(E),\quad i\Sigma_{2}^{P}(E)=-i \Omega H_{\mathrm{LO}}\Sigma_{2}(E) .
\end{align}
The NLO and \nnlo corrections to the triton propagator are given by the diagrams in Fig.~\ref{fig:NNLOTritonProp}.
Summing the NLO diagrams gives
\begin{equation}
\frac{i}{\Omega}\frac{1}{1-H_{\mathrm{LO}}\Sigma_{0}(E)}\left\{-i\Omega H_{\mathrm{LO}}\Sigma_{1}(E)-i\Omega \left(2\frac{\omega^{(1)}_{t0}}{\omega^{(0)}_{t0}}\right)H_{\mathrm{LO}}\Sigma_{0}(E)\right\}\frac{i}{\Omega}\frac{1}{1-H_{\mathrm{LO}}\Sigma_{0}(E)}
\end{equation}
for the NLO correction to the triton propagator.  The first (second) term comes from the first (second) diagram in the NLO box of Fig.~\ref{fig:NNLOTritonProp}.  The second diagram in the NLO box is $\Sigma_{0}^{P}(E)$, but with a $\omega^{(0)}_{t0}$ ($\omega^{(0)}_{s0}$) vertex replaced by $\omega^{(1)}_{t0}$ ($\omega^{(1)}_{s0}$).  A factor of two comes the fact the $\omega^{(1)}_{t0}$ ($\omega^{(1)}_{s0}$) vertex can be on the left or the right of Fig.~\ref{fig:Sigma0}.  Then using Eq.~(\ref{eq:HNLOHLOratio}) the NLO correction to the triton propagator reduces to
\begin{equation}
\frac{i}{\Omega}\frac{1}{1-H_{\mathrm{LO}}\Sigma_{0}(E)}\left\{-i\Omega H_{\mathrm{LO}}\Sigma_{1}(E)-i\Omega H_{\mathrm{NLO}}\Sigma_{0}(E)\right\}\frac{i}{\Omega}\frac{1}{1-H_{\mathrm{LO}}\Sigma_{0}(E)}.
\end{equation}
Carrying out a similar procedure gives the triton propagator up to and including \nnlo as
\begin{align}
i\Delta_{3}^{\mathrm{NNLO}}(E)=&\frac{i}{\Omega}\frac{1}{1-H_{\mathrm{LO}}\Sigma_{0}(E)}\left[1+\frac{H_{\mathrm{LO}}\Sigma_{1}(E)+H_{\mathrm{NLO}}\Sigma_{0}(E)}{1-H_{\mathrm{LO}}\Sigma_{0}(E)}\right.\\[.3cm]\nonumber
&\hspace{-3cm}+\frac{H_{\mathrm{LO}}\Sigma_{2}(E)+H_{\mathrm{NLO}}\Sigma_{1}(E)+H_{\mathrm{NNLO}}\Sigma_{0}(E)+\frac{4}{3}(M_{N}E+\gamma_{t}^{2})\widehat{H}_{2}/H_{\mathrm{LO}}}{1-H_{\mathrm{LO}}\Sigma_{0}(E)}\\\nonumber
&\left.+\frac{\left[H_{\mathrm{LO}}\Sigma_{1}(E)+H_{\mathrm{NLO}}\Sigma_{0}(E)\right]^{2}}{\left[1-H_{\mathrm{LO}}\Sigma_{0}(E)\right]^{2}}\right].
\end{align}
The $\widehat{H}_{2}/H_{\mathrm{LO}}$ term comes from the last NNLO diagram in Fig.~\ref{fig:NNLOTritonProp}.  Fitting the LO three-body force to the triton binding energy pole and ensuring that the pole is fixed at higher orders imposes the conditions
\begin{equation}
\label{eq:HNLOdef}
H_{\mathrm{LO}}\Sigma_{1}(B)+H_{\mathrm{NLO}}\Sigma_{0}(B)=0,
\end{equation}
and
\begin{equation}
\label{eq:HNNLOdef}
H_{\mathrm{LO}}\Sigma_{2}(B)+H_{\mathrm{NLO}}\Sigma_{1}(B)+\left(H_{\mathrm{NNLO}}+\frac{4}{3}(M_{N}B+\gamma_{t}^{2})\widehat{H}_{2}\right)\Sigma_{0}(B)=0.
\end{equation}
$H_{\mathrm{LO}}=1/\Sigma_{0}(B)$ has been used to rewrite the term with $\widehat{H}_{2}$.  These two conditions fix two higher-order three-body forces, and $H_{\mathrm{NNLO}}$ is fixed to the physical $nd$ doublet $S$-wave scattering length. It will be shown later how this is done in the new formalism.  The triton wavefunction renormalization is the residue about the triton pole, which up to \nnlo is given by

\begin{align}
Z_{\psi}=&-\frac{1}{\Omega}\frac{1}{H_{\mathrm{LO}}\Sigma_{0}'(B)}\left[1-\frac{\left[H_{\mathrm{LO}}\Sigma_{1}'(B)+H_{\mathrm{NLO}}\Sigma_{0}'(B)\right]}{H_{\mathrm{LO}}\Sigma_{0}'(B)}\right.\\[.3cm]\nonumber
&-\frac{\left[H_{\mathrm{LO}}\Sigma_{2}'(B)+H_{\mathrm{NLO}}\Sigma_{1}'(B)+H_{\mathrm{NNLO}}\Sigma_{0}'(B)\right]+\frac{4}{3}M_{N}\widehat{H}_{2}/H_{\mathrm{LO}}}{H_{\mathrm{LO}}\Sigma_{0}'(B)}\\\nonumber
&\left.+\frac{\left[H_{\mathrm{LO}}\Sigma_{1}'(B)+H_{\mathrm{NLO}}\Sigma_{0}'(B)\right]^{2}}{\left[H_{\mathrm{LO}}\Sigma_{0}'(B)\right]^{2}}\right].
\end{align}
Using Eqs.~(\ref{eq:Pole}), (\ref{eq:HNLOdef}), and (\ref{eq:HNNLOdef}) the dependence on $H_{\mathrm{LO}}$, $H_{\mathrm{NLO}}$, and $H_{\mathrm{NNLO}}$ can be removed yielding
\begin{align}
Z_{\psi}=&-\frac{1}{\Omega}\frac{1}{H_{\mathrm{LO}}\Sigma_{0}'(B)}\left[\vphantom{\left(\frac{\Sigma_{1}'(B)}{\Sigma_{0}'(B)}-\frac{\Sigma_{1}(B)}{\Sigma_{0}(B)}\right)^{2}}1-\left(\frac{\Sigma_{1}'(B)}{\Sigma_{0}'(B)}-\frac{\Sigma_{1}(B)}{\Sigma_{0}(B)}\right)\right.\\\nonumber
&-\left\{\frac{\Sigma_{2}'(B)}{\Sigma_{0}'(B)}-\frac{\Sigma_{1}(B)\Sigma_{1}'(B)}{\Sigma_{0}(B)\Sigma_{0}'(B)}+\left(\frac{\Sigma_{1}(B)}{\Sigma_{0}(B)}\right)^{2}-\frac{\Sigma_{2}(B)}{\Sigma_{0}(B)}\right.\\\nonumber
&\hspace{2cm}+\left.\left.\frac{4}{3}M_{N}\widehat{H}_{2}\Sigma_{0}(B)\left(\frac{\Sigma_{0}(B)}{\Sigma_{0}'(B)}-B-\frac{\gamma_{t}^{2}}{M_{N}}\right)\vphantom{\left(\frac{\Sigma_{1}(B)}{\Sigma_{0}(B)}\right)^{2}}\!\!\right\}+\left(\frac{\Sigma_{1}'(B)}{\Sigma_{0}'(B)}-\frac{\Sigma_{1}(B)}{\Sigma_{0}(B)}\right)^{2}\right].
\end{align}
For the triton vertex function there is only one external triton propagator, and therefore the square root of $Z_{\psi}$ must be taken.  Expanding the square root of $Z_{\psi}$ perturbatively to \nnlo gives
\begin{align}
&\sqrt{Z_{\psi}}=\sqrt{-\frac{1}{\Omega}\frac{1}{H_{\mathrm{LO}}\Sigma_{0}'}}\left[\underbrace{1\vphantom{\frac{1}{2}\left(\frac{\Sigma_{1}'}{\Sigma_{0}'}-\frac{\Sigma_{1}}{\Sigma_{0}}\right)}}_{\mathrm{LO}}-\underbrace{\frac{1}{2}\left(\frac{\Sigma_{1}'}{\Sigma_{0}'}-\frac{\Sigma_{1}}{\Sigma_{0}}\right)}_{\mathrm{NLO}}\right.\\[.3cm]\nonumber
&\left.\underbrace{-\frac{1}{2}\left[\frac{\Sigma_{2}'}{\Sigma_{0}'}+\frac{1}{2}\frac{\Sigma_{1}\Sigma_{1}'}{\Sigma_{0}\Sigma_{0}'}-\frac{\Sigma_{2}}{\Sigma_{0}}+\frac{1}{4}\left(\frac{\Sigma_{1}}{\Sigma_{0}}\right)^{2}-\frac{3}{4}\left(\frac{\Sigma_{1}'}{\Sigma_{0}'}\right)^{2}+\frac{4}{3}M_{N}\widehat{H}_{2}\Sigma_{0}\left(\frac{\Sigma_{0}}{\Sigma_{0}'}\!-\!B\!-\!\frac{\gamma_{t}^{2}}{M_{N}}\right)\right]}_{\mathrm{NNLO}}+\cdots\right].
\end{align}
Here the explicit energy dependence for all $\Sigma_{n}$ functions has been dropped with the understanding that all functions are evaluated at $E=B$.  The ``physical" triton vertex function is calculated using Eq.~(\ref{eq:GGPrelation}).  Using the definition of $\mathbf{B}_{0}$ and the triton wavefunction renormalization, the LO renormalization for the triton vertex function $\Gb_{0}(B,p)$ is
\begin{align}
\sqrt{Z_{\psi}^{\mathrm{LO}}}=&\sqrt{3}\omega^{(0)}_{t0}\sqrt{-\frac{1}{\Omega}\frac{1}{H_{\mathrm{LO}}\Sigma_{0}'(B)}}=\sqrt{-\frac{3(\omega^{(0)}_{t0})^{2}}{\pi\Omega}\frac{\pi}{H_{\mathrm{LO}}\Sigma_{0}'(B)}}=\sqrt{\frac{\pi}{\Sigma_{0}'(B)}}.
\end{align}
Eq.~(\ref{eq:H00def}) has been used to simplify the expression.  Thus the ``physical" LO triton vertex function is given by
\begin{equation}
\boldsymbol{\Gamma}_{0}(p)=\sqrt{Z_{\psi}^{\mathrm{LO}}}\Gb_{0}(B,p).
\end{equation}
This expression is equivalent to solving the homogeneous equation for the doublet $S$-wave channel with a nonzero three-body force and then normalizing the result using techniques in Refs. \cite{Konig:2011yq,Ji:2011qg}.  The NLO triton vertex function is given by $\Gb_{1}(B,p)$, $\Gb_{0}(B,p)$ with the $\omega_{t0}^{(0)}$ ($\omega_{s0}^{(0)}$) vertex replaced by $\omega_{t0}^{(1)}$ ($\omega_{s0}^{(1)}$), and the LO triton vertex function times the NLO triton wavefunction renormalization correction.  The $\omega_{t0}^{(1)}$ ($\omega_{s0}^{(1)}$) vertex can again be replaced by a ratio of three-body forces as in the calculation of the triton propagator, and then the ratio of three-body forces can be rewritten in terms of $\Sigma_{n}(B)$ using Eq. (\ref{eq:HNLOdef}).  With these simplifications the NLO triton vertex function is given by
\begin{equation}
\boldsymbol{\Gamma}_{1}(p)=\sqrt{Z^{\mathrm{LO}}_{\psi}}\left[\Gb_{1}(B,p)-\frac{1}{2}\frac{\Sigma_{1}'}{\Sigma_{0}'}\Gb_{0}(B,p)\right].
\end{equation}
The calculation of the \nnlo triton vertex function follows similarly and yields
\begin{align}
\boldsymbol{\Gamma}_{2}(p)&=\sqrt{Z^{\mathrm{LO}}_{\psi}}\left[\Gb_{2}(B,p)-\frac{1}{2}\frac{\Sigma_{1}'}{\Sigma_{0}'}\Gb_{1}(B,p)\right.\\\nonumber
&\hspace{2cm}\left.+\left\{\frac{3}{8}\left(\frac{\Sigma_{1}'}{\Sigma_{0}'}\right)^{2}-\frac{1}{2}\frac{\Sigma_{2}'}{\Sigma_{0}'}-\frac{2}{3}M_{N}\widehat{H}_{2}\frac{\Sigma_{0}^{2}}{\Sigma_{0}'}\right\}\Gb_{0}(B,p)\right].
\end{align}

\section{\label{sec:doubleSwave}Doublet $S$-wave scattering}

\begin{figure}[hbt]
\includegraphics[width=70mm]{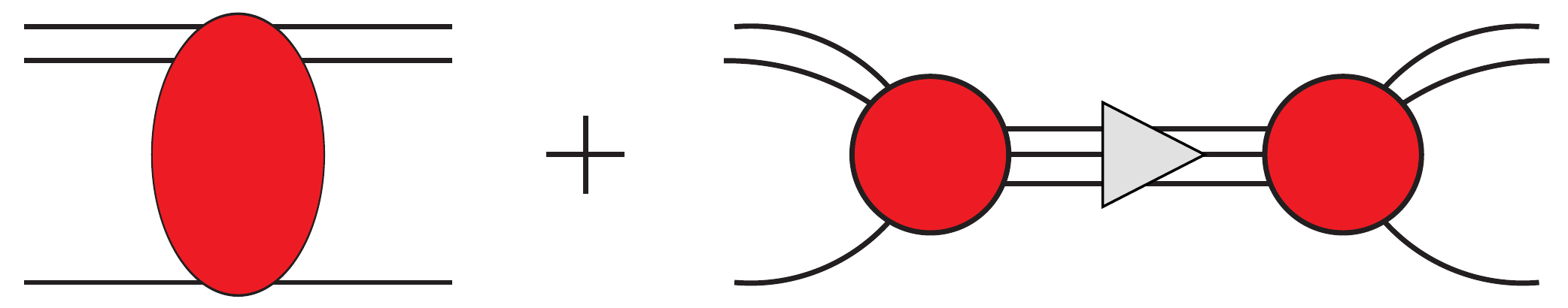}
\caption{Diagrams for the LO doublet $S$-wave $nd$ scattering amplitude.\label{fig:DoubletSLO}}
\end{figure}

In the formalism of this work the LO doublet $S$-wave on-shell $nd$ scattering amplitude is given by the sum of the two diagrams in Fig.~\ref{fig:DoubletSLO}.  The first diagram is the solution of Eq.~(\ref{eq:LOndScatt}) for $\ell=0$.  This diagram contains no three-body forces; all three-body force terms are contained in the second diagram.  
The sum of the two diagrams is given by
\begin{equation}
\label{eq:TLO}
T_{\mathrm{LO}}(k)=Z_{\mathrm{LO}}t_{0,Nt\to Nt}^{\ell=0}(k,k)+H_{\mathrm{LO}}\frac{1}{1-H_{\mathrm{LO}}\Sigma_{0}(E)}\pi Z_{\mathrm{LO}}\left[\G_{0,\psi\to Nt}(E,k)\right]^{2}.
\end{equation}
In the new formalism the LO three-body force $H_{\mathrm{LO}}$ is factored out of all numerically determined expressions.\footnote{The power of this formalism at LO lies in the fact that the triton pole contribution is contained solely in the second diagram of Fig.~\ref{fig:DoubletSLO}.  At higher orders contributions from poles are again clearly factored out in specific diagrams and can be easily read off.}  This is one advantage of this formalism. The LO three-body  force can be found algebraically in terms of numerically determined quantities by fitting to the scattering length, $a_{nd}=0.65$~fm~\cite{DILG1971208}, which  yields
\begin{equation}
H_{\mathrm{LO}}=\frac{x}{1+x\Sigma_{0}\left(-\frac{\gamma_{t}^{2}}{M_{N}}\right)},
\end{equation}
where
\begin{equation}
x=\frac{-\left(\frac{3\pi a_{nd}}{M_{N}}+Z_{\mathrm{LO}}t_{0,Nt\to Nt}^{\ell=0}(0,0)\right)}{\pi Z_{\mathrm{LO}}\left[\G_{0,\psi\to Nt}\left(-\frac{\gamma_{t}^{2}}{M_{N}},0\right)\right]^{2}}.
\end{equation}

\begin{figure}[hbt]
\includegraphics[width=100mm]{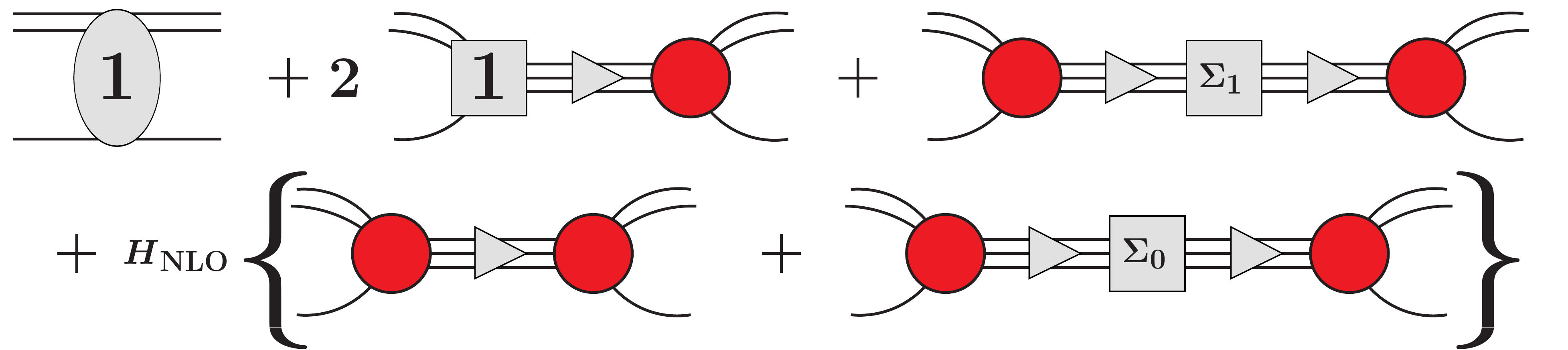}
\caption{Diagrams for the NLO correction to the doublet $S$-wave $nd$ scattering amplitude.  The factor of two takes into account the diagram related by time reversal symmetry that is not shown.\label{fig:DoubletSNLO}}
\end{figure}

\begin{figure}[hbt]
\begin{center}
\begin{tabular}{c}
\includegraphics[width=120mm]{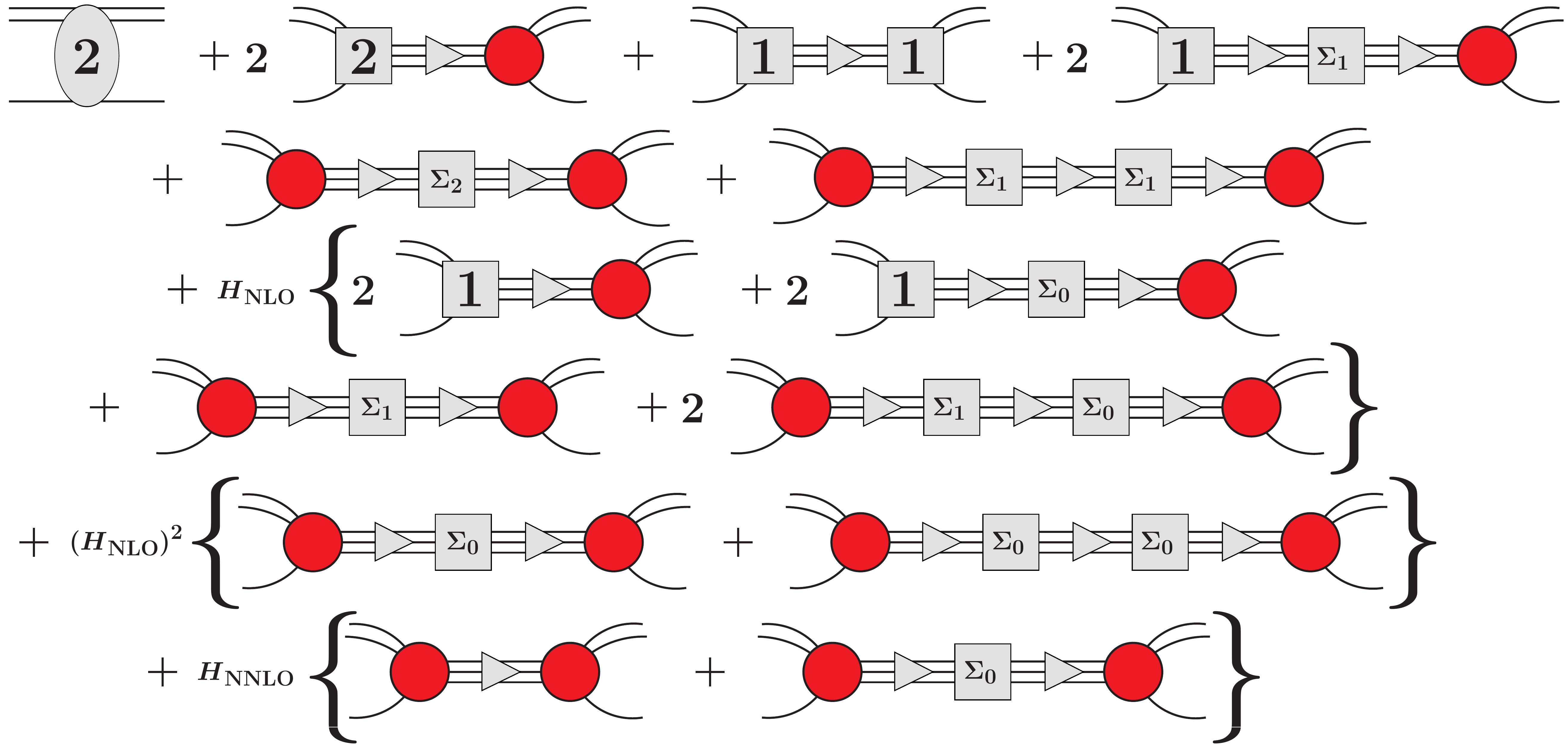}\\[-.8cm]
\includegraphics[width=50mm]{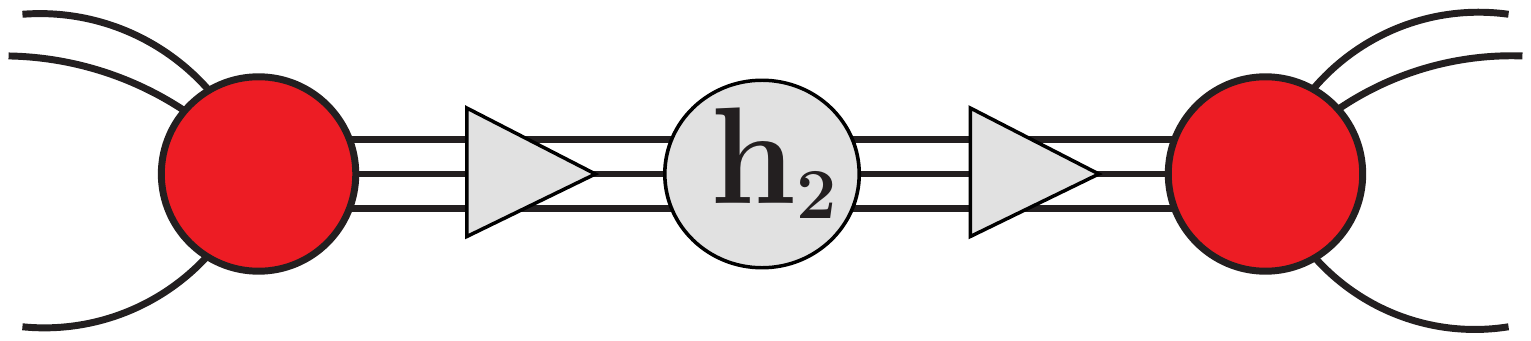}\\[-5.5cm]
\end{tabular}
\end{center}
\caption{Diagrams for the \nnlo correction to the doublet $S$-wave $nd$ scattering amplitude.  The factors of two take into account diagrams related by time reversal symmetry that are not shown.\label{fig:DoubletSNNLO}}
\end{figure}

The NLO $nd$ scattering amplitude is given by the sum of diagrams in Fig.~\ref{fig:DoubletSNLO}.  The factor of two for the second diagram comes from including the time reversed diagram not explicitly shown in Fig.~\ref{fig:DoubletSNLO}.
Summing these yields the NLO $nd$ scattering amplitude
\begin{align}
\label{eq:TNLO}
T_{\mathrm{NLO}}(k)=&Z_{\mathrm{LO}}t_{1,Nt\to Nt}^{\ell=0}(k,k)\\\nonumber
&+\frac{\pi Z_{\mathrm{LO}}}{1-H_{\mathrm{LO}}\Sigma_{0}(E)}\G_{0,\psi\to Nt}(E,k)\left[H_{\mathrm{NLO}}\G_{0,\psi\to Nt}(E,k)+2H_{\mathrm{LO}}\G_{1,\psi\to Nt}(E,k)\right]\\\nonumber
&+\frac{\pi H_{\mathrm{LO}}Z_{\mathrm{LO}}\left[H_{\mathrm{LO}}\Sigma_{1}(E)+H_{\mathrm{NLO}}\Sigma_{0}(E)\right]}{\left[1-H_{\mathrm{LO}}\Sigma_{0}(E)\right]^{2}}\left[\G_{0,\psi\to Nt}(E,k)\right]^{2}.
\end{align}
Again, the NLO three-body force is factored out of all numerically determined expressions and therefore can be algebraically fit to the doublet $S$-wave $nd$ scattering length.  The \nnlo $nd$ scattering amplitude is given by the sum of diagrams in Fig.~\ref{fig:DoubletSNNLO},
which gives
\begin{align}
\label{eq:TNNLO}
T_{\mathrm{N\mathrm{NLO}}}(k)=&Z_{\mathrm{LO}}t_{2,Nt\to Nt}^{\ell=0}(k,k)\\\nonumber
&+\frac{\pi Z_{\mathrm{LO}}}{1-H_{\mathrm{LO}}\Sigma_{0}(E)}\G_{0,\psi\to Nt}(E,k)\\\nonumber
&\hspace{.5cm}\times\left[H_{\mathrm{N\mathrm{NLO}}}\G_{0,\psi\to Nt}(E,k)+2H_{\mathrm{NLO}}\G_{1,\psi\to Nt}(E,k)+2H_{\mathrm{LO}}\G_{2,\psi\to Nt}(E,k)\right]\\\nonumber
&+\frac{\pi H_{\mathrm{LO}}Z_{\mathrm{LO}}\left[H_{\mathrm{LO}}\Sigma_{2}(E)+H_{\mathrm{NLO}}\Sigma_{1}(E)+H_{\mathrm{N\mathrm{NLO}}}\Sigma_{0}(E)\right]}{\left[1-H_{\mathrm{LO}}\Sigma_{0}(E)\right]^{2}}\left[\G_{0,\psi\to Nt}(E,k)\right]^{2}\\\nonumber
&+\frac{\pi H_{\mathrm{LO}}Z_{\mathrm{LO}}\left[H_{\mathrm{LO}}\Sigma_{1}(E)+H_{\mathrm{NLO}}\Sigma_{0}(E)\right]^{2}}{\left[1-H_{\mathrm{LO}}\Sigma_{0}(E)\right]^{3}}\left[\G_{0,\psi\to Nt}(E,k)\right]^{2}\\\nonumber
&+\frac{\pi H_{\mathrm{NLO}}Z_{\mathrm{LO}}\left[H_{\mathrm{LO}}\Sigma_{1}(E)+H_{\mathrm{NLO}}\Sigma_{0}(E)\right]}{\left[1-H_{\mathrm{LO}}\Sigma_{0}(E)\right]^{2}}\left[\G_{0,\psi\to Nt}(E,k)\right]^{2}\\\nonumber
&+\frac{2\pi H_{\mathrm{LO}}Z_{\mathrm{LO}}\left[H_{\mathrm{LO}}\Sigma_{1}(E)+H_{\mathrm{NLO}}\Sigma_{0}(E)\right]}{\left[1-H_{\mathrm{LO}}\Sigma_{0}(E)\right]^{2}}\G_{0,\psi\to Nt}(E,k)\G_{1,\psi\to Nt}(E,k)\\\nonumber
&+\frac{\pi H_{\mathrm{LO}}Z_{\mathrm{LO}}}{1-H_{\mathrm{LO}}\Sigma_{0}(E)}\left(\G_{1,\psi\to Nt}(E,k)\right)^{2}+\frac{\pi\frac{4}{3}(M_{N}E+\gamma_{t}^{2})\widehat{H}	_{2}Z_{\mathrm{LO}}}{\left[1-H_{\mathrm{LO}}\Sigma_{0}(E)\right]^{2}}\left[\G_{0,\psi\to Nt}(E,k)\right]^{2},
\end{align}
When $k=0$ the term with $\widehat{H}_{2}$ disappears and only one new three-body force $H_{\mathrm{NNLO}}$ is present, which can again be solved algebraically and fit to the $nd$ scattering length.  $\widehat{H}_{2}$ can then be fit to the triton binding energy.  In order to find the physical triton binding energy the scattering amplitude can be written in the form
\begin{align}
\label{eq:AmpForm}
\mathbf{t}_{0}(k,p,E)+\mathbf{t}_{1}(k,p,E)+\mathbf{t}_{2}(k,p,E)+\cdots&=\frac{\mathbf{Z}_{0}(k,p)+\mathbf{Z}_{1}(k,p)+\mathbf{Z}_{2}(k,p)}{E-B_{0}-B_{1}-B_{2}+\cdots}\\\nonumber
&+\boldsymbol{\mathcal{R}}_{0}(k,p,E)+\boldsymbol{\mathcal{R}}_{1}(k,p,E)+\boldsymbol{\mathcal{R}}_{2}(k,p,E)+\cdots,
\end{align}
as an expansion about the bound-state pole~\cite{Ji:2011qg,Vanasse:2014kxa}.  There is a pole at the physical triton binding energy $E_{(\jjvH)}=B_{0}+B_{1}+B_{2}+\cdots$, with smooth residue c.c.~space vector functions $\mathbf{Z}_{n}(k,p)$ and smooth remainder c.c.~space vector functions $\boldsymbol{\mathcal{R}}_{n}(k,p,E)$.  Expanding this expression perturbatively gives at LO
\begin{equation}
\mathbf{t}_{0}(k,p,E)=\frac{\mathbf{Z}_{0}(k,p)}{E-B_{0}}+\boldsymbol{\mathcal{R}}_{0}(k,p,E).
\end{equation}
Now the power of this formalism becomes clear because from Eq.~(\ref{eq:TLO}) it can clearly be seen that the pole contribution comes from the second term.  The location of the pole is given by Eq.~(\ref{eq:Pole}) and $\mathbf{Z}_{0}(k,p)$ is simply the residue about this pole, which is
\begin{equation}
\mathbf{Z}_{0}(k,k)=-\frac{\pi Z_{\mathrm{LO}}\left[\G_{0,\psi\to Nt}(B_{0},k)\right]^{2}}{\Sigma_{0}'(B_{0})}.
\end{equation}

At NLO the perturbative expansion of Eq.~(\ref{eq:AmpForm}) gives
\begin{equation}
\mathbf{t}_{1}(k,p,E)=\frac{\mathbf{Z}_{1}(k,p)}{E-B_{0}}+B_{1}\frac{\mathbf{Z}_{0}(k,p)}{(E-B_{0})^{2}}+\boldsymbol{\mathcal{R}}_{1}(k,p,E).
\end{equation}
Comparing to Eq.~(\ref{eq:TNLO}) and using the expression for $\mathbf{Z}_{0}(k,k)$, the contributions from the first and second order pole can easily be extracted, giving the NLO correction to the bound-state energy
\begin{equation}
\label{eq:NLObinding}
B_{1}=-\frac{H_{\mathrm{LO}}\Sigma_{1}(B_{0})+H_{\mathrm{NLO}}\Sigma_{0}(B_{0})}{H_{\mathrm{LO}}\Sigma_{0}'(B_{0})},
\end{equation}
and the NLO residue function 
\begin{equation}
\mathbf{Z}_{1}(k,k)=-\frac{\pi Z_{\mathrm{LO}}\G_{0,\psi\to Nt}(B_{0},k)\left[H_{\mathrm{NLO}}\G_{0,\psi\to Nt}(B_{0},k)+2 H_{\mathrm{LO}}\G_{1,\psi\to Nt}(B_{0},k)\right]}{H_{\mathrm{LO}}\Sigma_{0}'(B_{0})}.
\end{equation}
The \nnlo perturbative expansion of Eq.~(\ref{eq:AmpForm}) gives
\begin{equation}
\mathbf{t}_{2}(k,p,E)=\frac{\mathbf{Z}_{2}(k,p)}{E-B_{0}}+B_{2}\frac{\mathbf{Z}_{0}(k,p)}{(E-B_{0})^{2}}+B_{1}\frac{\mathbf{Z}_{1}(k,p)}{(E-B_{0})^{2}}+B_{1}^{2}\frac{\mathbf{Z}_{0}(k,p)}{(E-B_{0})^{3}}+\boldsymbol{\mathcal{R}}_{2}(k,p,E).
\end{equation}
Since $\mathbf{Z}_{1}(k,k)$ and $B_{1}$ are known, their second order pole contribution can be subtracted from Eq.~(\ref{eq:TNNLO}) leaving the contribution from $B_{2}$, which is given by
\begin{align}
\label{eq:NNLObinding}
B_{2}=-&\frac{H_{\mathrm{LO}}\Sigma_{2}(B_{0})+H_{\mathrm{NLO}}\Sigma_{1}(B_{0})+\left[H_{\mathrm{NNLO}}+\frac{4}{3}(M_{N}B_{0}+\gamma_{t}^{2})\widehat{H}_{2}\right]\Sigma_{0}(B_{0})}{H_{\mathrm{LO}}\Sigma_{0}'(B_{0})}\\\nonumber
&-B_{1}\frac{H_{\mathrm{LO}}\Sigma_{1}'(B_{0})+H_{\mathrm{NLO}}\Sigma_{0}'(B_{0})}{H_{\mathrm{LO}}\Sigma_{0}'(B_{0})}-\frac{1}{2}B_{1}^{2}\frac{\Sigma_{0}''(B_{0})}{\Sigma_{0}'(B_{0})}.
\end{align}
To fit $\widehat{H}_{2}$ to the bound-state energy one adjusts $\widehat{H}_{2}$ such that $E_{(\jjvH)}=B_{0}+B_{1}+B_{2}$.  Note that if one sets $B_{1}$ and $B_{2}$ to zero then the constraints on the three-body forces are equivalent to Eqs.~(\ref{eq:HNLOdef}) and (\ref{eq:HNNLOdef}) where the three-body forces were fit to the bound-state energy by fixing the pole position for the triton propagator. This formalism reproduces the results for three-body forces and doublet $S$-wave scattering amplitudes found in Ref.~\cite{Vanasse:2013sda} up to numerical accuracy.  But it is superior because it avoids iterative techniques for $H_{\mathrm{LO}}$ and numerical limiting procedures for $\widehat{H}_{2}$.


\section{\label{sec:chargeform} Triton Charge Form Factor}

\begin{figure}[hbt]
\includegraphics[width=100mm]{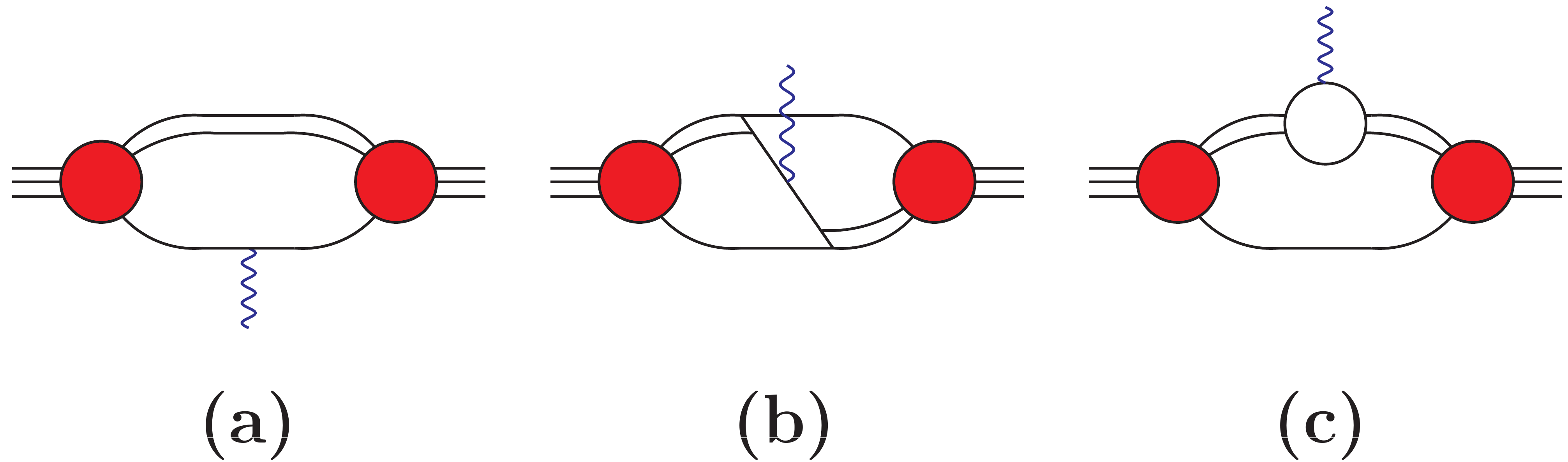}
\caption{Diagrams for the LO triton charge form factor.  The wavy blue lines represent minimally coupled $\hat{A}_{0}$ photons.\label{fig:FormFactorLO}}
\end{figure}

The LO triton charge form factor is given by the sum of diagrams in Fig.~\ref{fig:FormFactorLO}, where the wavy blue lines are minimally coupled $\hat{A}_{0}$ photons.  The form factor calculation is performed in the Breit frame in which the photon imparts no energy to the triton but only momentum.  In the Breit frame one chooses the initial (final) momentum of the triton to be $\vect{K}$ ($\vect{P}$).  The momentum imparted by the photon is $\vect{Q}=\vect{P}-\vect{K}$, and the form factor only depends on the value $\vect{Q}^{2}$.
Summing all three diagrams in the Breit frame gives
\begin{align}
\label{eq:LOformfactor}
Z_{\psi}^{\mathrm{LO}}\sum_{j=a,b,c}\int\!\!\frac{d^{4}k}{(2\pi)^{4}}\int\!\!\frac{d^{4}p}{(2\pi)^{4}}\Gb_{0}^{T}(E,\vect{P},p_{0},\vect{p})\boldsymbol{\chi}_{j}(E,\vect{K},\vect{P},p_{0},k_{0},\vect{p},\vect{k})\Gb_{0}(E,\vect{K},k_{0},\vect{k}),
\end{align}
where $\Gb_{0}(E,\vect{K},k_{0},\vect{k})$ is the LO triton vertex function in a frame boosted by momentum $\vect{K}$, and $E=B_{0}+\frac{1}{6M_{N}}K^{2}$, with $B_{0}=E_{(\jjvH)}$, is the total energy of the triton in this frame.  The functional forms of $\boldsymbol{\chi}_{j}(E,\vect{K},\vect{P},p_{0},k_{0},\vect{p},\vect{k})$ are listed in App.~\ref{app:chi}.  Choosing the four momentum of the dibaryon (nucleon) to be $[\frac{2}{3}E+k_{0},\vect{k}+\frac{2}{3}\vect{K}]$ ($[\frac{1}{3}E-k_{0},-\vect{k}+\frac{1}{3}\vect{K}]$) the triton vertex function in the boosted frame is related to the triton vertex function in the c.m.~frame via
\begin{align}
\label{eq:GLOboosted}
&\Gb_{0}(E,\vect{K},k_{0},\vect{k})=\widetilde{\mathbf{B}}_{0}\\\nonumber
&\hspace{2cm}+\left[\mathbf{R}_{0}\left(q,k,\frac{2}{3}B_{0}+k_{0}-\frac{\vect{K}\cdot\vect{k}}{3M_{N}}+\frac{\vect{k}^{2}}{2M_{N}}\right)\mathbf{D}^{(0)}\left(B_{0}-\frac{\vect{q}^{2}}{2M_{N}},\vect{q}\right)\right]\otimes\Gb_{0}(B_{0},q).
\end{align}
For diagram (a), $\boldsymbol{\chi}_{a}(\cdots)$ gives delta functions over momentum and energy that remove the integral over $d^{4}p$.  Then integrating over the energy $k_{0}$ and using Eq.~(\ref{eq:GLOboosted}) the LO contribution from diagram (a) can be written as
\begin{align}
\label{eq:FLOa}
&F_{0}^{(a)}(Q^{2})=Z_{\psi}^{\mathrm{LO}}\left\{\widetilde{\Gb}_{0}^{T}(p)\otimes \boldsymbol{\mathcal{A}}_{0}(p,k,Q)\otimes\widetilde{\Gb}_{0}(k)+2\widetilde{\Gb}_{0}^{T}(p)\otimes \boldsymbol{\mathcal{A}}_{0}(p,Q)+\mathcal{A}_{0}(Q)\right\}.
\end{align}
The subscript ``0" in the functions $F_{n}(Q^{2})$, $\boldsymbol{\mathcal{A}}_{n}(\cdots)$, $\mathcal{A}_{n}(Q)$,	 and $\widetilde{\Gb}_{n}(p)$ refer to LO. NLO and \nnlo contributions will be denoted by a ``1" and ``2" subscript respectively.  The function $\boldsymbol{\mathcal{A}}_{n}(p,k,Q)$ is a matrix function in c.c.~space, $\boldsymbol{\mathcal{A}}_{n}(p,Q)$ a vector function in c.c.~space, and $\mathcal{A}_{n}(Q)$ a scalar function.  Further details of this calculation and the form of the functions $\boldsymbol{\mathcal{A}}_{n}(\cdots)$ and $\mathcal{A}_{n}(Q)$ are given in App.~\ref{app:chi}.  The vector function $\widetilde{\Gb}_{n}(p)$ in c.c.~space is defined as
\begin{equation}
\widetilde{\Gb}_{n}(p)=\mathbf{D}^{(0)}\!\!\left(B_{0}-\frac{\vect{p}^{2}}{2M_{N}},\vect{p}\right)\Gb_{n}(B_{0},p).
\end{equation}
Diagram (b) of Fig.~\ref{fig:FormFactorLO} can be written as
\begin{align}
\label{eq:FLOb}
&F_{0}^{(b)}(Q^{2})=Z_{\psi}^{\mathrm{LO}}\widetilde{\Gb}_{0}^{T}(p)\otimes \boldsymbol{\mathcal{B}}_{0}(p,k,Q)\otimes\widetilde{\Gb}_{0}(k),
\end{align}
where $\boldsymbol{\mathcal{B}}_{0}(p,k,Q)$ is a matrix function in c.c.~space given in the App.~\ref{app:chi}.  For diagram (c) 
\begin{align}
\label{eq:FLOc}
&F_{0}^{(c)}(Q^{2})=Z_{\psi}^{\mathrm{LO}}\left\{\widetilde{\Gb}_{0}^{T}(p)\otimes \boldsymbol{\mathcal{C}}_{0}(p,k,Q)\otimes\widetilde{\Gb}_{0}(k)+\boldsymbol{\mathcal{C}}_{0}(k,Q)\otimes\widetilde{\Gb}_{0}(k)\right\},
\end{align}
where $\boldsymbol{\mathcal{C}}_{0}(p,k,Q)$ is a matrix function in c.c.~space and $\boldsymbol{\mathcal{C}}_{0}(k,Q)$ a vector function in c.c.~space\footnote{ Note that in Ref.~\cite{Hagen:2013xga} only the first term for $F_{0}^{(c)}(Q^{2})$ exists.  This is due to the difference in LO three-body forces between these two calculations.}.  Summing the contribution from all diagrams the LO triton charge form factor is given by
\begin{equation}
F_{0}(Q^{2})=F_{0}^{(a)}(Q^{2})+F_{0}^{(b)}(Q^{2})+F_{0}^{(c)}(Q^{2}).
\end{equation}
In the limit $Q^{2}\to 0$ $F_{0}(0)=1$ up to numerical accuracy.  It can be shown analytically that in the limit $Q^{2}\to 0$ the renormalization condition given in Ref.~\cite{Konig:2011yq} for the LO homogeneous solution of the doublet $S$-wave channel is recovered from $F_{0}(0)$.  This is shown in further detail in App.~\ref{app:norm}.

\begin{figure}[hbt]
\includegraphics[width=100mm]{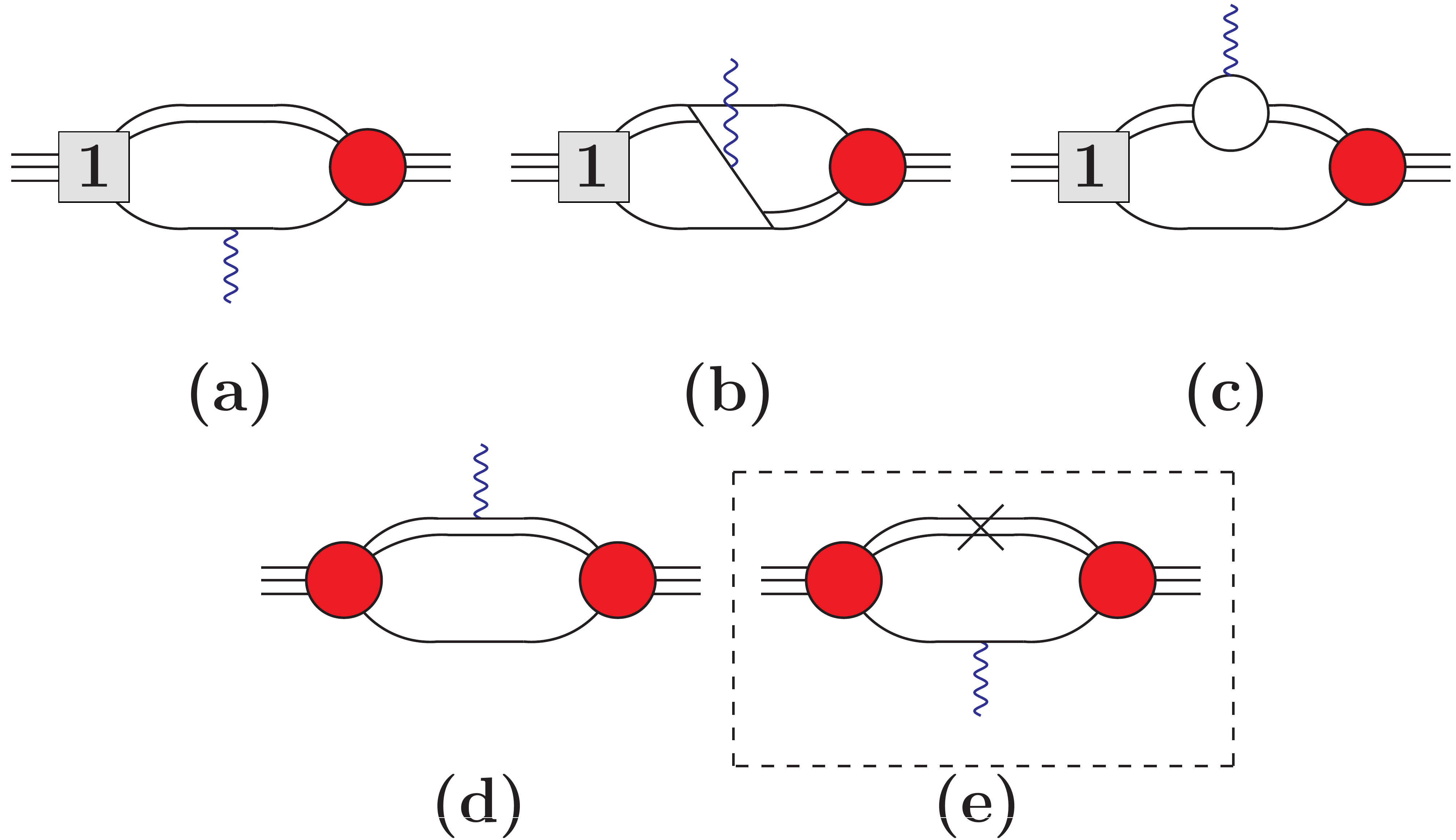}
\caption{Diagrams for the NLO correction to the triton charge form factor, where diagrams related by time reversal symmetry are not shown.  The diagram in the dashed box is subtracted from the other diagrams to avoid double counting.  The photon in diagram (d) is minimally coupled to the dibaryon.\label{fig:FormFactorNLO}}
\end{figure}

The NLO correction to the triton charge form factor is given by the diagrams in Fig.~\ref{fig:FormFactorNLO}. 
Diagrams (a) through (d) are added together while diagram (e) is subtracted to avoid double counting from diagram (a) and its time reversed version.  The photon in diagram (d) is minimally coupled via the dibaryon kinetic term.  Diagrams related by time reversal symmetry are not shown in Fig.~\ref{fig:FormFactorNLO}.  The sum of diagrams (a)-(d) and subtraction of diagram (e) is given by
\begin{align}
\label{eq:FormNLO}
&Z_{\psi}^{\mathrm{LO}}\sum_{j=a,b,c}\int\!\!\frac{d^{4}k}{(2\pi)^{4}}\int\!\!\frac{d^{4}p}{(2\pi)^{4}}\left\{\Gb_{1}^{T}(E,\vect{P},p_{0},\vect{p})\boldsymbol{\chi}_{j}(E,\vect{K},\vect{P},p_{0},k_{0},\vect{p},\vect{k})\Gb_{0}(E,\vect{K},k_{0},\vect{k})\right.\\\nonumber
&\left.\hspace{5cm}+\Gb_{0}^{T}(E,\vect{P},p_{0},\vect{p})\boldsymbol{\chi}_{j}(E,\vect{K},\vect{P},p_{0},k_{0},\vect{p},\vect{k})\Gb_{1}(E,\vect{K},k_{0},\vect{k})\right\}\\\nonumber
&+Z_{\psi}^{\mathrm{LO}}\sum_{d,-e}\int\!\!\frac{d^{4}k}{(2\pi)^{4}}\int\!\!\frac{d^{4}p}{(2\pi)^{4}}\Gb_{0}^{T}(E,\vect{P},p_{0},\vect{p})\boldsymbol{\chi}_{j}(E,\vect{K},\vect{P},p_{0},k_{0},\vect{p},\vect{k})\Gb_{0}(E,\vect{K},k_{0},\vect{k}).
\end{align}
Functions $\boldsymbol{\chi}_{j}(\cdots)$ for $j=a,b,c$ are the same as in the LO case.  At NLO there are new functions $\boldsymbol{\chi}_{d}(\cdots)$ and $\boldsymbol{\chi}_{e}(\cdots)$.  To obtain Eq.~(\ref{eq:FormNLO}) the LO expression Eq.~(\ref{eq:LOformfactor}) is replaced by NLO corrections wherever possible.  The NLO correction to the triton vertex function in a boosted frame is related to the NLO correction to the triton vertex function in the c.m.~frame by
\begin{align}
\label{eq:GNLOboosted}
&\Gb_{1}(E,\vect{K},k_{0},\vect{k})=\Gb_{0}(E,\vect{K},k_{0},\vect{k})\circ\mathbf{R}_{1}\left(\frac{2}{3}E+k_{0},\vect{k}+\frac{2}{3}\vect{K}\right)\\\nonumber
&\hspace{2cm}+\left[\mathbf{R}_{0}\left(q,k,\frac{2}{3}B_{0}+k_{0}-\frac{\vect{K}\cdot\vect{k}}{3M_{N}}+\frac{\vect{k}^{2}}{2M_{N}}\right)\mathbf{D}^{(0)}\!\!\left(B_{0}-\frac{\vect{q}^{2}}{2M_{N}},\vect{q}\right)\right]\otimes\Gb_{1}(B_{0},q).
\end{align}
Using Eq.~(\ref{eq:GLOboosted}) the NLO correction to the triton vertex function in a boosted frame can be written entirely in terms of c.m.~quantities.  The NLO contribution from diagram (a) minus diagram (e) is given by
\begin{align}
\label{eq:NLOatriton}
&F_{1}^{(a)}(Q^{2})=Z_{\psi}^{\mathrm{LO}}\left\{\widetilde{\Gb}_{0}^{T}(p)\otimes \boldsymbol{\mathcal{A}}_{1}(p,k,Q)\otimes\widetilde{\Gb}_{0}(k)+2\widetilde{\Gb}_{1}^{T}(p)\otimes \boldsymbol{\mathcal{A}}_{0}(p,k,Q)\otimes\widetilde{\Gb}_{0}(k)\right.\\\nonumber
&\left.\hspace{5cm}+2\widetilde{\Gb}_{0}^{T}(p)\otimes \boldsymbol{\mathcal{A}}_{1}(p,Q)+ 2\widetilde{\Gb}_{1}^{T}(p)\otimes \boldsymbol{\mathcal{A}}_{0}(p,Q)+\mathcal{A}_{1}(Q)\right\}.
\end{align}
To obtain this NLO expression one replaces all LO terms in Eq.~(\ref{eq:FLOa}) by their NLO counterparts.  The functions $\boldsymbol{\mathcal{A}}_{1}(\cdots)$ and $\mathcal{A}_{1}(Q)$ only differ from $\boldsymbol{\mathcal{A}}_{0}(\cdots)$ and $\mathcal{A}_{0}(Q)$ by the replacement of a LO dibaryon propagator by a NLO correction to the dibaryon propagator.  Again further details and their functional forms can be seen in App.~\ref{app:chi}.  The NLO contribution from diagram (b) is given by 
\begin{align}
&F_{1}^{(b)}(Q^{2})=Z_{\psi}^{\mathrm{LO}}\left\{2\widetilde{\Gb}_{1}^{T}(p)\otimes \boldsymbol{\mathcal{B}}_{0}(p,k,Q)\otimes\widetilde{\Gb}_{0}(k)\right\},
\end{align}
for diagram (c) by
\begin{align}
&F_{1}^{(c)}(Q^{2})=Z_{\psi}^{\mathrm{LO}}\left\{\widetilde{\Gb}_{0}^{T}(p)\otimes \boldsymbol{\mathcal{C}}_{1}(p,k,Q)\otimes\widetilde{\Gb}_{0}(k)+\widetilde{\Gb}_{1}^{T}(p)\otimes \boldsymbol{\mathcal{C}}_{0}(p,k,Q)\otimes\widetilde{\Gb}_{0}(k)\right.\\\nonumber
&\left.\hspace{2cm}+\widetilde{\Gb}_{0}^{T}(p)\otimes \boldsymbol{\mathcal{C}}_{0}(p,k,Q)\otimes\widetilde{\Gb}_{1}(k)+\boldsymbol{\mathcal{C}}_{1}(k,Q)\otimes\widetilde{\Gb}_{0}(k)+\boldsymbol{\mathcal{C}}_{0}(k,Q)\otimes\widetilde{\Gb}_{1}(k)\right\}.
\end{align}
and finally diagram (d) by
\begin{align}
&F_{1}^{(d)}(Q^{2})=Z_{\psi}^{\mathrm{LO}}\left\{\widetilde{\Gb}_{0}^{T}(p)\otimes \boldsymbol{\mathfrak{D}}_{1}(p,k,Q)\otimes\widetilde{\Gb}_{0}(k)+\boldsymbol{\mathfrak{D}}_{1}(k,Q)\otimes\widetilde{\Gb}_{0}(k)\right\}.
\end{align}
The function $\boldsymbol{\mathfrak{D}}_{n}(p,k,Q)$ is a matrix function in c.c.~space and $\boldsymbol{\mathfrak{D}}_{n}(k,Q)$ a vector function in c.c.~space.  For the functions $\boldsymbol{\mathfrak{D}}_{n}(\cdots)$ $n=0$ does not occur; its first contribution is at NLO.  The functions $\boldsymbol{\mathcal{B}}_{1}(p,k,Q)$ and $\boldsymbol{\mathcal{B}}_{2}(p,k,Q)$ also do not exist.  Summing all of the NLO contributions, replacing $\omega_{t0}^{(0)}$ and $\omega_{s0}^{(0)}$ by $\omega_{t0}^{(1)}$ and $\omega_{s0}^{(1)}$ in the LO contributions, and multiplying the LO contribution by the NLO triton wavefunction renormalization gives 
\begin{equation}
F_{1}(Q^{2})=\left(F_{1}^{(a)}(Q^{2})+F_{1}^{(b)}(Q^{2})+F_{1}^{(c)}(Q^{2})+F_{1}^{(d)}(Q^{2})\right)-\frac{\Sigma_{1}'}{\Sigma_{0}'}F_{0}(Q^{2}),
\end{equation}
for the NLO correction to the triton charge form factor.  In the limit $Q^{2}\to 0$ $F_{1}(0)=0$ up to numerical accuracy.

\begin{figure}[hbt]
\includegraphics[width=100mm]{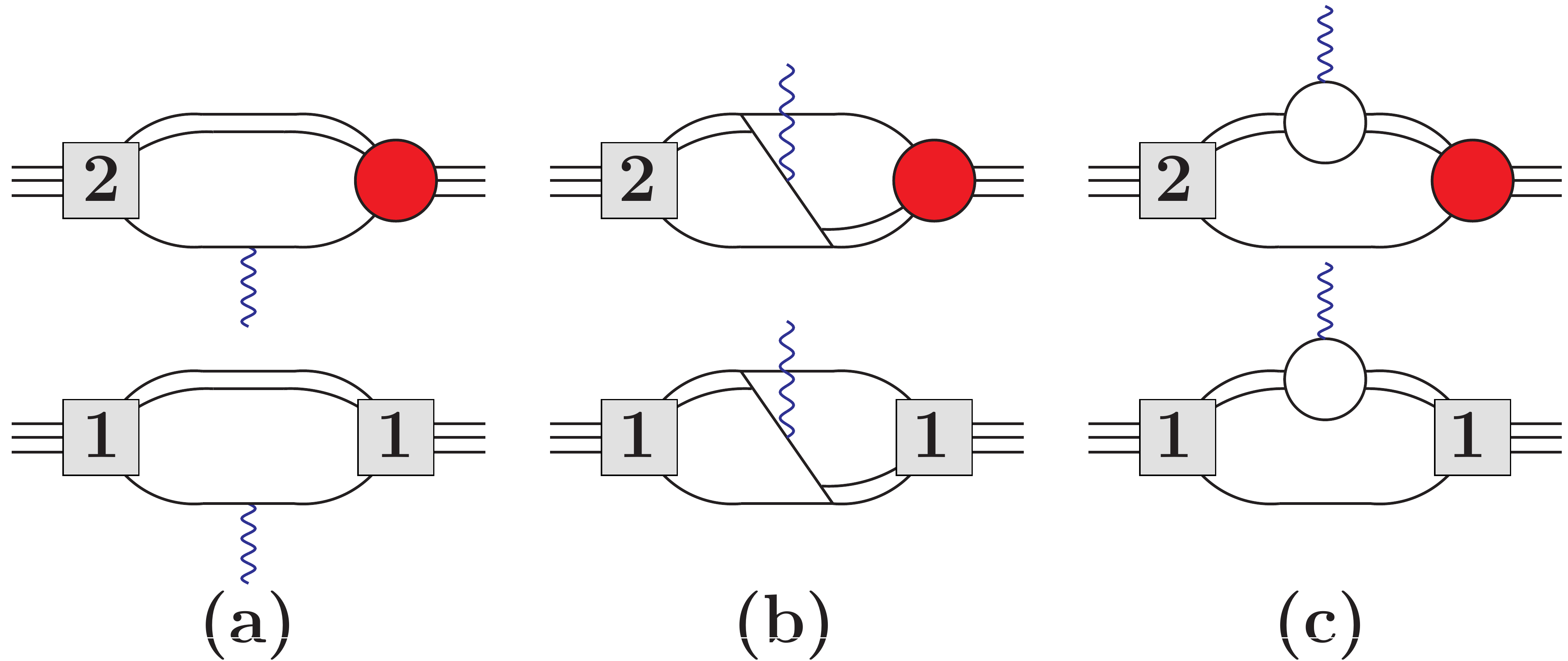}
\includegraphics[width=100mm]{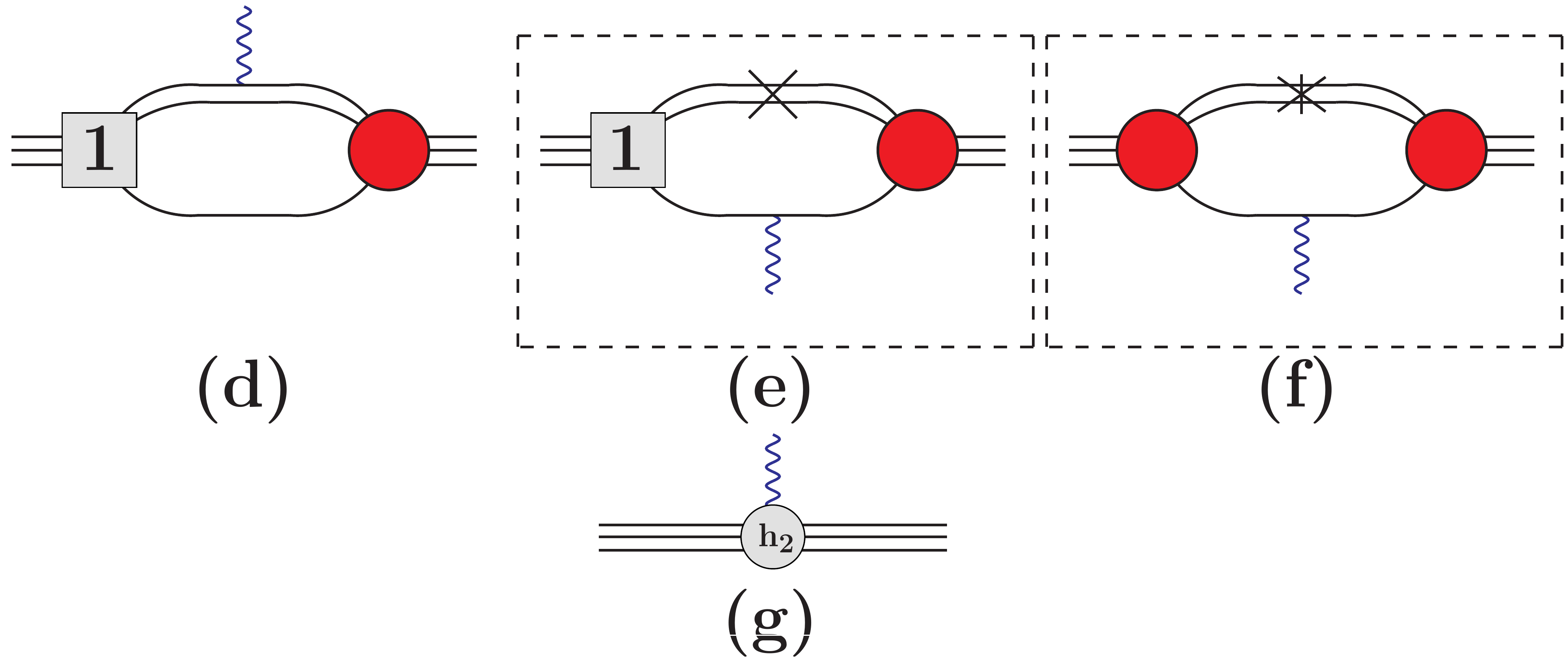}
\caption{Diagrams for the \nnlo correction to the triton charge form factor, where diagrams related by time reversal symmetry are not shown.  The diagrams in the dashed boxes are subtracted from the other diagrams to avoid double counting.\label{fig:FormFactorNNLO}}
\end{figure}

The \nnlo correction to the triton charge form factor is given by the diagrams in Fig.~\ref{fig:FormFactorNNLO}.  Diagrams of type (a) through (d) are added while diagrams (e) and (f) are subtracted to avoid double counting from (a) type diagrams and their time reversed versions.  Again diagrams related by time reversal symmetry are not shown.  Diagram (g) comes from gauging the kinetic term of the triton field.
Analogously to the NLO case the sum of diagrams (a) through (d) and subtraction of diagram (e) and (f) at \nnlo is given by
\begin{align}
&Z_{\psi}^{\mathrm{LO}}\sum_{j=a,b,c}\int\!\!\frac{d^{4}k}{(2\pi)^{4}}\int\!\!\frac{d^{4}p}{(2\pi)^{4}}\left\{\Gb_{2}^{T}(E,\vect{P},p_{0},\vect{p})\boldsymbol{\chi}_{j}(E,\vect{K},\vect{P},p_{0},k_{0},\vect{p},\vect{k})\Gb_{0}(E,\vect{K},k_{0},\vect{k})\right.\\[-.6cm]\nonumber
&\hspace{5cm}+\Gb_{0}^{T}(E,\vect{P},p_{0},\vect{p})\boldsymbol{\chi}_{j}(E,\vect{K},\vect{P},p_{0},k_{0},\vect{p},\vect{k})\Gb_{2}(E,\vect{K},k_{0},\vect{k})\\\nonumber
&\left.\hspace{5cm}+\Gb_{1}^{T}(E,\vect{P},p_{0},\vect{p})\boldsymbol{\chi}_{j}(E,\vect{K},\vect{P},p_{0},k_{0},\vect{p},\vect{k})\Gb_{1}(E,\vect{K},k_{0},\vect{k})\right\}\\\nonumber
&+Z_{\psi}^{\mathrm{LO}}\sum_{j=d,-e}\int\!\!\frac{d^{4}k}{(2\pi)^{4}}\int\!\!\frac{d^{4}p}{(2\pi)^{4}}\left\{\Gb_{1}^{T}(E,\vect{P},p_{0},\vect{p})\boldsymbol{\chi}_{j}(E,\vect{K},\vect{P},p_{0},k_{0},\vect{p},\vect{k})\Gb_{0}(E,\vect{K},k_{0},\vect{k})\right.\\\nonumber
&\left.\hspace{5cm}+\Gb_{0}^{T}(E,\vect{P},p_{0},\vect{p})\boldsymbol{\chi}_{j}(E,\vect{K},\vect{P},p_{0},k_{0},\vect{p},\vect{k})\Gb_{1}(E,\vect{K},k_{0},\vect{k})\right\}\\\nonumber
&+Z_{\psi}^{\mathrm{LO}}\sum_{j=-f}\int\!\!\frac{d^{4}k}{(2\pi)^{4}}\int\!\!\frac{d^{4}p}{(2\pi)^{4}}\Gb_{0}^{T}(E,\vect{P},p_{0},\vect{p})\boldsymbol{\chi}_{j}(E,\vect{K},\vect{P},p_{0},k_{0},\vect{p},\vect{k})\Gb_{0}(E,\vect{K},k_{0},\vect{k}).
\end{align}
The \nnlo correction to the triton vertex function in a boosted frame is related to the \nnlo correction to the triton vertex function in the c.m.~frame via
\begin{align}
&\Gb_{2}(E,\vect{K},k_{0},\vect{k})=\left[\Gb_{1}(E,\vect{K},k_{0},\vect{k})-\mathbf{c}_{1}\circ\Gb_{0}(E,\vect{K},k_{0},\vect{k})\right]\circ\mathbf{R}_{1}\left(\frac{2}{3}E+k_{0},\vect{k}+\frac{2}{3}\vect{K}\right)\\\nonumber
&\hspace{2cm}+\left[\mathbf{R}_{0}\left(q,k,\frac{2}{3}B_{0}+k_{0}-\frac{\vect{K}\cdot\vect{k}}{3M_{N}}+\frac{\vect{k}^{2}}{2M_{N}}\right)\mathbf{D}^{(0)}\!\!\left(B_{0}-\frac{\vect{q}^{2}}{2M_{N}},\vect{q}\right)\right]\otimes\Gb_{2}(B_{0},q).
\end{align}
Using Eqs.~(\ref{eq:GLOboosted}) and (\ref{eq:GNLOboosted}) the \nnlo correction to the triton vertex function in a boosted frame can be written in terms of c.m.~quantities.  The sum of type (a) diagrams minus diagrams (e) and (f) gives
\begin{align}
\label{eq:NNLOatriton}
&F_{2}^{(a)}(Q^{2})=Z_{\psi}^{\mathrm{LO}}\left\{\widetilde{\Gb}_{0}^{T}(p)\otimes \boldsymbol{\mathcal{A}}_{2}(p,k,Q)\otimes\widetilde{\Gb}_{0}(k)+2\widetilde{\Gb}_{1}^{T}(p)\otimes \boldsymbol{\mathcal{A}}_{1}(p,k,Q)\otimes\widetilde{\Gb}_{0}(k)\right.\\\nonumber
&\hspace{2cm}+2\widetilde{\Gb}_{2}^{T}(p)\otimes \boldsymbol{\mathcal{A}}_{0}(p,k,Q)\otimes\widetilde{\Gb}_{0}(k)+\widetilde{\Gb}_{1}^{T}(p)\otimes \boldsymbol{\mathcal{A}}_{0}(p,k,Q)\otimes\widetilde{\Gb}_{1}(k)\\\nonumber
&\hspace{2cm}\left.+2\widetilde{\Gb}_{0}^{T}(p)\otimes \boldsymbol{\mathcal{A}}_{2}(p,Q)+ 2\widetilde{\Gb}_{1}^{T}(p)\otimes \boldsymbol{\mathcal{A}}_{1}(p,Q)+ 2\widetilde{\Gb}_{2}^{T}(p)\otimes \boldsymbol{\mathcal{A}}_{0}(p,Q)+\mathcal{A}_{2}(Q)\right\}.
\end{align}
As in the NLO case all functions in Eq.~(\ref{eq:FLOa}) are replaced by their \nnlo counterparts.  In addition terms where two expressions are replaced by their NLO counterparts are included.  The functions $\boldsymbol{\mathcal{A}}_{2}(\cdots)$ and $\mathcal{A}_{2}(Q)$ are the same as $\boldsymbol{\mathcal{A}}_{0}(\cdots)$ and $\mathcal{A}_{0}(Q)$ respectively except with LO dibaryon propagators replaced by the \nnlo correction to the dibaryon propagators.  Diagrams of type (b) give the contribution
\begin{align}
&F_{2}^{(b)}(Q^{2})=Z_{\psi}^{\mathrm{LO}}\left\{2\widetilde{\Gb}_{2}^{T}(p)\otimes \boldsymbol{\mathcal{B}}_{0}(p,k,Q)\otimes\widetilde{\Gb}_{0}(k)+\widetilde{\Gb}_{1}^{T}(p)\otimes \boldsymbol{\mathcal{B}}_{0}(p,k,Q)\otimes\widetilde{\Gb}_{1}(k)\right\},
\end{align}
diagrams of type (c) give
\begin{align}
&F_{2}^{(c)}(Q^{2})=Z_{\psi}^{\mathrm{LO}}\left\{\widetilde{\Gb}_{0}^{T}(p)\otimes \boldsymbol{\mathcal{C}}_{2}(p,k,Q)\otimes\widetilde{\Gb}_{0}(k)+2\widetilde{\Gb}_{1}^{T}(p)\otimes \boldsymbol{\mathcal{C}}_{1}(p,k,Q)\otimes\widetilde{\Gb}_{0}(k)\right.\\\nonumber
&\hspace{3.5cm}+2\widetilde{\Gb}_{2}^{T}(p)\otimes \boldsymbol{\mathcal{C}}_{0}(p,k,Q)\otimes\widetilde{\Gb}_{0}(k)+\widetilde{\Gb}_{1}^{T}(p)\otimes \boldsymbol{\mathcal{C}}_{0}(p,k,Q)\otimes\widetilde{\Gb}_{1}(k)\\\nonumber
&\hspace{3.5cm}\left.+\boldsymbol{\mathcal{C}}_{2}(k,Q)\otimes\widetilde{\Gb}_{0}(k)+\boldsymbol{\mathcal{C}}_{1}(k,Q)\otimes\widetilde{\Gb}_{1}(k)+\boldsymbol{\mathcal{C}}_{0}(k,Q)\otimes\widetilde{\Gb}_{2}(k)\right\},
\end{align}
and diagram (d) gives
\begin{align}
&F_{2}^{(d)}(Q^{2})=Z_{\psi}^{\mathrm{LO}}\left\{\widetilde{\Gb}_{0}^{T}(p)\otimes \boldsymbol{\mathfrak{D}}_{2}(p,k,Q)\otimes\widetilde{\Gb}_{0}(k)+2\widetilde{\Gb}_{1}^{T}(p)\otimes \boldsymbol{\mathfrak{D}}_{1}(p,k,Q)\otimes\widetilde{\Gb}_{0}(k)\right.\\\nonumber
&\left.\hspace{7cm}+\boldsymbol{\mathfrak{D}}_{2}(k,Q)\otimes\widetilde{\Gb}_{0}(k)+\boldsymbol{\mathfrak{D}}_{1}(k,Q)\otimes\widetilde{\Gb}_{1}(k)\right\}.
\end{align}
Finally, the contribution from diagram (g) is given by the constant term
\begin{equation}
\frac{4}{3}M_{N}\widehat{H}_{2}\frac{\Sigma_{0}^{2}}{\Sigma_{0}'}
\end{equation}
Summing all of the \nnlo corrections to the triton charge form factor, replacing $\omega_{t0}^{(0)}$ and $\omega_{s0}^{(0)}$ by $\omega_{t0}^{(2)}$ and $\omega_{s0}^{(2)}$ and two factors of $\omega_{t0}^{(1)}$ and $\omega_{s0}^{(1)}$ in the LO contributions, replacing $\omega_{t0}^{(0)}$ and $\omega_{s0}^{(0)}$ by $\omega_{t0}^{(1)}$ and $\omega_{s0}^{(1)}$ in the NLO contributions, multiplying the NLO correction by the NLO triton wavefunction renormalization, and multiplying the LO term by the \nnlo triton wavefunction renormalization yields the \nnlo triton charge form factor
\begin{align}
&F_{2}(Q^{2})=\left(F_{2}^{(a)}(Q^{2})+F_{2}^{(b)}(Q^{2})+F_{2}^{(c)}(Q^{2})+F_{2}^{(d)}(Q^{2})\right)\\\nonumber
&\hspace{3cm}-\frac{\Sigma_{1}'}{\Sigma_{0}'}\left(F_{1}^{(a)}(Q^{2})+F_{1}^{(b)}(Q^{2})+F_{1}^{(c)}(Q^{2})+F_{1}^{(d)}(Q^{2})\right)\\\nonumber
&\hspace{3cm}+\left[\left(\frac{\Sigma_{1}'}{\Sigma_{0}'}\right)^{2}-\frac{\Sigma_{2}'}{\Sigma_{0}'}-\frac{4}{3}M_{N}\widehat{H}_{2}\frac{\Sigma_{0}^{2}}{\Sigma_{0}'}\right]F_{0}(Q^{2})+\frac{4}{3}M_{N}\widehat{H}_{2}\frac{\Sigma_{0}^{2}}{\Sigma_{0}'}
\end{align}

In the limit $Q^{2}\to0$ it should hold that $F_{2}(0)=0$.  However, it is found that $F_{2}(0)\sim10^{-8}$, which is only one order of magnitude smaller than the deviation of the LO value of the triton charge form factor from the value $F_{0}(0)=1$ for $Q^{2}\sim0.1$~MeV$^{2}$.  This is due to the fact that this qauntity is very fine tuned with respect to the three-body force $H_{\mathrm{\nnlo}}$: taking $H_{\mathrm{\nnlo}}$ fit to the triton binding energy and varying it by one part in $10^{12}$ it is found that $F_{2}(0)\sim10^{-15}$.  Despite the value of $F_{2}(0)$ being highly fine tuned with respect to $H_{\mathrm{\nnlo}}$ no such level of fine tuning is seen for the \nnlo correction to the triton point charge radius.  In other words the slope of the \nnlo correction to the triton charge form factor with respect to $Q^{2}$ is not fine tuned with respect to $H_{\mathrm{\nnlo}}$, but the y-intercept is.


\section{\label{sec:results}Triton point charge radius and Results}

The triton charge form factor can be expanded in powers of $Q^{2}$ yielding
\begin{equation}
F(Q^{2})=1-\frac{\left<r_{\jjvH}^{2}\right>}{6}Q^{2}+\cdots,
\end{equation}
where $\delta r_{C}=	\sqrt{\left<r_{\jjvH}^{2}\right>}$ is the triton point charge radius.  At LO the triton charge form factor is given by
\begin{equation}
F_{0}(Q^{2})=1-\frac{\left<r_{\jjvH}^{2}\right>_{0}}{6}Q^{2}+\cdots,
\end{equation}
where $\left<r_{\jjvH}^{2}\right>_{0}$ is the LO contribution to $(\delta r_{C})^{2}$.  The NLO correction to the triton charge form factor is given by
\begin{equation}
F_{1}(Q^{2})=-\frac{\left<r_{\jjvH}^{2}\right>_{1}}{6}Q^{2}+\cdots,
\end{equation}
and the \nnlo correction by
\begin{equation}
F_{2}(Q^{2})=-\frac{\left<r_{\jjvH}^{2}\right>_{2}}{6}Q^{2}+\cdots.
\end{equation}
$\left<r_{\jjvH}^{2}\right>_{1}$ is the NLO correction to $\delta r_{C}^{2}$ and $\left<r_{\jjvH}^{2}\right>_{2}$ is the \nnlo correction to $\delta r_{C}^{2}$, and the square of the triton point charge radius to \nnlo is simply given by
\begin{equation}
\left<\delta r_{C}^{2}\right>=\left<r_{\jjvH}^{2}\right>_{0}+\left<r_{\jjvH}^{2}\right>_{1}+\left<r_{\jjvH}^{2}\right>_{2}+\cdots.
\end{equation}
Taking the square root of this expression and expanding perturbatively the triton point charge radius $\delta r_{C}$ up to \nnlo is given by
\begin{equation}
\delta r_{c}=\sqrt{\left<r_{\jjvH}^{2}\right>_{0}}\left(\underbrace{\vphantom{\frac{1}{2}\frac{\left<r_{\jjvH}^{2}\right>_{1}}{\left<r_{\jjvH}^{2}\right>_{0}}}1}_{\mathrm{LO}}+\underbrace{\frac{1}{2}\frac{\left<r_{\jjvH}^{2}\right>_{1}}{\left<r_{\jjvH}^{2}\right>_{0}}}_{\mathrm{NLO}}+\underbrace{\frac{1}{2}\frac{\left<r_{\jjvH}^{2}\right>_{2}}{\left<r_{\jjvH}^{2}\right>_{0}}-\frac{1}{8}\left(\frac{\left<r_{\jjvH}^{2}\right>_{1}}{\left<r_{\jjvH}^{2}\right>_{0}}\right)^{2}}_{\mathrm{NNLO}}+\cdots\right).
\end{equation}

In order to calculate the point charge radius at each order the charge form factor can be calculated for low values of $Q^{2}$ and a linear fit with respect to $Q^{2}$ then performed to extract the point charge radius.  This procedure works well at LO, however, for higher cutoffs at NLO and \nnlo this approach quickly runs into numerical issues and the point charge radius cannot be reliably extracted.  In order to circumvent this one expands the functions $\boldsymbol{\mathcal{A}}_{n}(\cdots)$, $\mathcal{A}_{n}(Q)$, $\boldsymbol{\mathcal{B}}_{0}(\cdots)$, $\boldsymbol{\mathcal{C}}_{n}(\cdots)$, and $\boldsymbol{\mathfrak{D}}_{n}(\cdots)$ in powers of $Q^{2}$ and extracts their $Q^{2}$ pieces allowing for a direct calculation of the point charge radius contributions.  The $Q^{2}$ parts of these functions can be simplified further by analytical integrations of angular integrals, thereby reducing potential numerical issues and speeding up calculations.  The $Q^{2}$ parts of these functions are given in App.~\ref{app:QExpansion}.

The triton charge radius $r_{C}$ is related to the triton point charge radius $\delta r_{C}$ by
\begin{equation}
\label{eq:rpoint}
\left<\delta r_{C}^{2}\right>=\left<r_{C}^{2}\right>-\left<r_{p}^{2}\right>-2\left<r_{n}^{2}\right>,
\end{equation}
where $r_{p}=0.8783\pm0.0086$~fm~\cite{Angeli201369} is the proton charge radius, $r_{n}^{2}=-0.1149\pm0.0027$~fm$^{2}$~\cite{Angeli201369} is the neutron charge radius squared, and $r_{C}=1.7591\pm0.0363$~fm is the triton charge radius~\cite{Angeli201369}.  From this experimental data a triton point charge radius of $\delta r_{c}=1.5978\pm0.040$~fm is extracted.

\begin{figure}[hbt]
\includegraphics[width=100mm]{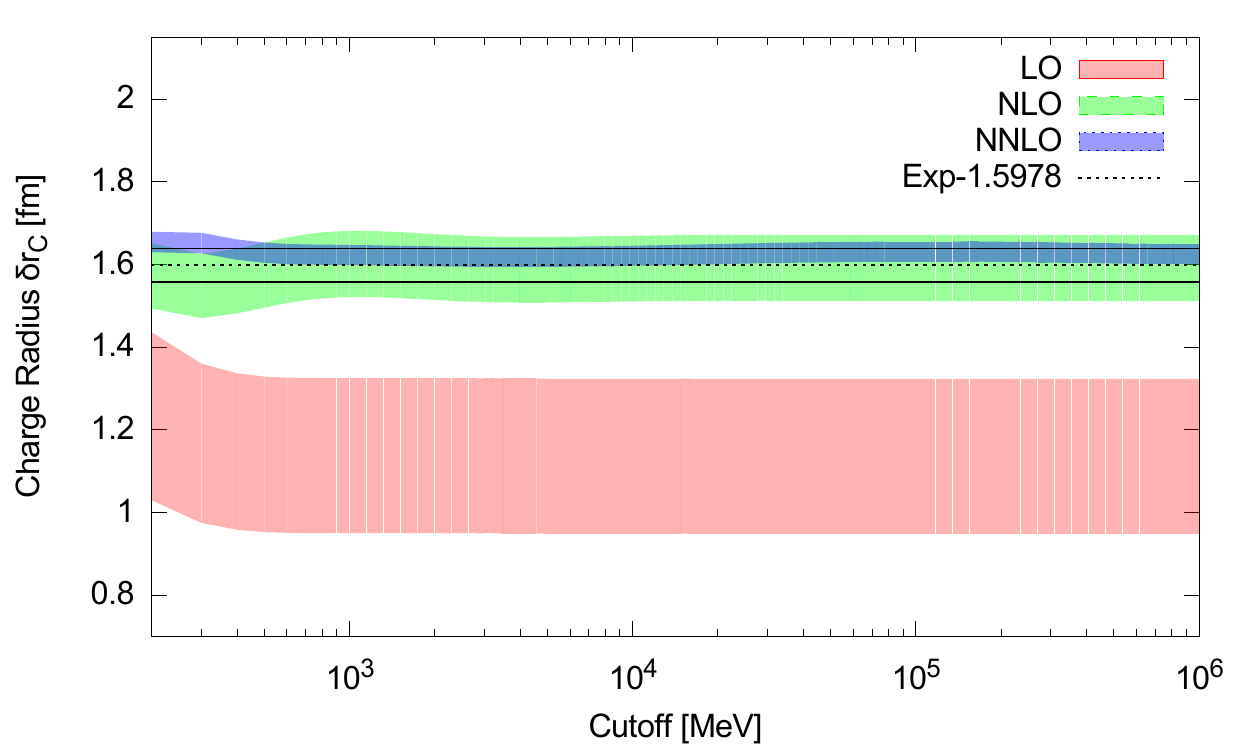}
\caption{ \label{fig:ChargeRadiusNLO}Cutoff dependence of the LO, NLO, and \nnlo predictions for the triton point charge radius.  The pink band is a 15\% error estimate for the LO triton point charge radius, the green band is a 5\% error estimate for the NLO triton point charge radius, and the blue band is a 1.5\% error estimate for the NNLO triton point charge radius.  The dotted line is the value extracted from experiment, $1.5978\pm0.040$~fm~\cite{Angeli201369}, and the black lines its error.}
\end{figure}

The cutoff dependence of the LO, NLO, and \nnlo triton point charge radius is given in Fig.~\ref{fig:ChargeRadiusNLO}.
Small values of the cutoff should be ignored since they are sensitive to shifts in the momentum in integrals from the finite cutoff regularization.  However, for sufficiently large cutoffs all terms that go like $1/\Lambda^{n}$ are suppressed and all integrals are effectively invariant under a shift in momentum. In Fig.~\ref{fig:ChargeRadiusNLO} the LO pink band corresponds to a 15\% error about the LO point charge radius prediction, the NLO green band corresponds to a 5\% error about the NLO point charge radius prediction, and the \nnlo blue band to a 1.5\% error about the \nnlo point charge radius prediction.\footnote{The usual \EFT error is 30\%, 10\%,  and 3\% for LO, NLO, and NNLO respectively.  Taking the square root to get the charge radius divides this percent error in half.}  The LO and NLO bands converge as a function of cutoff, while the NNLO band has a very slight cutoff variation.  The LO triton point charge radius converges to a value of 1.14~fm and the NLO value to 1.59~fm.  In the region of cutoffs from 1000 to $10^{6}$~MeV the NNLO point charge radius varies from 1.62~fm to 1.63~fm.  The NLO (\nnlo) value is within 5\% (1.5\%) of the experimental number for the triton point charge radius of  $1.5978\pm0.040$~fm~\cite{Angeli201369}.  From LO to NLO a large change is seen in the point charge radius.  This large change from LO to NLO is typical in the $Z$-parametrization where fixing the residue about the poles of the deuteron and ${}^{1}\!S_{0}$ virtual bound state makes a large correction from LO to NLO.  Further examples of this behavior can be seen in Ref. \cite{Phillips:1999hh} for the $np$ phase shift in the ${}^{3}\!S_{1}$ channel. 

The LO prediction for the triton point charge radius is more than 15\% away from the experimental error bars.  However, calculating the LO triton point charge radius in the unitary limit yields the numerical result $M_{N}E_{\jjvH}\left<r^{2}_{\jjvH}\right>_{0}=(1+s_{0}^{2})/9\approx 0.224$, which is in agreement with analytical techniques found in Ref.~\cite{Braaten:2004rn}.\footnote{The number $s_{0}=1.00624...$ is a universal number coming from the solution of the asymptotic form of the triton vertex function\cite{Bedaque:1998km,Bedaque:1998kg,Danilov:1963}.  Ref.~\cite{Braaten:2004rn} actually calculates the point matter radius in the unitary and equal mass limit, but this is equivalent to the point charge radius in this limit.}  This gives further confidence that the LO result, despite  perhaps seeming too small, is indeed correct.  At \nnlo a point charge radius of $1.62\pm0.03$~fm is predicted, which agrees with the experimental extraction within errors, where the error comes from a 1.5\% error estimate from \EFT and also a 1\% error from cutoff variation.  It is still an open question whether the NNLO result is strictly converging as $\Lambda\to\infty$.  In order to address this issue either a detailed asymptotic analysis must be carried out or a calculation to cutoffs large enough where signs of convergence or lack thereof can be clearly seen.  However, the \nnlo calculation suffers from numerical noise at large cutoffs ($\Lambda>10^{6}$ MeV) and new numerical techniques would be needed to deal with the fine tuning of three-body forces at large cutoffs.  Dealing with this fine tuning could also allow reliable calculations of the triton charge form factor and not just the triton point charge radius to higher cutoffs at NNLO.  Finally, a previous \EFT calculation using wavefunction methods obtained a LO prediction of $2.1\pm0.6$~fm for the triton point charge radius~\cite{Platter:2005sj}, and a coordinate space technique obtained the NLO \EFT prediction of $1.6\pm0.2$~fm~\cite{Kirscher:2009aj}.  Note, all of these techniques should find $M_{N}E_{\jjvH}\left<r^{2}_{\jjvH}\right>_{0}=(1+s_{0}^{2})/9\approx 0.224$ in the unitary and equal mass limit.

The point charge radius of the triton was obtained using Eq.~(\ref{eq:rpoint}) and the charge radius of the proton from electron scattering.  However, spectroscopy from muonic hydrogen finds a proton charge radius of 0.84087(39)~fm~\cite{Antognini:1900ns}, which is about seven standard deviations away from the averaged results of electron scattering and electronic hydrogen spectroscopy~\cite{Mohr:2012tt}.  This discrepancy is known as the ``proton radius puzzle".  An extensive review can be found in Ref.~\cite{Pohl:2013yb} and an overview of certain current and ongoing experimental efforts in Ref.~\cite{Pohl:2014bpa}.  Possible solutions lie in the way that functions are fit to electron scattering data to extract the charge radius~\cite{Griffioen:2015hta}.  However, this would not explain the discrepancy between muonic hydrogen and electronic hydrogen spectroscopy data.  Both experimental~\cite{Beyer:2013daa,Beyer:2013jla,ANDP:ANDP201300062} and theoretical~\cite{Karshenboim:2014baa} efforts are being carried out to reexamine the electronic hydrogen spectroscopy results.  Other possible theoretical explanations include using new muonic forces~\cite{Batell:2011qq,TuckerSmith:2010ra,Carlson:2013mya} and new proton structures~\cite{Hill:2011wy,Birse:2012eb,Miller:2012ne,Jentschura:2014ila,Hagelstein:2015yma,Hoferichter:2016duk}.  Using the value for the proton charge radius from muonic hydrogen gives a triton point charge radius of $1.6178\pm0.040$~fm.  The approximate 1\% difference between the experimental triton point charge radius from muonic hydrogen and electron scattering would require a N$^{3}$LO calculation in \EFT to distinguish them.  Note a N$^{3}$LO calculation does not give direct information about the fundamental interactions giving rise to the proton structure in the triton, but only to correlations within and between the triton and deuteron structures.

A comparison of various calculations of the triton point charge radius is shown in Table~\ref{tab:results}.  The results of Ref.~\cite{Dinur:2015vzv} use the Lanzcos sum rule and the effective interaction hyperspherical harmonics method with the two-body Argonne-v18 (AV18)~\cite{Wiringa:1994wb} and three-body Urbana IX (UIX)~\cite{Pudliner:1995wk} (AV18/UIX) potential to obtain a triton point charge radius of 1.593~fm and using a two-~\cite{Entem:2003ft} and three-body~\cite{Navratil:2007zn}  $\chi$EFT potential they find a triton point charge radius of 1.617~fm.  Ref.~\cite{Kievsky:2008es} uses the AV18/UIX potential with the hyperspherical harmonics (HH) method to get a triton point charge radius of 1.582~fm.  Using Green's function Monte Carlo (GFMC) with the AV18 and three-body Illinois 7 (IL7)~\cite{Pieper:2001ap} potential (AV18/IL7) a triton point charge radius of 1.58~fm is found~\cite{Pastore:2012rp}.  $\chi$EFT predicts a triton point charge radius of 1.594(8), where the error comes from looking at the cutoff dependence of the triton point charge radius~\cite{Piarulli:2012bn}.  The NNLO results of this work and other lower-order \EFT calculations are displayed as well.  Also shown in Table~\ref{tab:results} are predictions for the triton binding energy.  For EFT predictions the triton binding energy is fit to and therefore not shown\footnote{The three-body terms using the $\chi$EFT potential in Ref.~\cite{Dinur:2015vzv} are clearly not fit exactly to the triton binding energy.  For further details of how their three-body parameters are chosen consult Ref.~\cite{Navratil:2007zn}} .  Most techniques predict the triton binding energy reasonably well, but the GFMC seems to slightly overpredict it, and its error comes from Monte Carlo statistics.  All PMCs seem to predict roughly the same triton point charge radius, with the exception of the $\chi$EFT result from Ref.~\cite{Dinur:2015vzv}, which favors the triton point charge radius using the proton charge radius from muonic hydrogen.  None of the PMC values have any error estimates.  The \EFT predictions agree with the triton point charge radius within their respective errors.  $\chi$EFT seems to agree quite well with experiment and also has a small error.  However, estimating the error with cutoff variation should be done with caution~\cite{Furnstahl:2014xsa}.

\begin{table}
\begin{tabular}{|l|c|c|}
\hline
Method  & $B_{\jjvH}$~[MeV]  & $\delta r_{C}$~[fm] \\\hline
AV18/UIX~\cite{Dinur:2015vzv} & 8.473 & 1.593  \\
$\chi$EFT~\cite{Dinur:2015vzv} & 8.478 & 1.617\\
AV18/UIX HH~\cite{Kievsky:2008es} & 8.479 & 1.582 \\
AV18/IL7 GFMC\cite{Pastore:2012rp} & 8.50(1) & 1.58 \\
$\chi$EFT N3LO/N2LO \cite{Piarulli:2012bn} & -- & 1.594(8) \\
\EFT (LO)~\cite{Platter:2005sj} & -- & 2.1(6)\\
\EFT (NLO)~\cite{Kirscher:2009aj} & -- &  1.6(2)\\
\EFT (NNLO) & -- & 1.62(3)\\\hline
Experiment:  & 8.4818 & \\
Experiment: $e^{-}$ &  & 1.5978(40)~\cite{Angeli201369}\\
Experiment: $\mu^{-}$ & & \hspace{.65cm}1.6178(40)~\cite{Angeli201369,Antognini:1900ns} \\\hline
\end{tabular}
\caption{\label{tab:results}Different theoretical predictions for the triton point charge radius and the triton binding energy.  All EFT calculations fit to the experimental triton binding energy, with the exception of the $\chi$EFT calculation of Ref.~\cite{Dinur:2015vzv}.  The error for the triton binding energy for the GFMC results comes from statistical errors in Monte Carlo calculations.  All other errors are estimates from EFT or experimental errors.  The error for the $\chi$EFT value of $\delta r_{C}$ comes from varying the cutoff of the calculation~\cite{Piarulli:2012bn}.  Experimental numbers for the triton point charge radius are given using both the proton charge radius from electron scattering data and muonic hydrogen data.}
\end{table}

\section{\label{sec:conclusions} Conclusions}

Building upon the work of Hagen \emph{et al.}~\cite{Hagen:2013xga} I have introduced a technique to treat perturbative corrections to bound-state calculations for EFTs of short range interactions.  This work focused on the use of these techniques in \EFT, but they are equally useful for halo EFT or cold atom systems.  In addition, this new technique leads to numerical simplifications in calculating $nd$ scattering amplitudes and the LO three-body force in the doublet $S$-wave channel.  It also allows the \nnlo energy dependent three-body force to be fixed to the triton bound-state energy without the need for a limiting procedure~\cite{Ji:2012nj}.
  
Using this new technique the triton point charge radius was calculated to \nnlo in \EFT, giving a LO value of $1.14\pm0.19$~fm, a NLO value of $1.59\pm0.08$~fm, and a \nnlo value of $1.62\pm 0.03$~fm.  The LO value disagrees with the experimental extraction of $1.5978\pm0.040$~fm~\cite{Angeli201369} by about 40\%, which is more than the LO estimated \EFT error of 15\%.  However, it was found at LO that it agrees with analytical calculations in the unitary and equal mass limit~\cite{Braaten:2004rn}.  At NLO the value of $1.59\pm0.08$~fm agrees with the experimental extraction within the expected 5\% error.  The error for the \nnlo value comes from the expected 1.5\% error at \nnlo in \EFT and from the slight cutoff variation of the calculation.  Within these errors the NNLO prediction of $1.62\pm0.03$~fm agrees with the experimental extraction.  Future work should address the cutoff variation at NNLO, and see if the results actually converge as a function of cutoff.  In addition future work should carry out a more rigorous error analysis by means of Bayesian statistics~\cite{Furnstahl:2015rha}.

Fitting the three-body force to the triton binding energy in the unitary limit the triton point charge radius is $1.05$~fm.  Including the proper \NN scattering lengths gives the LO value $1.14$~fm, and including range corrections up to \nnlo gives the value $1.62\pm0.03$~fm.  Thus range corrections give significant contributions to the triton point charge radius with respect to the unitary limit.  Despite this, a controlled expansion in terms of a finite number of parameters from the unitary limit is observed, and therefore the triton can be thought of as being in the so called ``Efimov window"~\cite{Kievsky:2015dtk}.  

Future work will also consider the $\jjvHe$ point charge radius, which in the absence of Coulomb is the isospin mirror of the current calculation presented here.  Coulomb effects can be included in this formalism straightforwardly either perturbatively or nonperturbatively.  For a description of $\jjvHe$ it should be sufficient to treat Coulomb fully perturbatively~\cite{Konig:2015aka}.  In addition future work will consider the magnetic moments of the triton and $\jjvHe$ as well as their magnetic radii.  The magnetic radii are of interest because they will be measured to greater precision in upcoming experiments using spectroscopy of $\mu\jjvHe^{+}$~\cite{Antognini:2015vxo}.  \EFT offers a way to make precision calculations for these observables in a controlled expansion matched on to low energy nuclear observables.

\acknowledgments{I would like to thank Roxanne Springer, Thomas Mehen, Daniel Phillips, Hans-Werner Hammer, Bijaya Acharya, and Chen Ji for useful discussions during the course of this work.  In addition I would also like to thank the ExtreMe Matter Institute EMMI at the GSI Helmholtz Centre for Heavy Ion Research and the Institute for Nuclear Theory INT program 16-01: ``Nuclear Physics from Lattice QCD" for support during the completion of this work.  This material is based upon work supported by the U.S. Department of
Energy, Office of Science, Office of Nuclear Physics, under Award Number DE-FG02-05ER41368 and Award Number DE-FG02-93ER40756}

\appendix

\section{\label{app:chi}}
The function $(\boldsymbol{\chi}_{a}^{ji}(\cdots))_{\nu\beta}^{\mu\alpha}$ is given by
\begin{align}
&\left(\boldsymbol{\chi}_{a}^{ji}(E,\vect{K},\vect{P},p_{0},k_{0},\vect{p},\vect{k})\right)_{\nu\beta}^{\mu\alpha}=ie(2\pi)^{4}\delta\left(k_{0}-p_{0}\right)\boldsymbol{\delta}^{(3)}\left(\vect{k}-\vect{p}-\frac{2}{3}\vect{Q}\right)\\\nonumber
&\hspace{4cm}\times i\mathbf{D}^{(0)}\!\!\left(\frac{2}{3}E+k_{0},\vect{k}+\frac{2}{3}\vect{K}\right)\frac{i}{\frac{1}{3}E-k_{0}-\frac{(\vect{k}-\frac{1}{3}\vect{K})^{2}}{2M_{N}}+i\epsilon}\\\nonumber
&\hspace{4cm}\times\frac{i}{\frac{1}{3}E-k_{0}-\frac{(\vect{k}-\frac{2}{3}\vect{Q}-\frac{1}{3}\vect{P})^{2}}{2M_{N}}+i\epsilon}\left(\frac{1+\tau_{3}}{2}\right)^{\mu}_{\nu}\delta^{\alpha}_{\beta}\delta^{ij},
\end{align}
where $\alpha$ ($\beta$) is the initial (final) nucleon spin, $\mu$ ($\nu$) the initial (final) nucleon isospin, and $i$ ($j$) the initial (final) dibaryon polarization.  Using the projection operators as defined in Ref.~\cite{Griesshammer:2004pe} to project the c.c.~space spin-isospin operator into the doublet $S$-wave channel yields
\begin{equation}
\frac{1}{3}\left(\begin{array}{cc}
\sigma_{j}  & 0 \\[-1.5 mm]
0 & \tau_{B} 
\end{array}\right)
\left(\begin{array}{cc}
\left(\frac{1+\tau_{3}}{2}\right)\delta_{ij} & 0 \\[0 mm] 
0 & \left(\frac{1+\tau_{3}}{2}\right)\delta_{AB}
\end{array}\right)
\left(\begin{array}{cc}
\sigma_{i}  & 0 \\[-1.5 mm]
0 & \tau_{A} 
\end{array}\right)=
\left(\begin{array}{cc}
0  & 0 \\[-1.5 mm]
0 & \frac{2}{3}
\end{array}\right).
\end{equation}
Thus the function $\boldsymbol{\chi}_{a}(\cdots)$ is a matrix in c.c.~space given by
\begin{align}
&\boldsymbol{\chi}_{a}(E,\vect{K},\vect{P},p_{0},k_{0},\vect{p},\vect{k})=ie(2\pi)^{4}\delta\left(k_{0}-p_{0}\right)\boldsymbol{\delta}^{(3)}\left(\vect{k}-\vect{p}-\frac{2}{3}\vect{Q}\right)\\\nonumber
&\hspace{4cm}\times i\mathbf{D}^{(0)}\!\!\left(\frac{2}{3}E+k_{0},\vect{k}+\frac{2}{3}\vect{K}\right)\frac{i}{\frac{1}{3}E-k_{0}-\frac{(\vect{k}-\frac{1}{3}\vect{K})^{2}}{2M_{N}}+i\epsilon}\\\nonumber
&\hspace{4cm}\times\frac{i}{\frac{1}{3}E-k_{0}-\frac{(\vect{k}-\frac{2}{3}\vect{Q}-\frac{1}{3}\vect{P})^{2}}{2M_{N}}+i\epsilon}\left(\begin{array}{cc}
0 & 0 \\[-1.5 mm]
0 & \frac{2}{3}
\end{array}\right).
\end{align}
Plugging $\boldsymbol{\chi}_{a}(\cdots)$ into Eq.~(\ref{eq:LOformfactor}) the integration over $d^{4}p$ is removed by the delta functions.  Integrating over the energy pole the integration over $dk_{0}$ leaves only a $d^{3}k$ integration.  Next Eq.~(\ref{eq:GLOboosted}) is used to rewrite the triton vertex function in the boosted frame in terms of the triton vertex function in the c.m.~frame. The momentum $\vect{k}$ from Eq.~(\ref{eq:LOformfactor}) and momentum $\vect{q}$ from Eq.~(\ref{eq:GLOboosted}) are interchanged, and then $\vect{q}\to\vect{q}+\frac{1}{3}\vect{Q}$.  This shift makes the time reversal symmetry of the expressions manifest.  Finally, integrating over the azimuthal angle of $\vect{q}$ leaves a double integral for the analytical forms of the functions $\boldsymbol{\mathcal{A}}_{n}(\cdots)$ and $\mathcal{A}_{n}(Q)$ which are given by\footnote{Note all of the functions here should be similar to those found in  Hagen \emph{et al.}~\cite{Hagen:2013xga}, in the limit where the core mass equals the neutron mass.  However, where I find the term $Q^{2}/(12M_{N})$ in the dibaryon propagator for the functions $\boldsymbol{\mathcal{A}}_{n}(\cdots)$ and $\mathcal{A}_{n}(Q)$ they find $Q^{2}/(8M_{N})$.}
\begin{align}
\label{eq:A1}
&\boldsymbol{\mathcal{A}}_{n}(p,k,Q)=M_{N}\!\stackbin[0]{1}{\Big|}\!\int_{0}^{\Lambda}\!\!\!dqq^{2}\!\!\!\int_{-1}^{1}\!\!dx\frac{1}{qQx}\frac{1}{kp\sqrt{q^{2}+\frac{2}{3}qQx+\frac{1}{9}Q^{2}}\sqrt{q^{2}-\frac{2}{3}qQx+\frac{1}{9}Q^{2}}}\\\nonumber
&\times Q_{0}\left(\frac{k^{2}+q^{2}+\frac{1}{9}Q^{2}+(y-\frac{1}{3})qQx-M_{N}B_{0}}{k\sqrt{q^{2}+\frac{2}{3}qQx+\frac{1}{9}Q^{2}}}\right)Q_{0}\left(\frac{p^{2}+q^{2}+\frac{1}{9}Q^{2}+(y-\frac{2}{3})qQx-M_{N}B_{0}}{p\sqrt{q^{2}-\frac{2}{3}qQx+\frac{1}{9}Q^{2}}}\right)\\\nonumber
&\times D_{s}^{(n)}\!\!\left(B_{0}-\frac{q^{2}}{2M_{N}}-\frac{Q^{2}}{12M_{N}}+\left(\frac{1}{2}-y\right)\frac{qQx}{M_{N}},\vect{q}\right)
\left(\!\!\!\begin{array}{rr}
6 & -2 \\[-1.5 mm]
-2 & \frac{2}{3}
\end{array}\!\right),
\end{align}
\begin{align}
\label{eq:Avec}
&\boldsymbol{\mathcal{A}}_{n}(p,Q)=-\frac{M_{N}}{2\pi}\!\stackbin[0]{1}{\Big|}\!\int_{0}^{\Lambda}\!\!\!dqq^{2}\!\!\!\int_{-1}^{1}\!\!dx\frac{1}{qQx}\frac{1}{p\sqrt{q^{2}-\frac{2}{3}qQx+\frac{1}{9}Q^{2}}}\\\nonumber
&\hspace{3cm}\times Q_{0}\left(\frac{p^{2}+q^{2}+\frac{1}{9}Q^{2}+(y-\frac{2}{3})qQx-M_{N}B_{0}}{p\sqrt{q^{2}-\frac{2}{3}qQx+\frac{1}{9}Q^{2}}}\right)\\\nonumber
&\hspace{3cm}\times D_{s}^{(n)}\!\!\left(B_{0}-\frac{q^{2}}{2M_{N}}-\frac{Q^{2}}{12M_{N}}+\left(\frac{1}{2}-y\right)\frac{qQx}{M_{N}},\vect{q}\right)
\left(\!\!\begin{array}{r}
2 \\[-1.5 mm]
-\frac{2}{3}
\end{array}\right),
\end{align}
and
\begin{align}
\label{eq:Ascalar}
&\mathcal{A}_{n}(Q)=\frac{M_{N}}{4\pi^{2}}\!\stackbin[0]{1}{\Big|}\!\int_{0}^{\Lambda}\!\!\!dqq^{2}\!\!\!\int_{-1}^{1}\!\!dx\frac{1}{qQx}\frac{2}{3}D_{s}^{(n)}\!\!\left(B_{0}-\frac{q^{2}}{2M_{N}}-\frac{Q^{2}}{12M_{N}}+\left(\frac{1}{2}-y\right)\frac{qQx}{M_{N}},\vect{q}\right),
\end{align}
where 
\begin{equation}
\stackbin[0]{1}{\Big|}f(y)=f(1)-f(0).
\end{equation}
The matrix (vector) of the function $\boldsymbol{\mathcal{A}}_{n}(p,k,Q)$ ($\boldsymbol{\mathcal{A}}_{n}(p,Q)$) is defined in c.c.~space.  To obtain the c.c.~space matrix for $\boldsymbol{\mathcal{A}}_{n}(p,k,Q)$ the c.c.~space matrix from $\boldsymbol{\chi}_{a}(\cdots)$ is multiplied on either side by a c.c.~space matrix from the LO kernel leading to
\begin{equation}
\left(\!\!\!\begin{array}{rr}
1  & -3 \\[-1.5 mm]
-3 & 1 
\end{array}\!\right)
\left(\begin{array}{cc}
0  & 0 \\[-1.5 mm]
0 & \frac{2}{3} 
\end{array}\!\right)
\left(\!\!\!\begin{array}{rr}
1  & -3 \\[-1.5 mm]
-3 & 1 
\end{array}\!\right)=
\left(\!\!\!\begin{array}{rr}
6  & -2 \\[-1.5 mm]
-2 & \frac{2}{3} 
\end{array}\!\right),
\end{equation}
giving the c.c.~space matrix as defined in Eq. (\ref{eq:A1}). 

The function $(\boldsymbol{\chi}_{b}^{ji}(\cdots))_{\nu\beta}^{\mu\alpha}$ is given by
\begin{align}
&\left(\boldsymbol{\chi}^{ji}_{b}(E,\vect{K},\vect{P},p_{0},k_{0},\vect{p},\vect{k})\right)_{\nu\beta}^{\mu\alpha}=i\frac{2\pi e}{M_{N}}iD_{x}^{(0)}\!\!\left(\frac{2}{3}E+k_{0},\vect{k}+\frac{2}{3}\vect{K}\right)\frac{i}{\frac{1}{3}E-k_{0}-\frac{(\vect{k}-\frac{1}{3}\vect{K})^{2}}{2M_{N}}+i\epsilon}\\\nonumber
&\times \frac{i}{\frac{1}{3}E-p_{0}-\frac{(\vect{p}-\frac{1}{3}\vect{P})^{2}}{2M_{N}}+i\epsilon}\frac{i}{\frac{1}{3}E+k_{0}+p_{0}-\frac{(\vect{k}+\vect{p}-\frac{1}{3}\vect{Q}+\frac{1}{3}\vect{K})^{2}}{2M_{N}}+i\epsilon}\\\nonumber
&\times \frac{i}{\frac{1}{3}E+k_{0}+p_{0}-\frac{(\vect{k}+\vect{p}+\frac{1}{3}\vect{Q}+\frac{1}{3}\vect{P})^{2}}{2M_{N}}+i\epsilon}iD_{w}^{(0)}\!\!\left(\frac{2}{3}E+p_{0},\vect{p}+\frac{2}{3}\vect{P}\right)\left[{P_{i}^{(w)}}^{\dagger}\left(\frac{1+\tau_{3}}{2}\right)P_{j}^{(x)}\right]^{\alpha \mu}_{\beta \nu},
\end{align}
where $P_{j}^{(x)}=\sqrt{8}P_{j}$ ($P_{j}^{(x)}=\sqrt{8}\bar{P}_{j}$) for $x=t$ ($x=s$) in the spin-triplet iso-singlet (spin-singlet iso-triplet) channel.  Here the indices ``$i$" and ``$j$" are either spinor or isospinor indices depending on the values of ($x$) and ($w$).  The values of ($x$) and ($w$) pick out the matrix element of $(\boldsymbol{\chi}_{b}^{ji}(\cdots))_{\nu\beta}^{\mu\alpha}$ in c.c.~space.  Projecting $(\boldsymbol{\chi}_{b}^{ji}(\cdots))_{\nu\beta}^{\mu\alpha}$ onto the doublet $S$-wave channel gives
\begin{align}
&\boldsymbol{\chi}_{b}(E,\vect{K},\vect{P},p_{0},k_{0},\vect{p},\vect{k})=i\frac{2\pi e}{M_{N}}i\mathbf{D}^{(0)}\!\!\left(\frac{2}{3}E+k_{0},\vect{k}+\frac{2}{3}\vect{K}\right)\frac{i}{\frac{1}{3}E-k_{0}-\frac{(\vect{k}-\frac{1}{3}\vect{K})^{2}}{2M_{N}}+i\epsilon}\\\nonumber
&\times \frac{i}{\frac{1}{3}E-p_{0}-\frac{(\vect{p}-\frac{1}{3}\vect{P})^{2}}{2M_{N}}+i\epsilon}\frac{i}{\frac{1}{3}E+k_{0}+p_{0}-\frac{(\vect{k}+\vect{p}-\frac{1}{3}\vect{Q}+\frac{1}{3}\vect{K})^{2}}{2M_{N}}+i\epsilon}\\\nonumber
&\times \frac{i}{\frac{1}{3}E+k_{0}+p_{0}-\frac{(\vect{k}+\vect{p}+\frac{1}{3}\vect{Q}+\frac{1}{3}\vect{P})^{2}}{2M_{N}}+i\epsilon}\left(\!\!\!
\begin{array}{rr}
-1 & 1 \\[-1.5 mm]
1 & \frac{1}{3}
\end{array}\!\right)i\mathbf{D}^{(0)}\!\!\left(\frac{2}{3}E+p_{0},\vect{p}+\frac{2}{3}\vect{P}\right)
\end{align}
Plugging $\boldsymbol{\chi}_{b}(\cdots)$ into Eq.~(\ref{eq:LOformfactor}) and then integrating over the energy poles removes the $dp_{0}$ and $dk_{0}$ integrals.  After performing these integrations the LO triton vertex functions are already in the c.m.~frame, leaving only six integrations to be performed.  Integrating over one of the azimuthal angles and noting that Eq.~(\ref{eq:FLOb}) already has two integrations, the function $\boldsymbol{\mathcal{B}}_{0}(p,k,Q)$ has three remaining integrals and is defined by
\begin{align}
&\boldsymbol{\mathcal{B}}_{0}(p,k,Q)=-\frac{M_{N}}{4}\!\!\int_{-1}^{1}\!\!dx\!\!\int_{-1}^{1}\!\!dy\!\!\int_{0}^{2\pi}\!\!d\phi\\\nonumber
&\hspace{1cm}\times\frac{1}{k^{2}+p^{2}+kp\left(xy+\sqrt{1-x^{2}}\sqrt{1-y^{2}}\cos\phi\right)-\frac{1}{3}Q(kx+2py)+\frac{1}{9}Q^{2}-M_{n}B_{0}}\\\nonumber
&\hspace{1cm}\times\frac{1}{k^{2}+p^{2}+kp\left(xy+\sqrt{1-x^{2}}\sqrt{1-y^{2}}\cos\phi\right)+\frac{1}{3}Q(2kx+py)+\frac{1}{9}Q^{2}-M_{n}B_{0}}\\\nonumber
&\hspace{1cm}\times
\left(\!\!\!\begin{array}{rr}
-1 & 1 \\ [-1.5 mm]
1 & \frac{1}{3}
\end{array}\!\right).
\end{align}
Time reversal symmetry in this expression is immediately apparent as it is invariant under the transformation $k\longleftrightarrow p$, and $Q\to-Q$.

The function $(\boldsymbol{\chi}_{c}^{ji}(\cdots))_{\nu\beta}^{\mu\alpha}$ is given by
\begin{align}
&\left(\boldsymbol{\chi}_{c}^{ji}(E,\vect{K},\vect{P},p_{0},k_{0},\vect{p},\vect{k})\right)_{\nu\beta}^{\mu\alpha}=\\\nonumber
&i\frac{eM_{N}}{Q}(2\pi)^{4}\delta\left(k_{0}-p_{0}\right)\boldsymbol{\delta}^{(3)}\left(\vect{p}-\vect{k}-\frac{1}{3}\vect{Q}\right)\frac{i}{\frac{1}{3}E-k_{0}-\frac{(\vect{k}-\frac{1}{3}\vect{K})^{2}}{2M_{N}}+i\epsilon}\\\nonumber
&\times \arctan\left(\frac{Q}{2\sqrt{\frac{1}{4}(\vect{k}+\!\frac{2}{3}\vect{K})^{2}-\frac{2}{3}M_{N}E-M_{N}k_{0}}+2\sqrt{\frac{1}{4}(\vect{k}+\vect{Q}+\!\frac{2}{3}\vect{K})^{2}-\!\frac{2}{3}M_{N}E-M_{N}k_{0}}}\right)\\\nonumber
&\times iD_{w}^{(0)}\!\!\left(\frac{2}{3}E+k_{0},\vect{k}+\frac{2}{3}\vect{K}\right)iD_{x}^{(0)}\!\!\left(\frac{2}{3}E+k_{0},\vect{k}+\vect{Q}+\frac{2}{3}\vect{K}\right)\mathrm{Tr}\left[P_{j}^{(x)}\left(\frac{1+\tau_{3}}{2}\right){P_{i}^{(w)}}^{\dagger}\right]\delta^{\alpha}_{\beta}\delta^{\mu}_{\nu},
\end{align}
which projected onto the doublet $S$-wave channel gives
\begin{align}
&\boldsymbol{\chi}_{c}(E,\vect{K},\vect{P},p_{0},k_{0},\vect{p},\vect{k})=\\\nonumber
&i\frac{eM_{N}}{Q}(2\pi)^{4}\delta\left(k_{0}-p_{0}\right)\boldsymbol{\delta}^{(3)}\left(\vect{p}-\vect{k}-\frac{1}{3}\vect{Q}\right)\frac{i}{\frac{1}{3}E-k_{0}-\frac{(\vect{k}-\frac{1}{3}\vect{K})^{2}}{2M_{N}}+i\epsilon}\\\nonumber
&\times \arctan\left(\frac{Q}{2\sqrt{\frac{1}{4}(\vect{k}+\!\frac{2}{3}\vect{K})^{2}-\frac{2}{3}M_{N}E-M_{N}k_{0}}+2\sqrt{\frac{1}{4}(\vect{k}+\vect{Q}+\!\frac{2}{3}\vect{K})^{2}-\!\frac{2}{3}M_{N}E-M_{N}k_{0}}}\right)\\\nonumber
&\times i\mathbf{D}^{(0)}\!\!\left(\frac{2}{3}E+k_{0},\vect{k}+\frac{2}{3}\vect{K}\right)\left(
\!\begin{array}{cc}
2 & 0\\[-1.5 mm]
0 & \frac{2}{3}
\end{array}\!\right)i\mathbf{D}^{(0)}\!\!\left(\frac{2}{3}E+k_{0},\vect{k}+\vect{Q}+\frac{2}{3}\vect{K}\right),
\end{align}
where the analytical expression of the two-body sub-diagram of diagram (c) is included.  Integrating over the delta functions and the energy $dk_{0}$ leaves only the integration $d^{3}k$.  After this, one LO triton vertex function is in the c.m.~frame and the other is not and must be rewritten using Eq.~(\ref{eq:GLOboosted}).  Integrating over the azimuthal angle the functions $\boldsymbol{\mathcal{C}}_{n}(\cdots)$ are given by
\begin{align}
&\boldsymbol{\mathcal{C}}_{n}(p,k,Q)=-\frac{M_{N}\pi}{Q}\!\!\int_{-1}^{1}\!\!dx\\\nonumber
&\hspace{2cm}\times\arctan\left(\frac{Q}{2\sqrt{\frac{3}{4}k^{2}-M_{N}B_{0}}+2\sqrt{\frac{3}{4}k^{2}+\frac{1}{2}Qkx+\frac{1}{12}Q^{2}-M_{N}B_{0}}}\right)\\\nonumber
&\hspace{2cm}\times\frac{1}{p\sqrt{k^{2}+\frac{2}{3}kQx+\frac{1}{9}Q^{2}}}Q_{0}\left(\frac{p^{2}+k^{2}+\frac{2}{3}kQx+\frac{1}{9}Q^{2}-M_{N}B_{0}}{p\sqrt{k^{2}+\frac{2}{3}kQx+\frac{1}{9}Q^{2}}}\right)\\\nonumber
&\hspace{2cm}\times
\left(\!\!\!\begin{array}{rr}
2 & -2 \\[-1.5 mm]
-6 & \frac{2}{3}
\end{array}\!\right)\mathbf{D}^{(n)}\!\!\left(B_{0}-\frac{k^{2}}{2M_{N}}-\frac{Qkx}{2M_{N}}-\frac{Q^{2}}{12M_{N}},k\right),
\end{align}
and
\begin{align}
&\boldsymbol{\mathcal{C}}_{n}(k,Q)=\frac{M_{N}}{2Q}\!\!\int_{-1}^{1}\!\!dx\\\nonumber
&\hspace{2cm}\times\arctan\left(\frac{Q}{2\sqrt{\frac{3}{4}k^{2}-M_{N}B_{0}}+2\sqrt{\frac{3}{4}k^{2}+\frac{1}{2}Qkx+\frac{1}{12}Q^{2}-M_{N}B_{0}}}\right)\\\nonumber
&\hspace{2cm}\times\left(\!\!\begin{array}{r}
2  \\[-1.5 mm]
-\frac{2}{3}
\end{array}\right)^{T}
\mathbf{D}^{(n)}\!\!\left(B_{0}-\frac{k^{2}}{2M_{N}}-\frac{Qkx}{2M_{N}}-\frac{Q^{2}}{12M_{N}},k\right).
\end{align}
In the current form of the functions  $\boldsymbol{\mathcal{C}}_{n}(\cdots)$ time reversal invariance is  not immediately apparent.  Recasting these expressions into an immediately apparent time reversal invariant form requires shifting momentum before integrating out angles.  However, the gain in analytical insight is outweighed by the loss in numerical efficiency and the form above is kept.

Diagram (d) is essentially diagram (c) without the two-body sub-diagram and therefore $(\boldsymbol{\chi}_{d}^{ji}(\cdots))_{\nu\beta}^{\mu\alpha}$ is similar to $(\boldsymbol{\chi}_{c}^{ji}(\cdots))_{\nu\beta}^{\mu\alpha}$ and is given by
\begin{align}
&\left(\boldsymbol{\chi}_{d}^{ji}(E,\vect{K},\vect{P},p_{0},k_{0},\vect{p},\vect{k})\right)_{\nu\beta}^{\mu\alpha}=\\\nonumber
&ie(2\pi)^{4}\delta\left(k_{0}-p_{0}\right)\boldsymbol{\delta}^{(3)}\left(\vect{p}-\vect{k}-\frac{1}{3}\vect{Q}\right)\frac{i}{\frac{1}{3}E-k_{0}-\frac{(\vect{k}-\frac{1}{3}\vect{K})^{2}}{2M_{N}}+i\epsilon}\\\nonumber
&\times iD_{w}^{(0)}\!\!\left(\frac{2}{3}E+k_{0},\vect{k}+\frac{2}{3}\vect{K}\right)iD_{x}^{(0)}\!\!\left(\frac{2}{3}E+k_{0},\vect{k}+\vect{Q}+\frac{2}{3}\vect{K}\right)\mathbf{T}_{wx}^{ij}\delta^{\alpha}_{\beta}\delta^{\mu}_{\nu},
\end{align}
where $\mathbf{T}_{wx}^{ij}=\delta_{wx}(c_{0t}^{(0)}\delta_{wt}\delta_{ij}+c_{0s}^{(0)}\delta_{ws}\delta_{i3}\delta_{j3})$.  The function $\delta_{wt}$ picks out the contribution from the spin-triplet dibaryon and $\delta_{ws}$ from the spin-singlet dibaryon.  The indices $i$ and $j$ in $\delta_{i3}\delta_{j3}$ are isospin indices and correspond to the fact that only the the $np$ spin-singlet dibaryon is charged and not the $nn$ spin-singlet dibaryon.  Projecting $(\boldsymbol{\chi}_{d}^{ji}(\cdots))_{\nu\beta}^{\mu\alpha}$ onto the doublet $S$-wave channel yields
\begin{align}
&\boldsymbol{\chi}_{d}(E,\vect{K},\vect{P},p_{0},k_{0},\vect{p},\vect{k})=\\\nonumber
&ie(2\pi)^{4}\delta\left(k_{0}-p_{0}\right)\boldsymbol{\delta}^{(3)}\left(\vect{p}-\vect{k}-\frac{1}{3}\vect{Q}\right)\frac{i}{\frac{1}{3}E-k_{0}-\frac{(\vect{k}-\frac{1}{3}\vect{K})^{2}}{2M_{N}}+i\epsilon}\\\nonumber
&i\mathbf{D}^{(0)}\!\!\left(\frac{2}{3}E+k_{0},\vect{k}+\frac{2}{3}\vect{K}\right)\left(
\begin{array}{cc}
c_{0t}^{(0)} & 0 \\
0 & \frac{1}{3}c_{0s}^{(0)} \\
\end{array}\right)
i\mathbf{D}^{(0)}\!\!\left(\frac{2}{3}E+k_{0},\vect{k}+\vect{Q}+\frac{2}{3}\vect{K}\right)
\end{align}
The calculation of the functions $\boldsymbol{\mathfrak{D}}_{n}(\cdots)$ is analogous to the calculation of $\boldsymbol{\mathcal{C}}_{n}(\cdots)$ and yields
\begin{align}
&\boldsymbol{\mathfrak{D}}_{n}(p,k,Q)=\pi\!\!\int_{-1}^{1}\!\!dx\\\nonumber
&\hspace{2cm}\times\frac{1}{p\sqrt{k^{2}+\frac{2}{3}kQx+\frac{1}{9}Q^{2}}}Q_{0}\left(\frac{p^{2}+k^{2}+\frac{2}{3}kQx+\frac{1}{9}Q^{2}-M_{N}B_{0}}{p\sqrt{k^{2}+\frac{2}{3}kQx+\frac{1}{9}Q^{2}}}\right)\\\nonumber
&\hspace{2cm}\times\sum_{j=1}^{n}
\left(\begin{array}{cc}
c_{0t}^{(j-1)} & -c_{0s}^{(j-1)} \\
-3c_{0t}^{(j-1)} & \frac{1}{3}c_{0s}^{(j-1)}
\end{array}\right)\mathbf{D}^{(n-j)}\!\!\left(B_{0}-\frac{k^{2}}{2M_{N}}-\frac{Qkx}{2M_{N}}-\frac{Q^{2}}{12M_{N}},k\right),
\end{align}
and 
\begin{align}
&\boldsymbol{\mathfrak{D}}_{n}(k,Q)=-\frac{1}{2}\!\!\int_{-1}^{1}\!\!dx\\\nonumber
&\hspace{2cm}\times\sum_{j=1}^{n}
\left(\begin{array}{c}
c_{0t}^{(j-1)} \\[0 mm]
-\frac{1}{3}c_{0s}^{(j-1)}
\end{array}\right)^{T}\mathbf{D}^{(n-j)}\!\!\left(B_{0}-\frac{k^{2}}{2M_{N}}-\frac{Qkx}{2M_{N}}-\frac{Q^{2}}{12M_{N}},k\right).
\end{align}

The functions $\boldsymbol{\chi}_{e}(\cdots)$ and $\boldsymbol{\chi}_{f}(\cdots)$ are the same as $\boldsymbol{\chi}_{a}(\cdots)$, but with the LO dibaryon propagator replaced by its corresponding NLO and \nnlo correction.  The NLO and \nnlo results for type (a) diagrams Eqs.~(\ref{eq:NLOatriton}), (\ref{eq:NNLOatriton}), (\ref{eq:A1}), (\ref{eq:Avec}), and (\ref{eq:Ascalar}) already contain the subtraction of diagrams (e) and (f) and therefore $\boldsymbol{\chi}_{e}(\cdots)$ and $\boldsymbol{\chi}_{f}(\cdots)$ are not shown.

\section{\label{app:QExpansion}}

Expanding the scalar function $\mathcal{A}_{n}(Q)$ as a function of $Q^{2}$ and picking out the $Q^{2}$ contribution gives
\begin{align}
\frac{1}{2}\frac{\partial^{2}}{\partial Q^{2}}\mathcal{A}_{n}(Q)\Big{|}_{Q=0}=\frac{2}{3}\int_{0}^{\Lambda}dq q^{2}f_{n}(q),
\end{align}
where
\begin{align}
f_{0}(q)=\frac{M_{N}}{384\pi^{2}}\frac{1}{\Dwd^{5}D^{4}}\left\{q^{2}(D^{2}-2D\Dwd+2\Dwd^{2})+4D\Dwd^{2}(3\Dwd-\gamma_{s})\right\},
\end{align}
\begin{equation}
f_{1}(q)=(Z_{s}-1)f_{0}(q),
\end{equation}
and
\begin{equation}
f_{2}(q)=\left(\frac{Z_{s}-1}{2\gamma_{s}}\right)^{2}\left[\left(\Dwd^{2}-\gamma_{s}^{2}\right)f_{0}(q)+\frac{M_{N}}{192\pi^{2}\Dwd^{3}D^{3}}\left\{8\Dwd^{2}D-q^{2}(\gamma_{s}-3\Dwd)\right\}\right].
\end{equation}
The variables $D$ and $\Dwd$ are given by
\begin{equation}
\Dwd=\sqrt{\frac{3}{4}q^{2}-M_{N}E}\quad,\quad D=\gamma_{s}-\Dwd.
\end{equation}
Extracting the $Q^{2}$ part of the c.c.~space vector functions $\boldsymbol{\mathcal{A}}_{n}(p,Q)$ gives
\begin{equation}
\frac{1}{2}\frac{\partial^{2}}{\partial Q^{2}}\boldsymbol{\mathcal{A}}_{n}(p,Q)\Big{|}_{Q=0}=\int_{0}^{\Lambda}dq q^{2}f_{n}(p,q)
\left(\!\!\begin{array}{r}
2 \\[-1.5 mm]
-\frac{2}{3}
\end{array}\right),
\end{equation}
where
\begin{align}
f_{0}(p,q)=&-2\pi f_{0}(q)\frac{1}{p q}Q_{0}(a)\\\nonumber
&-\frac{M_{N}}{27\pi}\frac{1}{D}\frac{1}{(pq)^{3}}\left\{\frac{5a}{(1-a^{2})^2}+\left[\left(\frac{q}{p}+\frac{p}{q}\right)(1+3a^{2})-a(3+a^{2})\right]\frac{1}{(1-a^{2})^{3}}\right\}\\\nonumber
&-\frac{M_{N}}{432\pi}\frac{1}{\Dwd^{3}D^{3}}\frac{1}{(pq)^{2}}\left\{\Dwd^{2}D\left[\frac{38}{1-a^{2}}+\left(\left(20\frac{q}{p}+8\frac{p}{q}\right)a-4(1+a^{2})\right)\frac{1}{(1-a^{2})^{2}}\right]\right.\\\nonumber
&\left.-(\gamma_{s}-3\Dwd)\frac{9}{2}\frac{q^{2}}{1-a^{2}}\right\},
\end{align}
\begin{align}
f_{1}(p,q)=&\left(\frac{Z_{s}-1}{2\gamma_{s}}\right)\left[(\gamma_{s}+\Dwd)f_{0}(p,q)-2\pi Df_{0}(q)\frac{1}{pq}Q_{0}(a)\right.\\\nonumber
&-\frac{M_{N}}{432\pi}\frac{1}{\Dwd^{3}D^{2}}\frac{1}{(pq)^{2}}\left\{\left[38\Dwd^{2}D-\frac{9}{2}q^{2}(\gamma_{s}-3\Dwd)\right]\frac{1}{1-a^{2}}\right.\\\nonumber
&\left.\left.-\Dwd^{2}D\left[4(1+a^{2})-\left(20\frac{q}{p}+8\frac{p}{q}\right)a\right]\frac{1}{(1-a^{2})^{2}}\right\}\right],
\end{align}
and
\begin{align}
f_{2}(p,q)=&\left(\frac{Z_{s}-1}{2\gamma_{s}}\right)^{2}\left[(\Dwd^{2}-\gamma_{s}^{2})f_{0}(p,q)\right.\\\nonumber
&-\frac{M_{N}}{96\pi}\frac{1}{\Dwd^{3}D^{3}}\left\{8\Dwd^{2}D-q^{2}(\gamma_{s}-3\Dwd)\right\}\frac{1}{pq}Q_{0}(a)\\\nonumber
&-\frac{M_{N}}{216\pi}\frac{1}{\Dwd D^{2}}\frac{1}{(pq)^{2}}\left\{\left[38\Dwd D+9q^{2}\right]\frac{1}{1-a^{2}}\right.\\\nonumber
&\left.\left.-\Dwd D\left[4(1+a^{2})-\left(20\frac{q}{p}+8\frac{p}{q}\right)a\right]\frac{1}{(1-a^{2})^{2}}\right\}\right].
\end{align}
The variable $a$ is defined by
\begin{equation}
a=\frac{q^{2}+p^{2}-M_{N}E}{qp}.
\end{equation}
Pulling out the $Q^{2}$ part of the c.c.~space matrix functions $\boldsymbol{\mathcal{A}}_{n}(p,k,Q)$ gives
\begin{equation}
\frac{1}{2}\frac{\partial^{2}}{\partial Q^{2}}\boldsymbol{\mathcal{A}}_{n}(p,k,Q)\Big{|}_{Q=0}=\int_{0}^{\Lambda}dq q^{2}f_{n}(p,k,q)
\left(\!\!\begin{array}{rr}
6 & -2 \\[-1.5 mm]
-2 & \frac{2}{3}
\end{array}\right),
\end{equation}
where
\begin{align}
&f_{0}(p,k,q)=-2\pi\left\{f_{0}(k,q)\frac{1}{pq}Q_{0}(a)+f_{0}(p,q)\frac{1}{kq}Q_{0}(b)\right\}-4\pi^{2}f_{0}(q)\frac{1}{kq}Q_{0}(b)\frac{1}{pq}Q_{0}(a)\\\nonumber
&\hspace{.3cm}+\frac{M_{N}}{54}\frac{1}{\Dwd D^{2}}\frac{1}{q^{4}k^{2}p^{2}}\left\{2\Dwd D \left(\left[12(1-b^{2})(1-a^{2})+4\frac{q}{p}a(1-b^{2})+4\frac{q}{k}b(1-a^{2})\right]\right.\right.\\\nonumber
&\hspace{.3cm}+2ab\left[\frac{k}{p}(1-b^{2})+\frac{p}{k}(1-a^{2})\right]+2b\frac{k}{q}\left[2b^{2}-(1+a^{2})\right]+2a\frac{p}{q}\left[2a^{2}-(1+b^{2})\right]\\\nonumber
&\hspace{.3cm}\left.+2\frac{k}{q}\left(\frac{q}{p}a-2\right)(1-b^{2})^{2}Q_{0}(b)+2\frac{p}{q}\left(\frac{q}{k}b-2\right)(1-a^{2})^{2}Q_{0}(a)\right)\frac{1}{(1-b^{2})^{2}(1-a^{2})^{2}}\\\nonumber
&\hspace{.3cm}+q^{2}\left(\left[4+\frac{k}{q}b+\frac{p}{q}a-2\frac{k}{q}\frac{p}{q}ab\right]+\frac{k}{q}(1-b^{2})\left(1-2a\frac{p}{q}\right)Q_{0}(b)\right.\\\nonumber
&\hspace{.3cm}\left.\left.+\frac{p}{q}(1-a^{2})\left(1-2b\frac{k}{q}\right)Q_{0}(a)-2\frac{k}{q}\frac{p}{q}(1-b^{2})(1-a^{2})Q_{0}(b)Q_{0}(a)\right)\frac{1}{(1-b^{2})^{2}(1-a^{2})^{2}}\right\},
\end{align}
\begin{align}
f_{1}(p,k,q)=&\left(\frac{Z_{s}-1}{2\gamma_{s}}\right)(\gamma_{s}+\Dwd)f_{0}(p,k,q)-2\pi f_{1}(k,q)\frac{1}{pq}Q_{0}(a)-2\pi f_{1}(p,q)\frac{1}{kq}Q_{0}(b)\\\nonumber
&+\left(\frac{Z_{s}-1}{2\gamma_{s}}\right)\frac{M_{N}}{54}\frac{1}{\Dwd D}\frac{1}{q^{2}k^{2}p^{2}}\left\{\left[4+\frac{k}{q}b+\frac{p}{q}a-2\frac{k}{q}\frac{p}{q}ab\right]\right.\\\nonumber
&+\frac{k}{q}(1-b^{2})\left(1-2a\frac{p}{q}\right)Q_{0}(b)+\frac{p}{q}(1-a^{2})\left(1-2b\frac{k}{q}\right)Q_{0}(a)\\\nonumber
&\left.\left.-2\frac{k}{q}\frac{p}{q}(1-b^{2})(1-a^{2})Q_{0}(b)Q_{0}(a)\right\}\frac{1}{(1-b^{2})(1-a^{2})}\right.\\\nonumber
&+2\pi \left(\frac{Z_{s}-1}{2\gamma_{s}}\right)(\gamma_{s}+\Dwd)\left[f_{0}(k,q)\frac{1}{pq}Q_{0}(a)+f_{0}(p,q)\frac{1}{kq}Q_{0}(b)\right]\\\nonumber
&-4\pi^{2}\left(f_{1}(q)-\left(\frac{Z_{s}-1}{2\gamma_{s}}\right)(\gamma_{s}+\Dwd)f_{0}(q)\right)\frac{1}{pq}Q_{0}(a)\frac{1}{kq}Q_{0}(b),
\end{align}
and
\begin{align}
f_{2}(p,k,q)=&\left(\frac{Z_{s}-1}{2\gamma_{s}}\right)^{2}(\Dwd^{2}-\gamma_{s}^{2})f_{0}(p,k,q)-2\pi f_{2}(k,q)\frac{1}{pq}Q_{0}(a)-2\pi f_{2}(p,q)\frac{1}{kq}Q_{0}(b)\\\nonumber
&+\left(\frac{Z_{s}-1}{2\gamma_{s}}\right)^{2}\frac{M_{N}}{27}\frac{1}{D}\frac{1}{q^{2}k^{2}p^{2}}\left\{\left[4+\frac{k}{q}b+\frac{p}{q}a-2\frac{k}{q}\frac{p}{q}ab\right]\right.\\\nonumber
&+\frac{k}{q}(1-b^{2})\left(1-2a\frac{p}{q}\right)Q_{0}(b)+\frac{p}{q}(1-a^{2})\left(1-2b\frac{k}{q}\right)Q_{0}(a)\\\nonumber
&\left.-2\frac{k}{q}\frac{p}{q}(1-b^{2})(1-a^{2})Q_{0}(b)Q_{0}(a)\right\}\frac{1}{(1-b^{2})(1-a^{2})}\\\nonumber
&+2\pi\left(\frac{Z_{s}-1}{2\gamma_{s}}\right)^{2}(\Dwd^{2}-\gamma_{s}^{2})\left[f_{0}(k,q)\frac{1}{pq}Q_{0}(a)+f_{0}(p,q)\frac{1}{kq}Q_{0}(b)\right]\\\nonumber
&-4\pi^{2}\left(f_{2}(q)-\left(\frac{Z_{s}-1}{2\gamma_{s}}\right)^{2}(\Dwd^{2}-\gamma_{s}^{2})f_{0}(q)\right)\frac{1}{pq}Q_{0}(a)\frac{1}{kq}Q_{0}(b).
\end{align}
The variable $b$ is defined as
\begin{equation}
b=\frac{q^{2}+k^{2}-M_{N}E}{qk}.
\end{equation}

Extracting the $Q^{2}$ part of the c.c.~space matrix function $\boldsymbol{\mathcal{B}}_{0}(p,k,Q)$ gives
\begin{align}
&\frac{1}{2}\frac{\partial^{2}}{\partial Q^{2}}\boldsymbol{\mathcal{B}}_{0}(p,k,Q)\Big{|}_{Q=0}=-\frac{2M_{N}\pi}{9}\frac{1}{p^{3}k^{3}}\frac{1}{(1-a^{2})^{2}}\\\nonumber
&\hspace{4cm}\times\left\{\frac{4}{3}\frac{a}{1-a^{2}}-2a-\frac{1}{3}\frac{p^{2}+k^{2}}{pk}\frac{1+3a^{2}}{1-a^{2}}\right\}\left(\!\!\begin{array}{rr}
-1 & 1\\[-1.5 mm]
1 & \frac{1}{3}
\end{array}\!\right),
\end{align}
where
\begin{equation}
\label{eq:avaleq}
a=\frac{p^{2}+k^{2}-M_{N}E}{pk}.
\end{equation}

The $Q^{2}$ part of the c.c.~space vector function $\boldsymbol{\mathcal{C}}_{n}(k,Q)$ is
\begin{align}
\frac{1}{2}\frac{\partial^{2}}{\partial Q^{2}}\boldsymbol{\mathcal{C}}_{n}(k,Q)\Big{|}_{Q=0}=
\left(\begin{array}{c}
2 g_{t}^{(n)}(k) \\
-\frac{2}{3}g_{s}^{(n)}(k)
\end{array}\right)^{T},
\end{align}
where
\begin{align}
g_{\{t,s\}}^{(0)}(k)=\frac{M_{N}}{384\Dwd^{5}D_{\{t,s\}}^{3}}\left\{4\Dwd^{2}D_{\{t,s\}}(2\Dwd-\gamma_{\{t,s\}})+k^{2}(\gamma_{\{t,s\}}-3\Dwd)D_{\{t,s\}}+2k^{2}\Dwd^{2}\right\},
\end{align}
\begin{align}
g_{\{t,s\}}^{(1)}(k)=&\left(\frac{Z_{\{t,s\}}-1}{2\gamma_{\{t,s\}}}\right)\left[\vphantom{+\frac{M_{N}}{192\Dwd^{4}D_{\{t,s\}}^{2}}\left\{2\Dwd^{2} D_{\{t,s\}}+k^{2}(\Dwd-D_{\{t,s\}})\right\}}(\gamma_{\{t,s\}}+\Dwd)g_{\{t,s\}}^{(0)}(k)\right.\\\nonumber
&\hspace{4cm}\left.+\frac{M_{N}}{192\Dwd^{4}D_{\{t,s\}}^{2}}\left\{2\Dwd^{2} D_{\{t,s\}}+k^{2}(\Dwd-D_{\{t,s\}})\right\}\right],
\end{align}
and 
\begin{align}
g_{\{t,s\}}^{(2)}(k)=&\left(\frac{Z_{\{t,s\}}-1}{2\gamma_{\{t,s\}}}\right)^{2}\left[\vphantom{+\frac{M_{N}}{96\Dwd^{3}D_{\{t,s\}}^{2}}\left\{2\Dwd^{2}D_{\{t,s\}}+k^{2}\left(\Dwd-\frac{1}{2}D_{\{t,s\}}\right)\right\}}(\Dwd^{2}-\gamma_{\{t,s\}}^{2})g_{\{t,s\}}^{(0)}(k)\right.\\\nonumber
&\hspace{4cm}\left.+\frac{M_{N}}{96\Dwd^{3}D_{\{t,s\}}^{2}}\left\{2\Dwd^{2}D_{\{t,s\}}+k^{2}\left(\Dwd-\frac{1}{2}D_{\{t,s\}}\right)\right\}\right].
\end{align}
For these functions and all functions below in this appendix, $a$ is given by Eq.~\ref{eq:avaleq} and the variables $\Dwd$, $D_{t}$, and $D_{s}$ are defined as
\begin{equation}
\Dwd=\sqrt{\frac{3}{4}k^{2}-M_{N}E}\quad,\quad D_{t}=\gamma_{t}-\Dwd\quad,\quad D_{s}=\gamma_{s}-\Dwd.
\end{equation}

Note the notation $\{t,s\}$ is a shorthand for two different functions one with subscript $t$ and the other with subscript $s$.  The $Q^{2}$ dependence of the c.c.~space matrix function $\boldsymbol{\mathcal{C}}_{n}(p,k,Q)$ is given by
\begin{align}
&\frac{1}{2}\frac{\partial^{2}}{\partial Q^{2}}\boldsymbol{\mathcal{C}}_{n}(p,k,Q)\Big{|}_{Q=0}=\left(
\begin{array}{cc}
2g_{t}^{(n)}(p,k) & -2g_{s}^{(n)}(p,k)\\
-6g_{t}^{(n)}(p,k) & \frac{2}{3}g_{s}^{(n)}(p,k)
\end{array}\right),
\end{align}
where
\begin{align}
g_{\{t,s\}}^{(0)}(p,k)=&-2\pi g_{\{t,s\}}^{(0)}(k)\frac{1}{pk}Q_{0}(a)\\\nonumber
&-\frac{M_{N}\pi}{54\Dwd D_{\{t,s\}}}\frac{1}{pk}\left\{\frac{1}{pk}\frac{1}{1-a^{2}}+\frac{1}{p^{2}}\left(4a+a\left(\frac{p}{k}\right)^{2}-2\frac{p}{k}(1+a^{2})\right)\frac{1}{(1-a^{2})^{2}}\right\}\\\nonumber
&-\frac{M_{N}\pi}{144}\frac{k}{p}\frac{1}{\Dwd^{3}D_{\{t,s\}}^{2}}\left\{\frac{1}{k^{2}}Q_{0}(a)-\frac{1}{pk}\frac{2-\frac{p}{k}a}{1-a^{2}}\right\}\left[\gamma_{\{t,s\}}-3\Dwd\right],
\end{align}
\begin{align}
g_{\{t,s\}}^{(1)}(p,k)=&\left(\frac{Z_{\{t,s\}}-1}{2\gamma_{\{t,s\}}}\right)\left[\vphantom{-\frac{k}{p}\frac{M_{N}\pi}{72\Dwd^{2}D_{\{t,s\}}}\left\{\frac{2}{pk}\frac{1}{1-a^{2}}-\frac{1}{k^{2}}\frac{a}{1-a^{2}}-\frac{1}{k^{2}}Q_{0}(a)\right\}}(\gamma_{\{t,s\}}+\Dwd)g_{\{t,s\}}^{(0)}(p,k)\right.\\\nonumber
&-\frac{M_{N}\pi}{96\Dwd^{4} D_{\{t,s\}}^{2}}\frac{1}{pk}Q_{0}(a)\left\{2\Dwd^{2} D_{\{t,s\}}+k^{2}(\Dwd-D_{\{t,s\}})\right\}\\\nonumber
&\left.-\frac{k}{p}\frac{M_{N}\pi}{72\Dwd^{2}D_{\{t,s\}}}\left\{\frac{2}{pk}\frac{1}{1-a^{2}}-\frac{1}{k^{2}}\frac{a}{1-a^{2}}-\frac{1}{k^{2}}Q_{0}(a)\right\}\right],
\end{align}
and
\begin{align}
g_{\{t,s\}}^{(2)}(p,k)=&\left(\frac{Z_{\{t,s\}}-1}{2\gamma_{\{t,s\}}}\right)^{2}\left[(\Dwd^{2}-\gamma_{\{t,s\}}^{2})g_{\{t,s\}}^{(0)}(p,k)\right.\\\nonumber
&-\frac{M_{N}\pi}{48\Dwd^{3} D_{\{t,s\}}^{2}}\frac{1}{pk}Q_{0}(a)\left\{2\Dwd^{2}D_{\{t,s\}}+k^{2}\left(\Dwd-\frac{1}{2}D_{\{t,s\}}\right)\right\}\\\nonumber
&\left.-\frac{k}{p}\frac{M_{N}\pi}{36\Dwd D_{\{t,s\}}}\left\{\frac{2}{pk}\frac{1}{1-a^{2}}-\frac{1}{k^{2}}\frac{a}{1-a^{2}}-\frac{1}{k^{2}}Q_{0}(a)\right\}\right].
\end{align}

Extracting the $Q^{2}$ term of the c.c.~space vector function $\boldsymbol{\mathfrak{D}}_{n}(k,Q)$ gives
\begin{align}
\frac{1}{2}\frac{\partial^{2}}{\partial Q^{2}}\boldsymbol{\mathfrak{D}}_{n}(k,Q)\Big{|}_{Q=0}=\left(\begin{array}{c}h_{t}^{(n)}(k)c_{0t}^{(0)} \\[0mm]
-\frac{1}{3}h_{s}^{(n)}(k)c_{0s}^{(0)}
\end{array}\right)^{T},
\end{align}
where
\begin{align}
h_{\{t,s\}}^{(1)}(k)=-\frac{1}{96\Dwd^{3}D_{\{t,s\}}^{3}}\left\{4\Dwd^{2}D_{\{t,s\}}+k^{2}(3\Dwd-\gamma_{\{t,s\}})\right\}
\end{align}
and
\begin{align}
h_{\{t,s\}}^{(2)}(k)=0
\end{align}
Note there is no $n=0$ value for the $\boldsymbol{\mathfrak{D}}_{n}(\cdots)$ functions.  Finally, the $Q^{2}$ piece of the c.c.~space matrix function $\boldsymbol{\mathfrak{D}}_{n}(p,kQ)$ is given by
\begin{align}
&\frac{1}{2}\frac{\partial^{2}}{\partial Q^{2}}\boldsymbol{\mathfrak{D}}_{n}(p,k,Q)\Big{|}_{Q=0}=
\left(\begin{array}{cc}
h_{t}^{(n)}(p,k)c_{0t}^{(0)} & -h_{s}^{(n)}(p,k)c_{0s}^{(0)}\\
-3h_{t}^{(n)}(p,k)c_{0t}^{(0)} & \frac{1}{3}h_{s}^{(n)}(p,k)c_{0s}^{(0)}\\
\end{array}\right),
\end{align}
where
\begin{align}
h_{\{t,s\}}^{(1)}(p,k)=&-2\pi h_{\{t,s\}}^{(1)}(k)\frac{1}{pk}Q_{0}(a)\\\nonumber
&+\frac{2\pi}{27 D_{\{t,s\}}}\frac{1}{(pk)^{2}}\left[\left(4\frac{k}{p}+\frac{p}{k}\right)a-3a^{2}-1\right]\frac{1}{(1-a^{2})^{2}}\\\nonumber
&-\frac{\pi}{18\Dwd D_{\{t,s\}}^{2}}\frac{1}{pk}\left\{Q_{0}(a)+\frac{a-2\frac{k}{p}}{1-a^{2}}\right\},
\end{align}
and
%

\begin{align}
h_{\{t,s\}}^{(2)}(p,k)=&-\left(\frac{Z_{\{t,s\}}-1}{2\gamma_{\{t,s\}}}\right)\left[D_{\{t,s\}}h_{\{t,s\}}^{(1)}(p,k)+2\pi D_{\{t,s\}}h_{\{t,s\}}^{(1)}(k)\frac{1}{pk}Q_{0}(a)\right.\\\nonumber
&\hspace{4cm}\left.-\frac{\pi}{18\Dwd D_{\{t,s\}}}\frac{1}{pk}\left\{\left[2\frac{k}{p}-a\right]\frac{1}{1-a^{2}}-Q_{0}(a)\right\}\right]
\end{align}

\section{\label{app:norm}}

Taking the limit $Q^{2}\to0$ the contribution from the LO diagram (a) is given by

\begin{align}
&-ieF_{0}^{(a)}(0)=-ie\pi^{2}M_{N}\left(\widetilde{\boldsymbol{\Gamma}}_{0}(q)\right)^{T}\otimes\frac{1}{q^{2}}\frac{\delta(q-\ell)}{\sqrt{\frac{3}{4}q^{2}-M_{N}B_{0}}}
\left(\begin{array}{cc}
0 & 0 \\[-1.5 mm]
0 & \frac{2}{3}
\end{array}\right)\otimes\widetilde{\boldsymbol{\Gamma}}_{0}(\ell)\\\nonumber
&+i2\pi eM_{N}\left(\widetilde{\boldsymbol{\Gamma}}_{0}(q)\right)^{T}\otimes
\frac{1}{q^{2}\ell^{2}-(q^{2}+\ell^{2}-M_{N}B_{0})^{2}}
\left(\!\!\!\begin{array}{rr}
0 & -2\\[-1.5 mm]
-2 & \frac{4}{3}
\end{array}\!\right)\otimes
\widetilde{\boldsymbol{\Gamma}}_{0}(\ell),
\end{align}
where
\begin{equation}
\widetilde{\boldsymbol{\Gamma}}_{0}(q)=\mathbf{D}^{(0)}\!\!\left(B_{0}-\frac{q^{2}}{2M_{N}},q\right)\boldsymbol{\Gamma}_{0}(q).
\end{equation}
In order to obtain the expression for $F_{0}^{(a)}(0)$ it is easiest to take the limit $Q^{2}\to0$ before carrying out the integration over energy.  Doing this creates a double pole that is then integrated out to lead to the expression above.  Evaluating the LO diagram (b) in the limit $Q^{2}\to0$ yields
\begin{align}
&-ieF_{0}^{(b)}(0)=-i2\pi eM_{N}\left(\widetilde{\boldsymbol{\Gamma}}_{0}(q)\right)^{T}\otimes
\frac{1}{q^{2}\ell^{2}-(q^{2}+\ell^{2}-M_{N}B_{0})^{2}}\left(\!\!\!\begin{array}{rc}
-1 & 1\\[-1.5 mm]
1 & \frac{1}{3}
\end{array}\!\right)\otimes\widetilde{\boldsymbol{\Gamma}}_{0}(\ell),
\end{align}
and for the LO diagram (c)
\begin{align}
&-ieF^{(c)}_{0}(0)=-ie\pi^{2}M_{N}\left(\widetilde{\boldsymbol{\Gamma}}_{0}(q)\right)^{T}\otimes\frac{1}{q^{2}}\frac{\delta(q-\ell)}{\sqrt{\frac{3}{4}q^{2}-M_{N}B_{0}}}
\left(\begin{array}{cc}
1 & 0 \\[-1.5 mm]
0 & \frac{1}{3}
\end{array}\right)\otimes\widetilde{\boldsymbol{\Gamma}}_{0}(\ell).
\end{align}
Combining all these terms the total LO triton charge form factor in the limit $Q^{2}\to0$ is given by
\begin{align}
&F_{0}(0)=2\pi M_{N}\left(\widetilde{\boldsymbol{\Gamma}}_{0}(q)\right)^{T}\otimes\left\{\frac{\pi}{2}\frac{1}{q^{2}}\frac{\delta(q-\ell)}{\sqrt{\frac{3}{4}q^{2}-M_{N}B_{0}}}
\left(\begin{array}{cc}
1 & 0 \\[-1.5 mm]
0 & 1
\end{array}\right)\right.\\\nonumber
&\hspace{6cm}\left.-\frac{1}{q^{2}\ell^{2}-(q^{2}+\ell^{2}-M_{N}B_{0})^{2}}
\left(\!\!\begin{array}{rr}
1 & -3 \\[-1.5 mm]
-3 & 1
\end{array}\!\right)\right\}\otimes\widetilde{\boldsymbol{\Gamma}}_{0}(\ell).
\end{align}
The normalization expression for the triton vertex function in Ref.~\cite{Konig:2011yq} is equivalent to the expression for $F_{0}(0)$ derived here.  Therefore, it automatically follows that $F_{0}(0)=1$ if the triton vertex function is properly renormalized.

\section{}
The method used to derive the corrections to the bound-state energy are rigorous but cumbersome.  An elegant way to obtain the same corrections to the bound-state energy is shown here.  The condition that the triton propagator have a bound-state pole at the triton binding energy is given by
\begin{equation}
1-H\Sigma(B)=0.
\end{equation}
In this formula $H$, $\Sigma(B)$, and $B$ represent the full non-perturbative expressions that contain corrections from all orders in \EFT.  Expanding each of these expressions perturbatively gives
\begin{align}
&1-(H_{0}+H_{1}+H_{2}+\cdots)\\\nonumber
&\hspace{2cm}\times\left[\Sigma_{0}(B_{0}+B_{1}+B_{2}+\cdots)+\Sigma_{1}(B_{0}+B_{1}+B_{2}+\cdots)\right.\\\nonumber
&\hspace{6.95cm}\left.+\,\Sigma_{2}(B_{0}+B_{1}+B_{2}+\cdots)+\cdots\right]=0,
\end{align}
where the subscript $n=0$ is LO, $n=1$ is NLO, and so on.  The term $H_{2}$ contains contributions from both $H_{\mathrm{\nnlo}}$ and the energy dependent three-body force $\widehat{H}_{2}$.  Collecting expressions order by order and solving for the bound-state energy reproduces Eqs.~(\ref{eq:NLObinding}) and (\ref{eq:NNLObinding}).  This same technique can also be used to derive the expressions in Eqs.~(\ref{eq:TNLO}) and (\ref{eq:TNNLO}).


\end{document}